\documentclass[fleqn,usenatbib]{mnras}

\usepackage{newtxtext,newtxmath}
\usepackage{xcolor}
\usepackage[T1]{fontenc}

\DeclareRobustCommand{\VAN}[3]{#2}
\let\VANthebibliography\thebibliography
\def\thebibliography{\DeclareRobustCommand{\VAN}[3]{##3}\VANthebibliography}

\usepackage{graphicx}
\usepackage{amsmath}

\title{\textit{3D code for MAgneto-Thermal evolution in Isolated Neutron Stars, MATINS:} The Magnetic Field Formalism}

\author[C. Dehman et al.]{
Clara Dehman,$^{1,2}$\thanks{E-mail: c.dehman@csic.es}
Daniele Viganò,$^{1,2,3}$
José A. Pons,$^{4}$ Nanda Rea$^{1,2}$ \\
$^{1}$Institute of Space Sciences (ICE-CSIC), Campus UAB, Carrer de Can Magrans s/n, 08193, Barcelona, Spain\\
$^{2}$Institut d'Estudis Espacials de Catalunya (IEEC), Carrer Gran Capità 2–4, 08034 Barcelona, Spain\\
$^{3}$Institute of Applied Computing \& Community Code (IAC3), University of the Balearic Islands, Palma, 07122, Spain\\
$^{4}$Departament de Física Aplicada, Universitat d'Alacant, 03690 Alicante, Spain \\ }

\date{Accepted 2022 September 22. Received 2022 September 20; in original form 2022 July 21}

\begin{document}
\label{firstpage}
\maketitle

\begin{abstract}

The long-term evolution of the internal, strong magnetic fields of neutron stars needs a specific numerical modeling. The diversity of the observed phenomenology of neutron stars indicates that their magnetic topology is rather complex and three-dimensional simulations are required, for example, to explain the observed bursting mechanisms and the creation of surface hotspots. We present \textit{MATINS}, a new three dimensions numerical code for magneto-thermal evolution in neutron stars, based on a finite-volume scheme that employs the cubed-sphere system of coordinates. 
In this first work, we focus on the crustal magnetic evolution, with the inclusion of realistic calculations for the neutron star structure, composition and electrical conductivity assuming a simple temperature evolution profile.
\textit{MATINS} follows the evolution of strong fields $(10^{14}-10^{15}$ Gauss$)$ with complex non-axisymmetric topologies and dominant Hall-drift terms, and it is suitable for handling sharp current sheets. After introducing the technical description of our approach and some tests, we present long-term simulations of the non-linear field evolution in realistic neutron star crusts. The results show how the non-axisymmetric Hall cascade redistributes the energy over different spatial scales. Following the exploration of different initial topologies, we conclude that during a few tens of kyr, an equipartition of energy between the poloidal and toroidal components happens at small-scales. However, the magnetic field keeps a strong memory of the initial large-scales, which are much harder to be restructured or created. 
This indicates that large-scale configuration attained during the neutron star formation is crucial to determine the field topology at any evolution stage.

\end{abstract}

\begin{keywords}
stars: neutron -- stars: magnetars -- stars: interiors -- stars: magnetic field -- stars: evolution
\end{keywords}

%%%%%%%%%%%%%%%%%%%%%%%%%%%%%%%%%%%%%%%%%%%%%%%%%%

%%%%%%%%%%%%%%%%% BODY OF PAPER %%%%%%%%%%%%%%%%%%

\section{Introduction}

Understanding the long-term (Myr) evolution of the strong magnetic fields embedding neutron stars (NSs), particularly magnetars, is key to unraveling the physical processes at the origin of most of their observational phenomenology. However, 
performing MHD-like simulations in stars is a difficult task, where the step from two-dimensions to three-dimensions simulations is far from trivial. 

%Finite volume methods are conservative since the flux entering a given volume is identical to that leaving the adjacent volume. Another advantage of the finite volume method is that it is easily formulated to allow for unstructured meshes. Finite volume methods can be compared and contrasted with the finite difference methods, which approximate derivatives using nodal values, or finite element methods, which create local approximations of a solution using local data, and construct a global approximation by stitching them together. In contrast a finite volume method evaluates exact expressions for the average value of the solution over some volume, and uses this data to construct approximations of the solution within cells.
The internal magnetic field evolution of isolated NSs has been largely explored in 2D simulations \citep{pons2007}, later coupled to the temperature evolution \citep{aguilera2008,pons09,vigano2012,vigano2021}. The models successfully explained the general properties of the isolated NS population \citep{vigano12b,vigano13,pons13,gullon14,gullon15}. Recent efforts were devoted to investigate the magnetic evolution without the restrictions of axial symmetry. \cite{wood2015} and \cite{gourgouliatos2016} presented the first 3D simulations of crustal-confined fields, using a pseudo-spectral code, adapted from the geo-dynamo code \textit{PARODY} \citep{dormy1998} to the NS scenario.
These simulations show new dynamics and the creation of long-living magnetic structures at a wide range of spatial scales. Even using initial axisymmetric conditions, the growth of initially tiny perturbations breaks the symmetry and non-axisymmetric modes quickly grow \citep{gourgouliatos2020}. These have typical length scales of the order of the crust thickness.

Generally speaking, for high enough magnetic fields ($B\gtrsim 10^{14}$ G), the Hall cascade keeps transferring energy to small scales \citep{gourgouliatos2016}, which in turn enhances Ohmic dissipation and eventually keeps the star hot and X-ray visible for longer timescales, as seen in 2D simulations \citep{vigano13}. Another interesting result is the formation of magnetic spots on the surface of NS \citep{gourgouliatos2018}, using extreme initial configurations previously explored in 2D \citep{geppert2014}. Very recently, \cite{grandis2020} presented the first 3D magneto-thermal evolution code with increasing physical self-consistency, applied to different sub-classes of NSs  \citep{igoshev21a,igoshev21,degrandis21}. See also \cite{pons2019} for a review of magneto-thermal evolution models.

Classically there are several approaches to the problem: pseudo-spectral methods which use spherical coordinates; finite volume/finite difference schemes which prefer Cartesian coordinates to avoid difficulties with the axis (star-in-a-box); or using a restricted grid that does not include the axis or the central region of the star. However, in many cases none of these techniques is an optimal choice for several reasons. First, fields and physical quantities usually vary more rapidly in the radial direction, so it is more convenient to distinguish the radial coordinate separated from the other two coordinates. Secondly, the surface is spherical (possible deviations from sphericity are much smaller than any other relevant scale here), and its description in Cartesian coordinates is not convenient, since this choice implies a much higher computational cost, compared to systems of coordinates that include a radial direction. This is because, on one side, in order to solve the strong radial gradients, one needs to refine all directions; on the other side, the discretization of the spherical boundaries onto the Cartesian grid introduces more spurious noise, leading to artificial modes and partially curable by increasing the resolution (see Appendix A of \cite{vigano2021} for more details). One natural choice would then be to use the spherical coordinates, as in two dimensions. However, the coordinate system does not behave regularly on the axis, resulting in a number of (sometimes compelling) numerical limitations. 

Here we employ the cubed-sphere coordinates, originally introduced by \cite{ronchi96}. Codes based on such a grid have been used to simulate many physical scenarios, such as: general circulation models for Earth or planets \citep{breitkreuz18,ding19}, general relativity \citep{lehner05,hebert18,carrasco18,carrasco19}, MHD accretion \citep{koldoba02,fragile09,hossein18}, solar wind \citep{wang19}, seismic waves \citep{vandriel21}, or dynamo in a shell \citep{yin22}.
% See  \url{https://www.gfdl.noaa.gov/fv3/fv3-documentation-and-references/}
In this paper, we use this peculiar coordinate system, adapted to the Schwarschild metric, to develop a new code designed to handle the Hall term in the induction equation for low physical resistivity.

We introduce \textit{MATINS} a new {\it three-dimensional code for MAgneto-Thermal evolution in Isolated Neutron Stars} based on a finite-volume scheme. In this work, we only present the magnetic evolution part, considering  crustal-confined magnetic fields (thus neglecting the core). As a first step, we evolve the crustal temperature with a simplified treatment, adopted from \cite{yakovlev2011}. This is at contrast with the coupled thermal and magnetic evolution, but with a simplified microphysical prescription described in the \textit{PARODY}-based published works \citep{grandis2020,degrandis21,igoshev21a,igoshev21}. Compared to those studies,
\textit{MATINS} has some distinctive features: (i) the use of the most recent temperature-dependent microphysical calculations, (ii) the use of a star structure coming from a realistic equation of state (EOS) and the inclusion of the corresponding relativistic factors in the evolution equations, (iii) the use of finite-volume numerical schemes discretized over a cubed-sphere grid.
% For what concerns points (i) and (ii), a reading of the literature about PARODY reveals that the code is in principle adaptable to deal with both the most recent microphysical calculations and realistic equations of state. However, these capabilities have not been explored in the recent works by \cite{grandis2020,degrandis21} and \cite{igoshev21a,igoshev21}, due essentially to numerical limitations.

More specifically, we implement the state-of-the-art calculations for the temperature-dependent electrical conductivity at each point of the star using Potekhin's public codes\footnote{\url{http://www.ioffe.ru/astro/conduct/}} \citep{potekhin2015}. We build the background NS model using different models of EOS at zero temperature, taken from the online public database CompOSE\footnote{\url{https://compose.obspm.fr/}} (CompStar Online Supernovae Equations of State). In particular, here we will show results that employ a Skyrme-type model of EOS, SLy4 \citep{douchin2001}. Here we will consider only one model, leaving a different choice of EOSs and masses for future studies.
% For a recent comprehensive discussions on the the core evolution, we invite the reader to check \cite{graber2015,ofengeim2017,castillo2017,gusakov2019,castillo2020,dommes2020}.

This paper is structured as follows. In section \ref{sec: cubed sphere formalism}, we briefly prescribe the NS model, the Hall induction equation, the cubed-sphere formalism applied to a Schwarschild metric, and the numerical scheme used in the three dimensions magnetic evolution code. In section \ref{sec: EMHD limit and Hall Induction Equation}, we display the 
inner and outer magnetic boundary conditions used in this study. The numerical tests and the comparison with 2D axisymmetric models are presented in section \ref{sec: numerical test}. Finally, we illustrate in section \ref{sec: results and simulations} the results obtained considering different non-axisymmetric initial configurations. We conclude and state our future lines of research in section \ref{sec: conclusion}.

\section{The Cubed Sphere Formalism with the Schwarzschild Interior Metric}
\label{sec: cubed sphere formalism}

\subsection{Background star's structure}
\label{subsec: NS structure}

Our aim is to study the global evolution of the magnetic field in isolated NSs, which are relativistic stars in which general relativity corrections are important. The structure is provided by the Tolman-Oppenheimer-Volkoff equations \citep{oppenheimer1939} which solve the hydrostatic equilibrium assuming a static interior Schwarzschild metric
\begin{equation}
    ds^2= -c^2 e^{2 \nu(r)} dt^2 + e^{2 \lambda(r)} dr^2 + r^2 d\Omega^2, 
    \label{eq: standard static metric}
\end{equation}
where $e^{2\nu(r)}$ is the lapse function that accounts for redshift corrections and it is determined by the equation
\begin{equation}
\frac{d \nu (r) }{d r } = \frac{G }{c^2}  \frac{m(r)}{r^2} \bigg(1+ \frac{4 \pi r^3 P}{C^2 m(r)} \bigg) \bigg(1- \frac{2G}{c^2} \frac{m(r)}{r} \bigg)^{-1},
  \label{eq: lapse function}  
\end{equation}
with the boundary condition $e^{2 \nu(R)} = 1- 2GM/c^2 R$ at the stellar radius $r=R$. $G$ is the gravitational constant, $c$ is the speed of light, $m(r)$ is the enclosed gravitational mass within radius r, $P(r)$ is the pressure profile and it is determined by the Tolman-Oppenheimer-Volkoff equation and
$e^{\lambda(r)}= (1 - 2Gm(r)/c^2r)^{-1/2} $ is the space curvature factor. 
The relativistic length correction $e^{\lambda(r)}$ is hereafter included in the definition of the line and surface elements of the integrals and in the operators $\boldsymbol{\nabla}$ containing the radial derivatives.

We can either prescribe a simple shell, or obtain the NS structure by using realistic EOS. In particular, we make use of the online public database CompOSE, which allows one to interpolate the provided tables using different schemes to obtain the relevant quantities, selected by the user.

By default, we build the background NS model using the Skyrme-type EOS at zero temperature, describing both the star crust and the liquid core, based on the effective nuclear interaction SLy4 \citep{douchin2001}. Considering the SLy4 EOS, we build a NS model with a radius $R_\star=11.7$ km and a mass of 1.4 $M_\odot$. The central pressure is $1.36 \times 10^{35}$ in c.g.s unit. The solution of the TOV equation determines, among other quantities, the electron number density profile and the composition, essential for our simulations.

Our computational domain covers the range from $R_c=10.9$ km to $R=11.6$ km, i.e. from the crust-core interface up to a density $\rho \sim 10^{10}$ g$/$cm$^3$, which we label as the crust-envelope interface. The envelope extends about 100 meters more, through which the diffusivity steeply increases. Therefore, the dynamical timescales get very short and computationally expensive to follow. The common assumption, that we also follow, is to assume that anyway the currents can live too shortly in the envelope. Therefore, we take the crust-envelope interface as the numerical surface, $R$.

\subsection{Patches and coordinates}\label{subsec: definition of patches}

In the cubed sphere formalism originally introduced by \cite{ronchi96}, one of the three coordinates is the radial direction, like in spherical coordinates: the volume is composed of multiple radial layers. Each layer is covered by six non-overlapping patches, which are topologically identical. The patches can be thought as the result of inflating the six faces of a cube, until it reaches a spherical shape. Therefore, each patch is bordered by four patches and is naturally described by two angular-like coordinates that play the same role of the spherical coordinates $\theta$ and $\phi$. Here we use the same notation of the original paper: the patch coordinates are $\xi$ and $\eta$, both in the range $[-\pi/4:\pi/4]$. The two coordinates are orthogonal to the radial direction, but they are non-orthogonal to each other, except at the patch centers. They cover the two directions in the same way, i.e. the patch shape is invariant for any $n\pi/2$ ($n$ integer) rotation around the center of the patch. The transformation relations between the cubes sphere, spherical and Cartesian coordinate systems are reported in Appendix \ref{app:coordinates}.

\begin{figure*}
	\centering
	\includegraphics[width=0.8\textwidth]{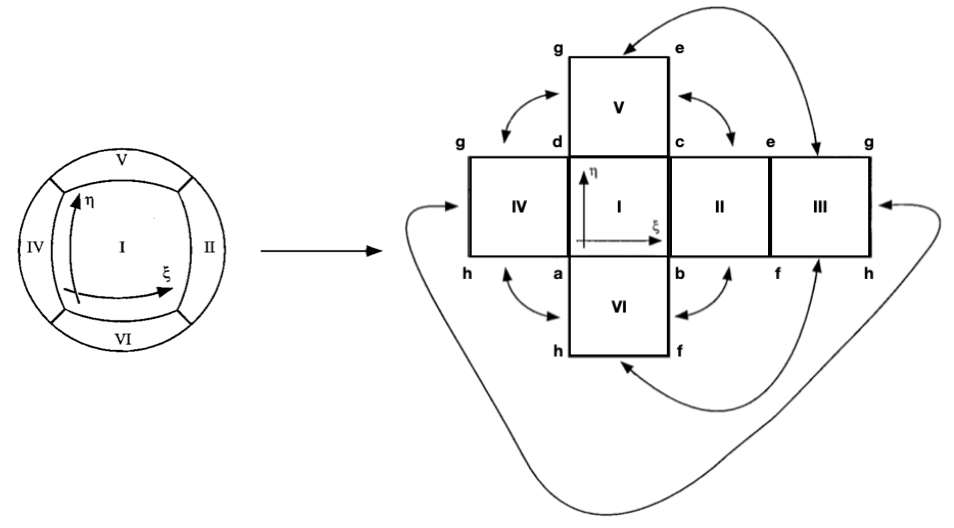}
	\includegraphics[width=0.8\textwidth]{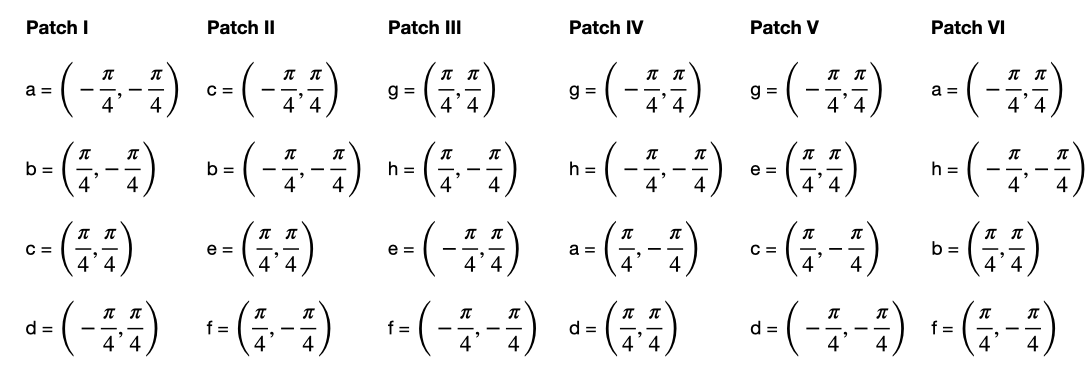}
	\caption{Exploded, cubed view of the patches \citep{ronchi96}. Each patch is identical and is described by the coordinates $\xi$ and $\eta$, both spanning the range $[-\pi/4;\pi/4]$. In the exploded view $\xi$ and $\eta$ grow to the right and upward, respectively, for all patches (only patch I is explicitly drawn here). Arrows identify the 12 edges between patches. The coordinate values $(\xi,\eta)$ of the corners for each of these patches are written in the bottom part as well.}
	\label{fig:full_grid}
\end{figure*}

\subsection{Metric}
\label{subsec: Metric}

We follow the same approach as in \cite{ronchi96}, but using Schwarzschild interior metric solution of the TOV equation. We introduce the auxiliary variables that will be used in our formalism
\begin{eqnarray}
   X &\equiv& \tan(\xi), 
   \nonumber\\    
   Y &\equiv& \tan(\eta), 
    \nonumber\\  
   \delta &\equiv& 1+ X^2 + Y^2,
     \nonumber\\
     C &\equiv& (1+ X^2)^{1/2} \equiv  \frac{1}{\cos (\xi)},
 \nonumber\\
 D &\equiv& (1+ Y^2)^{1/2} \equiv  \frac{1}{\cos (\eta)}.
   \label{eq: cubed sphere variables}
\end{eqnarray}
The metric tensor has in all patches the same functional dependence on the auxiliary variables: in the unit vector basis, it reads 
\begin{equation} 
\begin{pmatrix}
1 &0  & 0\\
0 & 1 &  - \frac{XY}{CD} \\
0 &  - \frac{XY}{CD} & 1
\end{pmatrix}
\label{eq: metric tensor}
\end{equation}

% The inverse of the tensor metric reads
% \begin{equation} g_{ij}^{-1}=
% \begin{pmatrix}
% 1 &0  & 0\\
% 0 & \frac{C^2 D^2}{\delta} &   \frac{CDXY}{\delta} \\
% 0 &  \frac{CDXY}{\delta} & \frac{C^2 D^2}{\delta}
% \end{pmatrix}
% \label{eq: inverse metric tensor}
% \end{equation}
Note that, since $X$ and $Y$ are defined differently in each patch, the metric and its inverse are of course different. In all patches, the radial versor $\hat{e}_r$ is orthogonal to the plane formed by $\hat{e}_\xi$ and $\hat{e}_\eta$ unit vectors, which are not in general orthogonal to each other.

Below, we will employ vectors using either their covariant components, denoted by lower indices, or their contravariant components, denoted by upper indices.
Let us focus first on the geometrical elements. The contravariant components of the infinitesimal length element\footnote{Note that the factor two difference with respect to \cite{ronchi96} arises because the geometrical elements used in the circulation extend twice the size of the cell (once per each side around a central point, see as an example the red solid lines in Fig.~\ref{fig: overlap region}).} at a given position $\{r,\xi,\eta\}$ are 

\begin{eqnarray}
     dl^r(r) &=& e^{\lambda(r)}dr, \nonumber\\
      dl^\xi(r,\xi,\eta) &= & \frac{2rC^2D}{\delta} d\xi ,\nonumber\\
       dl^\eta(r,\xi,\eta) &=&  \frac{2rCD^2}{\delta} d\eta .
     \label{eq: length elements contravariant}
\end{eqnarray}
We define the covariant components of the surface elements in terms of the contravariant length element: 
\begin{eqnarray}
   dS_r(r,\xi,\eta)& =& \frac{4r^2}{\delta^{3/2}} C^2D^2 d\eta d\xi,  \nonumber\\
    dS_\xi(r,\xi,\eta) & =&   \frac{2re^{\lambda(r)} D }{\delta^{1/2}}dr d\eta, \nonumber\\
    dS_\eta(r,\xi,\eta) &=&  \frac{2re^{\lambda(r)}C }{\delta^{1/2}} dr d\xi.
   \label{eq: surface elements covariant}
\end{eqnarray}
For further details on the derivation of eqs. (\ref{eq: surface elements covariant}) we refer to the Appendix, in particular eqs. (\ref{eq: covariant cross product} - \ref{eq: covariant cross product metric}). Last, the infinitesimal volume element is obtained by doing the mixed product between the three geometrical lengths:
 \begin{equation}
     dV(r,\xi,\eta) = e^{\lambda(r)} \frac{4r^2 C^2D^2}{\delta^{3/2}} dr d\xi d\eta 
     \label{eq: mixed product}
 \end{equation}

\subsection{Induction equation in neutron star crust}
\label{subsec: finite Volume schemes}
 
We study the non-linear evolution of magnetic fields in NS crusts with special attention to the influence of the Hall drift. The evolution of magnetic fields in the crust of a NS is governed by the induction equation, in short form: 
\begin{eqnarray}
\frac{\partial \boldsymbol{B}}{\partial t} = -c \boldsymbol{\nabla} \times \big( e^{\nu} \boldsymbol{E} \big).
\label{eq: induction equation compact}
\end{eqnarray} 

For our purposes (magnetic field evolution in a NS crust), the electric field resulting from a generalized Ohm's law, will be given by \citep{pons2019}:
\begin{eqnarray}
\boldsymbol{E} = \eta_b \left( \boldsymbol{J} + R_m \frac{\boldsymbol{J} \times \boldsymbol{B}}{B} \right)  ,
\label{eq: induction equation}
\end{eqnarray}
where $\eta_b = \frac{c^2}{4\pi \sigma_e}$ is the temperature- and density-dependent magnetic diffusivity, $\sigma_e$ is the electric conductivity and $R_m =  \frac{f_h B}{\eta_b}$ is the so-called magnetic Reynolds number or magnetization parameter, with $f_h = \frac{c}{4\pi e n_e}$ is the Hall-prefactor. Here $B=|\boldsymbol{B}|$, $e$ is the elementary electric charge and $n_e$ is the electron number density.
We have defined the electric current, $\boldsymbol{J}$ by
\begin{eqnarray}
\boldsymbol{J} =  e^{-\nu} \frac{c}{4\pi}
\boldsymbol{\nabla} \times (e^{\nu}\boldsymbol{B})~,
\end{eqnarray}
i.e., with the effective current being $e^{-\nu} 
\boldsymbol{\nabla} \times (e^{\nu}\boldsymbol{B})~.$

% We have defined an effective current, $\boldsymbol{J}$ by
% \begin{eqnarray}
% \boldsymbol{J} =  e^{-\nu} \frac{c}{4\pi}
% \boldsymbol{\nabla} \times (e^{\nu}\boldsymbol{B})~,
% \end{eqnarray}
% i.e., $4\pi/c$ times the electric current.

% \DV{throughout the paper, there is an inconsistency with the current defined like this: if you define J only as the curl, without the $4\pi/c$, then the $Q_j$ has to change, and also $\eta_b$, that right now is not explicitly defined (it seems it's $=1/\sigma$, which is different from the usual definition of magnetic diffusivity. I strongly suggest not to reinvent the wheel and use the usual notation, with the prefactor included in the current, so we don't need to change all definition and mess up with other notation that everybody uses.} 

The first term is the Ohmic (dissipative) term and the second is the non-linear Hall term which is the effect of the Lorentz force acting on the electrons.  
The magnetic Reynolds number is an indicator of the relative importance between the Ohmic and the Hall terms. The Hall drift dominates when the magnetic Reynolds number greatly exceeds unity, and in this case the purely parabolic diffusion equation changes its character to hyperbolic. 

The curl operator, needed to compute $\boldsymbol{J}$ and to advance $\boldsymbol{B}$, can be written in the following concise form in our non-orthogonal metric (applied to a given vector $\boldsymbol{A}$): 
    \begin{equation} 
      (\boldsymbol{\nabla} \times \boldsymbol{A}) = \frac{1}{\sqrt{g} dl^r dl^\xi dl^\eta } \begin{vmatrix}
      dl^r \boldsymbol{e_r} &  dl^\xi \boldsymbol{e_\xi} &   dl^\eta \boldsymbol{e_\eta} \\ 
     dr \frac{\partial}{\partial r} & d\xi \frac{\partial}{\partial \xi} & d\eta \frac{\partial}{\partial \eta} \\ 
      dl^r \boldsymbol{A} \cdot \boldsymbol{e_r} &  dl^\xi \boldsymbol{A} \cdot \boldsymbol{e_\xi}  &  dl^\eta \boldsymbol{A} \cdot \boldsymbol{e_\eta} 
      \end{vmatrix},
      \label{eq: curl operator cubed-sphere}
  \end{equation}
  where $\sqrt{g} =
%   \boldsymbol{e_\eta . e_r \times e_\xi}=
  \sqrt{\delta}/CD$.
%   is the orthogonality indicator.
  Explicitly, the components read:

      \begin{eqnarray}
      \big(\boldsymbol{\nabla} \times \boldsymbol{A}\big)^r &=& 
      \frac{d\xi}{\sqrt{g} dl^\xi dl^\eta}  \bigg( \frac{\partial}{\partial \xi} \big(dl^\eta \boldsymbol{A} \cdot \boldsymbol{ e_\eta}  \big) -  \frac{d\eta}{d\xi} \frac{\partial}{\partial \eta} \big(dl^\xi \boldsymbol{A} \cdot \boldsymbol{e_\xi}  \big) \bigg) 
      \nonumber\\
      &=& 
      \frac{d\xi}{dS_r} \Bigg( \frac{\partial }{\partial \xi}\bigg(dl^\eta A^\eta\bigg)    - \frac{Y}{D} \frac{\partial }{\partial \xi} \bigg( \frac{X}{C} dl^\eta A^\xi  \bigg) 
    \nonumber\\ &&   -  \frac{d\eta}{d\xi} \frac{\partial}{\partial \eta}\bigg( dl^\xi  A^\xi\bigg)  +\frac{X d\eta}{C d\xi} \frac{\partial}{\partial \eta} \bigg( \frac{Y}{D}dl^\xi  A^\eta  \bigg)\Bigg)
    \label{eq: stokes theorem radial component}
  \end{eqnarray}

 \begin{eqnarray}
           \big(\boldsymbol{\nabla} \times \boldsymbol{A}\big)^\xi &=& \frac{dr}{\sqrt{g} dl^r dl^\eta}  \bigg(\frac{d\eta}{dr} \frac{\partial}{\partial \eta} \big(dl^r \boldsymbol{A} \cdot \boldsymbol{ e_r}  \big) -  \frac{\partial}{\partial r} \big(dl^\eta \boldsymbol{A}  \cdot \boldsymbol{ e_\eta}  \big) \bigg) \nonumber\\
      &=& \frac{dr}{dS_\xi} \bigg( dl^r \frac{d\eta}{dr} \frac{\partial A^r}{\partial \eta}
       -  \frac{\partial}{\partial r}\big( dl^\eta A^\eta \big) \nonumber\\ && + \frac{XY }{CD} \frac{\partial}{ \partial r} \big(dl^\eta A^\xi \big)  \bigg)
       \label{eq: stokes theorem xi component}
  \end{eqnarray}
  
    \begin{eqnarray}
    \big(\boldsymbol{\nabla} \times \boldsymbol{A}\big)^\eta &=& \frac{dr}{\sqrt{g} dl^r dl^\xi}  \bigg( \frac{\partial}{\partial r} \big(dl^\xi \boldsymbol{A} \cdot \boldsymbol{ e_\xi}  \big) -  \frac{d\xi}{dr} \frac{\partial}{\partial \xi} \big(dl^r \boldsymbol{A} \cdot \boldsymbol{e_r}  \big) \bigg) \nonumber\\
      &=& \frac{dr}{dS_\eta} \bigg( \frac{\partial}{\partial r} \big( dl^\xi A^{\xi}\big)  - \frac{XY}{CD} \frac{\partial}{\partial r}\big( dl^\xi A^{\eta} \big) \nonumber\\ && - dl^r \frac{d\xi}{dr} \frac{\partial A^r}{\partial \xi}  \bigg) 
      \label{eq: stokes theorem eta component}
  \end{eqnarray}
where in the second equivalences we apply the Stokes theorem on an infinitesimal surface.

For any field, for output and plotting purposes we calculate the $\theta$ and $\phi$ components, using the transformations detailed in Appendix \ref{app:coordinates}.
 
%The electric field is then computed using the current and the magnetic field at each time step.

\subsection{Numerical schemes and computational features}
\label{subsec: discretization}

We use an equally spaced grid in the two angular coordinates of each patch (steps $d\xi=d\eta$), and a uniform step in the radial coordinate, $dr$, fine enough to sample the large density and field gradients in the crust. 

To evolve the magnetic field, we discretize the induction equation in the cubed-sphere coordinates, in our shell domain. Using the geometrical elements of Section \ref{subsec: Metric}, we calculate the eqs. (\ref{eq: stokes theorem radial component})-(\ref{eq: stokes theorem eta component}) in our discretized scheme. We compute the circulation as a second-order accurate line integral along the edges of a cell face and divide it by the corresponding area, like in our previous 2D codes \citep{vigano2012,vigano2021}. The surface around which the circulation is performed includes the area of the four grid cells surrounding each point (therefore, all geometrical elements related to a given point extend one cell size at both sides along the considered direction). A detailed sketch of the circulation is illustrated in red on the left hand side of Fig. \ref{fig: overlap region}. As noted in previous works (see Appendix A of \cite{vigano19}), rising the accuracy of the line integral (for instance, considering the values at the corners of the face) tends to create more numerical instabilities. Therefore, we stick to this second-order recipe.

To advance in time, we use an explicit fourth-order Runge-Kutta scheme. Other Runge-Kutta schemes are implemented, but the results are not shown here. In explicit algorithms, the stability of the method is limited by the Courant condition, which limits the timestep to ensure that the fastest wave cannot travel more than one cell length in each time step. An estimate of the maximum allowed timestep for this non-linear system can be written as:
\begin{equation}
    dt^h= k_c {\rm min}\bigg[ \frac{(\Delta l)^2 }{f_h B + \eta}\bigg]_{{\rm points}}
    \label{eq: magnetic timestep}
\end{equation}
where $k_c$ is the Courant number and it is a factor $<1$ (typically $10^{-2}-10^{-1}$), and $(\Delta l)^2 = [(dl^r)^{-2}+(dl^\xi)^{-2}+(dl^{\eta})^{-2}]^{-1}$ represents the square of the shortest resolved length scale, and the minimum is calculated over all the numerical points of the domain.

The numerical stability of the magnetic evolution in the two codes ({\it MATINS} and the 2D), for a given initial setup, seems comparable: numerical instabilities start to appear at late times, when the star cools down and consequently the dynamics become largely Hall-dominated (see \cite{vigano2021} for a more detailed discussion). This similarity with the 2D is surprising: here we don't employ the upwind-like scheme, the Burgers-like treatment for the toroidal field and the hyper-resistivity, which were all helping the numerical stability in 2D. As discussed in \cite{vigano2012,vigano2021}, in 2D all of them can be formulated and implemented in a compact way, without violating the field divergence and exploiting the axial symmetry, which allows a separation by components of the toroidal and poloidal field. In 3D, applying the same schemes is not possible by construction, and analogous more sophisticated ways to stabilize the code have not been developed so far.

{\it MATINS} is written in  {\tt Fortran90} in a modular way, with a logic and flow substantially similar to its 2D (i.e., axisymmetric) version (see \cite{vigano2021}). The microsphysics and star's structure modules, with different choices of EOSs, are indeed the same as there.

The code uses OpenMP to optimize the main loops. The computation bottlenecks are represented by the spherical harmonic decomposition needed in the boundary conditions and by the calculation of the circulation (done twice per each time sub-step). Among the two, the former takes more weight as the resolution increases. The code is faster when compiled with Intel compilers, compared to GNU. To give an idea, for the magnetic evolution simulations starting with $\sim 10^{14}-10^{15}$ G, here presented, and the typical resolution used, e.g., $N_r=40$ and $N_\xi=N_\eta=43$ per patch, the total computational time for a run of $100$ kyr is of about 7 days using six i9-10900 processors (2.80 GHz). For such a simulation, about $\sim 2.5$ million iterations are needed to reach $100$ kyr of evolution and it takes about $0.24$ s per iteration. The computational time goes up to 16 days if one utilizes one processor instead of six (i.e., scalability efficiency $16/(7\times 6)\sim 0.4$). Due to the relatively low number of points ($<10^6$ in total for the resolutions used here), the scalability with openMP is decent only up to 6 processors. Therefore, we usually use 6 processors, a number that also takes advantage of the division by 6 patches. The computational cost of the simulations is set by the large number of iterations needed (${\cal O}(10^6)$ for 100 kyr at the resolution here employed), which is in turn limited by the maximum timestep allowed, eq.~(\ref{eq: magnetic timestep}). The latter scales with the square of the resolution $\Delta l$: our computational cost rises then with $\sim (\Delta l)^5$.

Further optimization of the code is still possible and would potentially improve the performance, but will not affect the physical results shown here.

%  CLARA: DETAILS OF PROCESSORS, NOT GENERIC DESKTOP. 

% At each time step, we estimate the Courant time ($t_c$) as 
% \begin{equation}
%     t_c = min\bigg(\frac{\delta l}{u}
%     \bigg)_{i,j,k,p},
%     \label{eq: courant time step}
% \end{equation}
% where $\delta l$ is the length between mesh elements, $u$ is the velocity magnitude of the propagation of the wave and (i,j,k,p) denote the numerical cells in cubed-sphere coordinates.

% The Courant condition is then
% \begin{equation}
%     \Delta t = k_c t_c,
%     \label{eq: Courant condition}
% \end{equation}
% where $k_c$ is the Courant number and it is a factor $\leq 1$ and $t_c$ is the minimum Courant time calculated among all the numerical cells. For smaller Courant condition $\Delta t$, the code is more stable and capable of handling extreme and stiffer structures, at the cost of becoming computationally more expensive.

\subsection{Treatment of the edges between patches}
\label{subsec: edges of the patches}
% Across the edges between patches, we consider a one-cell layer of ghost cells. There, the values of the fields are obtained by applying an interpolation and considering the change of coordinates among patches. This is described in detail in App. A4

When computing the curl operator introduced in eqs. (\ref{eq: stokes theorem radial component}-\ref{eq: stokes theorem eta component}) at the edges (corners) of the patch, one needs information about the values of the functions in some points which lie in the coordinate system(s) of the neighbouring patch(es). A way to deal with this issue is to extend one layer of ghost cells in each direction, for each patch. The field components at the ghost cells are obtained by interpolating the vectors in the neighbouring patch coordinates.

\begin{figure}
     \centering
     \includegraphics[width=0.55\textwidth]{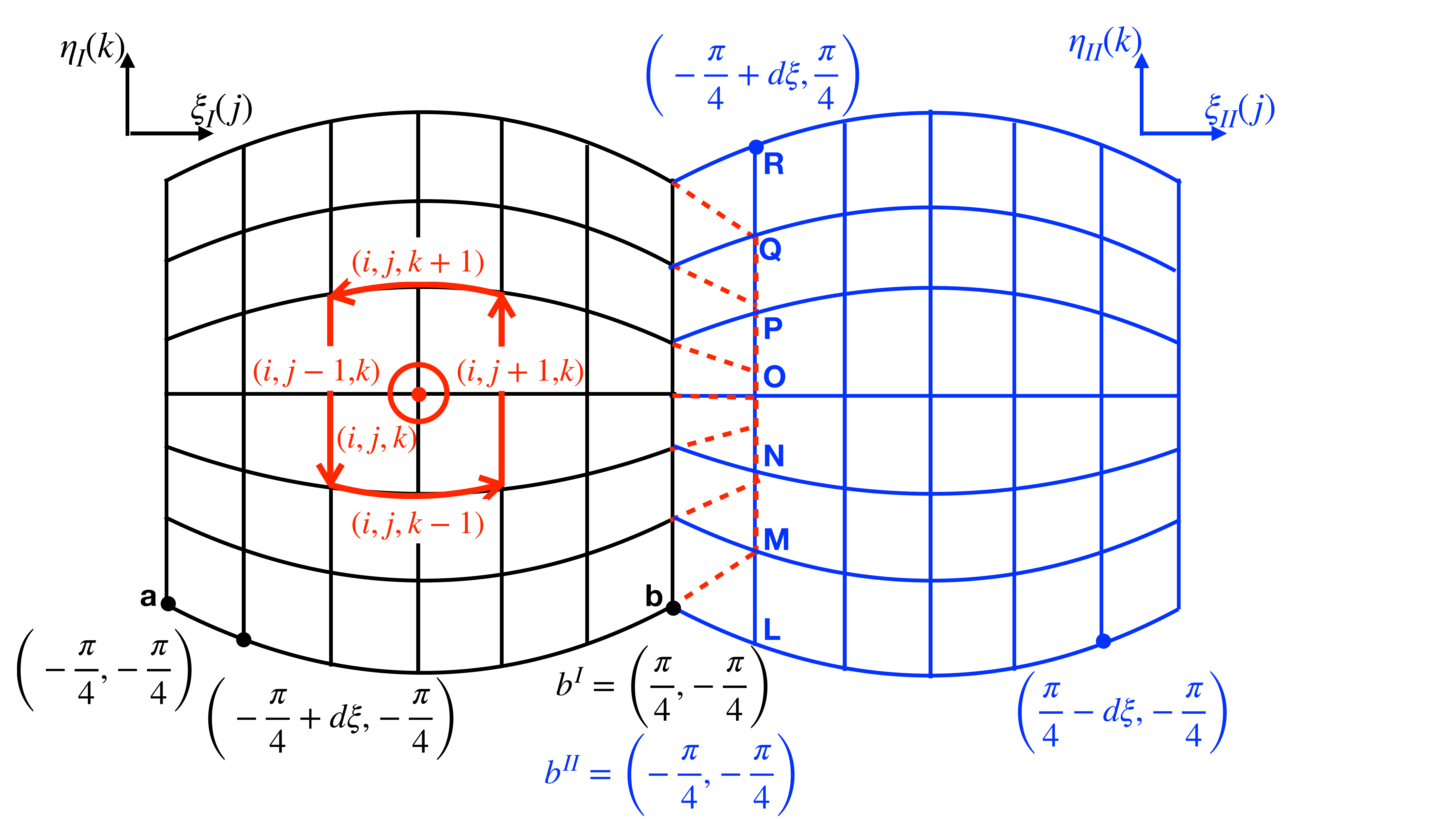}
     \caption{Schematic view of two contiguous equatorial blocks, e.g., patch I (black) and patch II (blue), and the ghosts cells of patch I (endpoints of the red dashes). The view is centered on the common vertical boundary line. The pseudo-horizontal coordinates $\xi$ of the ghost points of one grid, i.e., patch I, coincide with the second points along the $\xi$ coordinates of the last one interior grid points of the contiguous block, i.e., patch II. The ghost points are traced by the red line, and the values of the fields along the pseudo-vertical coordinate, $\eta$, are obtained by interpolations among the adjacent patch points (blue letters). Note that for other pairs of patches, the correspondence of coordinates may be less trivial (see Table \ref{tab:edges}). A sketch of a centered discretized circulation which extends twice the size of the cell (once per each side around a central point $(i,j,k)$) is displayed on the left hand side of this plot, in red. The circulation shown here is applied to calculate the radial component of the curl operator for a given vector A, i.e., $(\boldsymbol{\nabla} \times \boldsymbol{A})^r$. }
     \label{fig: overlap region}
 \end{figure}

%this paragraph is an example between patch I and II
 Fig. \ref{fig: overlap region} illustrates the mapping between two contiguous patches. Using the same regular grid size in both patches, we notice that the ghost vertical grid line in one patch (e.g., patch I in Fig. \ref{fig: overlap region} (vertical dotted red line)) coincides with the interior vertical grid line of the contiguous one (e.g., patch II in Fig. \ref{fig: overlap region}, blue vertical line). Consequently, only a one-dimensional interpolation along the vertical $\eta$ direction will be required. Note that, since $\xi$ and $\eta$ have the same grid spacing $d\xi =d\eta = \Delta$ and the same range $[-\pi/4; \pi/4]$, this idea can be applied in both vertical and horizontal directions.

We now formalise the mapping of coordinates between two different patches. Let us consider a point close to the edge between two patches. We shall call, for each patch, $p$ the value of the point coordinate parallel to the interface, and $q$ the one pseudo-perpendicular to it (since the coordinates are not orthogonal except along the central axes of each patch). We shall use the superscript $^o$ to indicate the original coordinate system (for which we know $(p^o,q^o)$). The mapping of the point in the adjacent patch (superscript $^m$) is then given by
\begin{eqnarray}
  && q^m = {\rm sgn}(q^{o}_{edge}) q^\circ - \frac{\pi}{2} \\
  && \tan (p^m) = {\rm sgn}(q^{o}_{edge})\frac{\tan(p^o)}{\tan(q^o)}
\end{eqnarray}
where ${\rm sgn}(q^{o}_{edge})$ is here used to identify the two edges in the original patch coordinate system, $q^{o}_{edge} = \pm \pi/4$, respectively. In the case of mapping ghost points to the adjacent patch where they fall into, we have a set of points with different $\{p^o\}=\{-\pi/4 + (i/(N-1))\pi/2\}$ ($i=0,N-1$) and the same $q^o = {\rm sgn}(q^{o}_{edge}) (\pi/4 + \Delta)$. Therefore:

\begin{eqnarray}
&& q^m = - {\rm sgn}(q^{o}_{edge}) \left(\frac{\pi}{4} - \Delta\right) \label{eq: mapping_edges_q} \\
&& \tan (p^m) = \frac{\tan(p^o)}{\tan \left(\frac{\pi}{4} + \Delta\right)} \label{eq: mapping_edges_p}
\end{eqnarray}
 In Table \ref{tab:edges}, we provide the correspondence of the direction for each edge, which involves two patches. The sign indicates the direction of growth of the coordinate: if they have the same (opposite) sign, the two coordinates $p$ increase in the same (opposite) way. 

\begin{table}
\centering
	\begin{tabular}{ | c | c | c | c | }			
		\hline
		edge & patches & $q$'s & $p$'s \\ \hline
        1  & I-II & $\xi^I=\pi/4$, $\xi^{II}=-\pi/4$ & $\eta^I$, $\eta^{II}$ \\
        2  & II-III & $\xi^{II}=\pi/4$, $\xi^{III}=-\pi/4$ & $\eta^{II}$, $\eta^{III}$ \\
        3  & III-IV & $\xi^{III}=\pi/4$, $\xi^{IV}=-\pi/4$ & $\eta^{III}$, $\eta^{IV}$ \\
        4  & IV-I & $\xi^{IV}=\pi/4$, $\xi^{I}=-\pi/4$ & $\eta^{IV}$, $\eta^{I}$ \\
        5  & I-V & $\eta^I=\pi/4$, $\eta^V=-\pi/4$ & $\xi^I$, $\xi^V$ \\
        6  & II-V & $\eta^{II}=\pi/4$, $\xi^V=\pi/4$ & $\xi^{II}$, $\eta^{V}$ \\
        7  & III-V & $\eta^{III}=\pi/4$, $\eta^V=\pi/4$ & $\xi^{III}$, $-\xi^{V}$ \\
        8  & IV-V & $\eta^{IV}=\pi/4$, $\xi^V=-\pi/4$ & $\xi^{IV}$, $-\eta^{V}$ \\
        9  & I-VI & $\eta^I=-\pi/4$, $\eta^{VI}=\pi/4$ & $\xi^I$, $\xi^{VI}$ \\
        10 & II-VI & $\eta^{II}=-\pi/4$, $\xi^{VI}=\pi/4$ & $\xi^{II}$, $-\eta^{VI}$ \\
        11 & III-VI & $\eta^{III}=-\pi/4$, $\eta^{VI}=-\pi/4$ & $\xi^{III}$, $-\xi^{VI}$ \\
        12 & IV-VI & $\eta^{IV}=-\pi/4$, $\xi^{VI}=-\pi/4$ & $\xi^{IV}$, $\eta^{VI}$ \\
        \hline
	\end{tabular}
	\caption{\emph{Coordinates at the 12 edges.} Identification of coordinates at the twelve edges of the cubed sphere: pair of patch numbers, pair of values of the pseudo-perpendicular coordinate $q$ identifying the interface, parallel coordinate $p$ (the one to be mapped from one patch to the other when ghost points are defined).  }
	\label{tab:edges}
\end{table}

% In order to map from one patch to the other at a given edge you apply eqs. (\ref{eq: mapping_edges_q})-(\ref{eq: mapping_edges_p}), where $q^o$ and $p^o$ are the positions in the coordinate system of the original patch, i.e., the standard grid points along $\xi$ and $\eta$ directions and $q^m$ and $p^m$ will be the mapped coordinate in the second patch. For instance, in the example above of mapping ghost cell of patch I onto patch II, you would identify the values of $(\xi^I,\eta^I)$ as $(q^o,p^o)$, in order to obtain $ (q^m,p^m)\equiv(\xi^{II},\eta^{II})$. If you instead want to map points known in patch II onto patch VI, then $(q^o,p^o)\equiv (\eta^{II},\xi^{II})$ and $(q^m,p^m)\equiv (\xi^{VI},-\eta^{VI})$.Note that ghost nodes tend to bend toward the center of the adjacent patch compared to the corresponding $\eta$ (or $\xi$) points in the adjacent patch, with a shift not larger than $\Delta$, as we can see in Fig. \ref{fig: overlap region}.

% This generalization can be used in a common subroutine with $q^o$ (one value) and $\{p^o\}$ ($N$ values) as an input (careful that the sign matters, since it will determine the indexes at play!) and the corresponding $q^m,\{p^m\}$ as outputs. The subroutine is called only once at the beginning for each edge, every time providing in input the values according to Table \ref{tab:edges}. 

Once the position of the ghost points is determined, we define a set of relative distances to the first neighbours, needed to linearly interpolate the vectors:
\begin{equation}
  W = \frac{p^m - p^o}{ \Delta} \in [0:1].
 \label{eq: interpolation weights}
\end{equation}
At the center of the edge, the distance $W$ is zero since the ghost point coincides with a point of the adjacent patch (point "O" of Fig. \ref{fig: overlap region}). Note that the set of distances is universal, valid for any pair of patches.

The vector components at the ghost points are calculated in the coordinate system of the adjacent patch as follows:
\begin{equation}
    A_\text{gAdj}^{r, \xi, \eta}  = F1_\text{Adj}^{r, \xi, \eta}\big(1 -W \big) + F2_\text{Adj}^{r, \xi, \eta} \hspace{0.5mm}W, 
\end{equation}
where $F1_\text{Adj}^{r, \xi, \eta}$ and $F2_\text{Adj}^{r, \xi, \eta}$ are the vector components at the corresponding grid points in the adjacent patch surrounding the ghost point. Importantly, the angular components of the vector $\boldsymbol{A}$, need a change of coordinates from the adjacent to the original patch by using the Jacobians detailed in Appendix \ref{Appendix: Jacobians}: 
\begin{eqnarray}
     A^{\xi} &=&  \text{JAC}(1,1) A_\text{gAdj}^{\xi} + \text{JAC}(1,2) A_\text{gAdj}^{\eta}  \nonumber\\ 
       A^{\eta} &=& \text{JAC}(2,1) A_\text{gAdj}^{\xi}  + \text{JAC}(2,2) A_\text{gAdj}^{\eta}. 
       \nonumber
\end{eqnarray}
At the edges (corners) between two (three) contiguous patches, 
there are two (three) coexisting coordinate systems, each one assigning slightly different values to the vector components. To guarantee identical field components at the egdes/corners between the patches, and to reduce numerical noise, after each timestep
we average the values of the electric currents and electric fields obtained from each patch. The appropriate change of coordinates is needed for the angular components to perform such a correction. 

% At the interface between two patches, and at the corner between three contiguous patches, we apply an averaging of the field values obtained from each patch. Each of the patch has its own coordinate system, therefore, we apply a change of coordinates from one patch to the other. Then, we average the values at the interface/corner between two/three patches in the coordinate system of one of the patches. That allows us to determine the averaged field value at the interface/corner in the coordinate system of one of the patches among the two/three patches sharing the same interface/corner. Finally, we apply the change of coordinates from one patch to the other(s) in order to determine the averaged value in the coordinate system of the other patch(es) sharing the same interface/corner. 

% \DV{If this long paragraph is just to say that you do an average, you can say in one line, stressing that the average of values at edges and corners decreases the numerical noise at the edges.}

\section{Boundary Conditions}
\label{sec: EMHD limit and Hall Induction Equation}

 \subsection{Inner boundary conditions}
 \label{subsec: inner BC}

In this paper we focus on the magnetic evolution in the crust. For simplicity, the inner boundary conditions are imposed by demanding that the normal (radial) component of the magnetic field has to vanish at $r=R_c$. Physically, this mimics the 
transition from normal to superconducting matter. We will also impose the vanishing of the tangential components of the electric field to avoid the formation of current sheets. Under such assumptions, the Poynting flux at $r=R_c$ is zero and no energy is allowed to flow into/from the core. 

We note that, when using a second-order central difference scheme for the second derivative of a function, combined with our choice of the inner boundary conditions causes a numerical problem known as odd-even decoupling or checkerboard oscillations. This results in the numerical decoupling of two slightly different solutions, one for the odd grid points, and another one for the even grid points.
In order to relieve this, we increase the radial coupling among the nearest neighbours (found at a distance $dr$), as follows: 
\begin{eqnarray}
&& E^\xi(R_c)= \frac{1}{2}  E^\xi(R_c+dr),\nonumber\\
 && E^\eta(R_c)= \frac{1}{2}  E^\eta(R_c+dr) ,\nonumber\\
&& B^\xi(R_c-dr) = \frac{R_c}{R_c-dr} B^\xi(R_c), \nonumber\\
  && B^\eta(R_c-dr) = \frac{R_c}{R_c-dr} B^\eta(R_c). 
  \label{eq: odd-even decoupling solution inner BC}
\end{eqnarray}
In the equations above, we omit the angular dimensions for clarity.
This choice reduces the tangential current at the crust-core interface and improves the stability during the evolution. 

\begin{figure}
	\centering
	\includegraphics[width=0.45\textwidth]{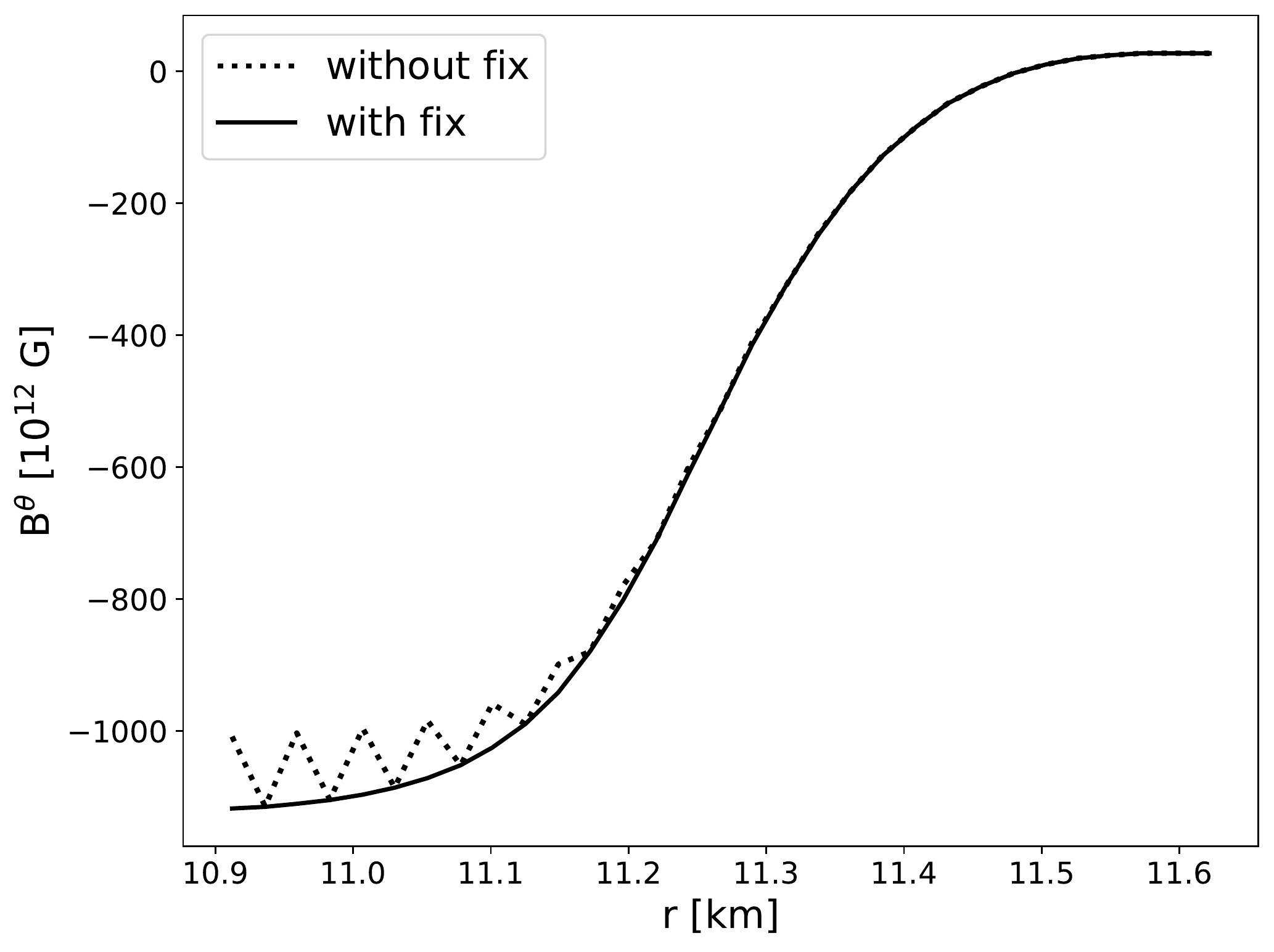}
	\caption{Difference between fixing the odd-even decoupling (eq. \ref{eq: odd-even decoupling solution inner BC}, solid line) and not (dots), for a representative evolved radial profile of a magnetic field component at a given angle. As a representative example, we show $B^{\theta}$ in the upper right corner of patch II.}
	\label{fig: odd-even decoupling}
\end{figure}

In Fig. \ref{fig: odd-even decoupling}, we illustrate a representative case of the difference in radial profile of a component with (solid line) and without the prescription above (dots).

% This choice of the inner boundary conditions causes an odd-even decoupling of cells. The odd-even decoupling problem occurs when considering a second order scheme in space, i.e., central difference scheme. 
%We couple the grid cells by tying the first two points in the radial direction to avoid odd-even decoupling, like in \cite{vigano2021}.

\subsection{Outer boundary conditions: potential field }
\label{subsec: outer B.C}

%  \clara{rephrase this section and mention that you use the Laplace spherical harmonics and cite the Blanco paper.}

The magnetosphere of a NS plays an important role in explaining several observational properties \citep{beloborodov2009,akgun2017}. Connecting the magnetic evolution in the interior of the star with its magnetosphere is outside the scope of this paper. If surface currents sheets are excluded, all components of the magnetic field are continuous through the outer boundary. 

We impose an external potential (current free) solution for the magnetic field at the surface of the star, determined by $\nabla \times B = 0$ and $\nabla \cdot B=0$. The magnetic field can be then expressed as the gradient of the magneto-static potential $\chi_m$ that satisfies the Laplace equation:
\begin{eqnarray}
   && \boldsymbol{B}= \boldsymbol{\nabla} \chi_m, \nonumber\\
   && \nabla^2 \chi_m = 0.
\end{eqnarray}
The spherical harmonics expansion of the scalar potential $\chi_m$ reads: 
 \begin{equation}
\chi_m = - B_0 R \sum_{l=0}^{\infty} \sum_{m=-l}^{m=+l} Y_{lm}(\theta,\phi) \bigg( b^m_l \bigg(\frac{R}{r} \bigg)^{l+1} + c^m_l\bigg(\frac{r}{R} \bigg)^l \bigg) 
\label{eq: magnetostatic potential}
\end{equation}
where $B_0$ is a normalization, $b^m_{l}$ corresponds to the weight of the multipoles, and $Y_{lm}$ are the spherical harmonics. In this study, we use the $Y_{lm}$ decomposition introduced by \cite{blanco1997}, since we are interested in working with the real set of spherical harmonics (Laplace spherical harmonics). The latter forms an orthonormal and complete set. One can choose real functions by combining complex conjugate functions, corresponding to opposite values of $m$. 
%  The sum is over $l$ and $m$, for $l\geq 1$ and $-l \leq m \leq l$.
 Note that we exclude $l=0$ since it corresponds to a magnetic monopole and it violates $\boldsymbol{\nabla} \cdot \boldsymbol{B}= 0$. The dimensionless weights $b_{l}^{m}$ and $c_{l}^m$ are associated to $l$ and $m$ multipoles of two branches of solutions. The second branch, $\propto \big(r/R \big)^l$, diverges for a domain extending to $r ~ \rightarrow \infty$, like the magnetosphere, therefore we set $c_l^m=0$. 
 
The normal components of the magnetic field $B^r$ are evolving and known at the surface of the star at each timestep. But to impose potential boundary conditions we need to determine the angular components of the magnetic field at the surface and one cell above the surface of the star. We proceed as follows.

Continuity of $B^r$ across the surface allows one to write it in terms of the magneto-static potential as:
 \begin{equation}
     B^r= \frac{1}{e^{\lambda(r)}}\frac{\partial \chi_m}{\partial r} = \frac{B_0}{e^{\lambda(r)}} \sum_{l=0}^{\infty} \sum_{m=-l}^{m=+l} (l+1) Y_{lm}(\theta,\phi)  b^m_l \bigg(\frac{R}{r} \bigg)^{l+2}.
     \label{eq: spectral decomposition of Br}
 \end{equation}

Then, we evaluate the weights of the multipoles $b_l^m$ by applying the orthogonality properties of spherical harmonics to eq. \ref{eq: spectral decomposition of Br}, obtaining: 
 \begin{equation}
     b^m_l = \frac{e^{\lambda(R)}}{B_0(l+1)} \int \frac{dS^r}{r^2} B^r Y_{lm}(\theta,\phi).
 \end{equation}
From this, the angular components of the magnetic field for $r\geq R$ can be evaluated: 
 \begin{equation}
    B^{\theta} = - B_0 \sum_{l=0}^{\infty} \sum_{m=-l}^{l} b^m_l \bigg(\frac{R}{r}\bigg)^{l+2} \frac{\partial Y_{lm}(\theta,\phi)}{\partial \theta}.
 \end{equation}
 
\begin{equation}
    B^{\phi} =   - \frac{B_0}{sin(\theta)} \sum_{l=0}^{\infty} \sum_{m=-l}^{l} b^m_l \bigg(\frac{R}{r}\bigg)^{l+2} \frac{\partial Y_{lm}{\theta,\phi}}{\partial \phi},
\end{equation}
which are then converted into the $B^\xi$ and $B^\eta$ components in the code.

Finally, analogously to what described for the inner boundary (section \ref{subsec: inner BC}), we prevent the radial odd-even decoupling at the surface by setting the values of the tangential components of the magnetic field as the average between the values one point above and below the surface.
%  As described in section \ref{subsec: inner BC}, a solution for the decoupling of the even and odd grid's cells in the radial direction is needed. For this purpose the following solution is applied at the surface of the star ($r(nr)=R$) to avoid odd-even decoupling.
%  \begin{eqnarray}
%     B^{\xi} (nr) &=& \frac{1}{2} \bigg( B^{\xi} (nr-1) + B^{\xi} (nr+1)  \bigg),    \nonumber\\
% B^\eta(nr) &=& \frac{1}{2} \bigg( B^{\eta} (nr-1) + B^{\eta} (nr+1)  \bigg).  
% \label{eq: odd-even decoupling outer BC}
%  \end{eqnarray}
%  Note that in eq. \ref{eq: odd-even decoupling outer BC}, we omit the angular dimensions for illustration purposes. 

\section{Numerical Tests}
\label{sec: numerical test}

\subsection{Diagnostics}
\label{subsec: diagnostics}

A necessary test for any numerical code is to check the instantaneous (local and global) energy balance. Any type of numerical instability usually results in the violation of the energy conservation, or any other physical constraint (the divergence condition). Therefore a careful monitoring of the energy balance is performed.
The magnetic energy balance equation for Hall eMHD can be expressed as :
\begin{equation}
    \frac{\partial }{\partial t} \bigg(e^{\nu} \frac{B^2}{8\pi} \bigg) = - e^{2\nu}Q_j - \boldsymbol{\nabla} \cdot \big( e^{2\nu} \boldsymbol{S} \big),
    \label{eq: energy balance}
    \end{equation}
    where $Q_j= 4\pi \eta_b J^2/c^2$ is the Joule dissipation rate and $\boldsymbol{S} = c\boldsymbol{E} \times \boldsymbol{B}/4\pi$ is the Poynting vector.

Integrating eq. (\ref{eq: energy balance}) over the whole volume of the numerical domain, we obtain the balance between the time variation of the total magnetic energy $E_{mag} = \int_V (e^{\nu} B^2/8\pi) dV$, the Joule dissipation rate $Q_{tot} = \int_V e^{2\nu} Q_j dV$, and the Poynting flux through the boundaries $S_{tot} = \int_S e^{2\nu} \boldsymbol{S} \cdot \hat{n}dS $. In our case, the boundaries are the star surface and the crust-core interface, so that $S_{tot}$ is given by the integration of $S_r$ over them. Thus, the volume-integrated energy balance is
\begin{equation}
    \frac{d}{dt} E_{mag} + Q_{tot} + S_{tot} = 0.
    \label{eq: volume-integrated energy balance}
\end{equation}

We also calculate the local magnetic field divergence in the cubed sphere coordinates by using Gauss' theorem:
  \begin{equation}
        \boldsymbol{\nabla} \cdot \boldsymbol{B} 
       = \frac{1}{dV}\bigg[\frac{\partial}{\partial r} (dS_r B^r ) + \frac{\partial}{\partial \xi} (dS_\xi B^\xi)  + \frac{\partial}{\partial \eta} (dS_\eta B^\eta )  \bigg].
      \label{eq: div(B)}
  \end{equation}
Starting from an initial divergence-free magnetic field (see section \ref{appendix: initial conditions} for more details), we monitor that indeed the divergence of the magnetic field does not grow in time above some tolerable error. To measure this, we compare the volume integral of $(\nabla\cdot\boldsymbol{B})^2$
\begin{equation}
    D_d=\int \big(\boldsymbol{\nabla} \cdot \boldsymbol{B} \big)^2 dV
    \label{eq: divB volume}
\end{equation}
to a physical quantity with the same units and scaling, e.g., the integrated values of the square of the effective current 
\begin{equation}
   D_J=\int [\boldsymbol{\nabla}\times (e^{\nu}\boldsymbol{B})]^2 dV~,
    \label{eq: J2-star}
\end{equation}
or to $(B/<dl>)^2$, where $<dl>$ is the geometrical mean of the cell's edge lengths (see also \cite{vigano19} for a related discussion). 
We verify that during the evolution, the divergence of the magnetic field always keeps several orders of magnitude smaller than the other quantities, throughout the star. 

A detailed analysis of the spectral energy distribution is performed in this study. The explicit calculation of this quantity is done using the poloidal and toroidal decomposition of the magnetic field described (see Appendix \ref{appendix: Poloidal and toroidal decomposition}).
The magnetic energy content in each mode, including the relativistic corrections, can be written as 
 \begin{equation}
    %  E_\text{spectral} = \frac{1}{8\pi} \int e^{\nu} dr  \sum_{lm}  l(l+1) \bigg[  \frac{l(l+1)}{r^2} \Phi_{lm}^2 + \bigg( \frac{\partial \Phi_{lm}}{\partial r}\bigg)^{*2} +\Psi_{lm}^2  \bigg]
         E_{lm} = \frac{1}{8\pi} \int  e^{\lambda+\nu} dr ~l(l+1) \bigg[  \frac{l(l+1)}{r^2} \Phi_{lm}^2 +\big(\Phi'_{lm}\big)^2   +\Psi_{lm}^2  \bigg],
     \label{eq: spectral magnetic energy}
 \end{equation}
where $\Phi'_{lm}$ is the radial derivative of $\Phi_{lm}$, explicitly given by eq. (\ref{eq: radial derivative of the Phi scalar function}).
%  \begin{equation}
% \Phi'_{lm} = e^{-\lambda} \frac{\partial \Phi_{lm}}{\partial r} + \frac{1 - e^{-\lambda}}{r} \Phi_{lm}.
%   \label{eq: radial derivative of the Phi scalar function}
%  \end{equation}
 The first two terms in eq. (\ref{eq: spectral magnetic energy}) account for the poloidal magnetic energy and the last term accounts for the toroidal energy.  The total energy density is simply $\sum_{lm}  E_{lm}$. 

\subsection{The purely resistive benchmark}
\label{subsec: bessel test}

A classical benchmark test that admits analytical solutions to compare with is the evolution of axisymmetric modes under Ohmic dissipation only (zero magnetic Reynolds number) and  constant magnetic diffusivity $\eta_b$. The induction equation in this limit reads 
\begin{equation}
    \frac{\partial \boldsymbol{B}}{\partial t} = - \eta_b \boldsymbol{\nabla} \times ( \boldsymbol{\nabla}  \times \boldsymbol{B} ).
    \label{eq: ohmic induction eq}
 \end{equation}
 The Ohmic eigenmodes consist of force-free solutions  satisfying $\boldsymbol{\nabla} \times \boldsymbol{B} = \alpha \boldsymbol{B}$, where $\alpha$ is a constant parameter. Then, we have 
\begin{equation}
    \frac{\partial \boldsymbol{B}}{\partial t} = - \eta_b \alpha^2  \boldsymbol{B}.
    \label{eq: bessel induction eq}
\end{equation}
which shows that each component of the magnetic field decays exponentially with the same diffusion timescale $\tau_d= 1/(\eta_b \alpha^2)$.
\begin{equation}
    \boldsymbol{B}(t) = e^{-t/{\tau_d}}\boldsymbol{B}(t=0).
    \label{eq: bessel decay in time}
\end{equation}
Note that the evolution of each component is decoupled in this case. A solution of eq. (\ref{eq: bessel induction eq}) are the spherical Bessel functions. For more details, we refer the reader to section 5.4 of \cite{pons2019}. 

% In spherical coordinates, the solutions of eq. \ref{eq: bessel induction eq} are described by factorized functions,
% which radial parts involve the spherical Bessel functions. The regularity condition
% at the center selects only one branch of the spherical Bessel functions (of the
% first kind), which, for the $(l,m) = (1,0)$ mode are
% \begin{eqnarray}
%     B^r &=& \frac{B_0 R}{r} cos\theta \bigg( \frac{sin x}{x^2}  - \frac{cos x}{x}     \bigg), \nonumber\\
%       B^\theta &=& \frac{B_0 R}{2r} sin\theta \bigg( \frac{sin x}{x^2}  - \frac{cos x}{x} - sin x    \bigg), \nonumber\\
%          B^\phi &=& \frac{\alpha R B_0 }{2} sin\theta \bigg( \frac{sin x}{x^2}  - \frac{cos x}{x} \bigg),
%          \label{eq: B bessel fct}
% \end{eqnarray}
%   where $B_0$ is the normalization factor, $x= \alpha r$ is a dimensionless quantity, and $R=10$ corresponds to the maximum radius of the spherical shell.

% In the limit where $x$ $-> 0$, we recover the solution corresponding to a homogeneous field aligned with the magnetic axis 
% \begin{eqnarray}
% %    \boldsymbol{B} -> \frac{k}{3} B_0 \hat{z}
%         B^r &=& k B_0 \frac{cos\theta}{3}, \nonumber\\
%          B^\theta &=& - k B_0 \frac{sin\theta}{3}, \nonumber\\
%           B^\phi &=& 0
%           \label{eq: bessel x->0}
% \end{eqnarray}

\begin{figure*}
\centering
\includegraphics[width=0.33\textwidth]{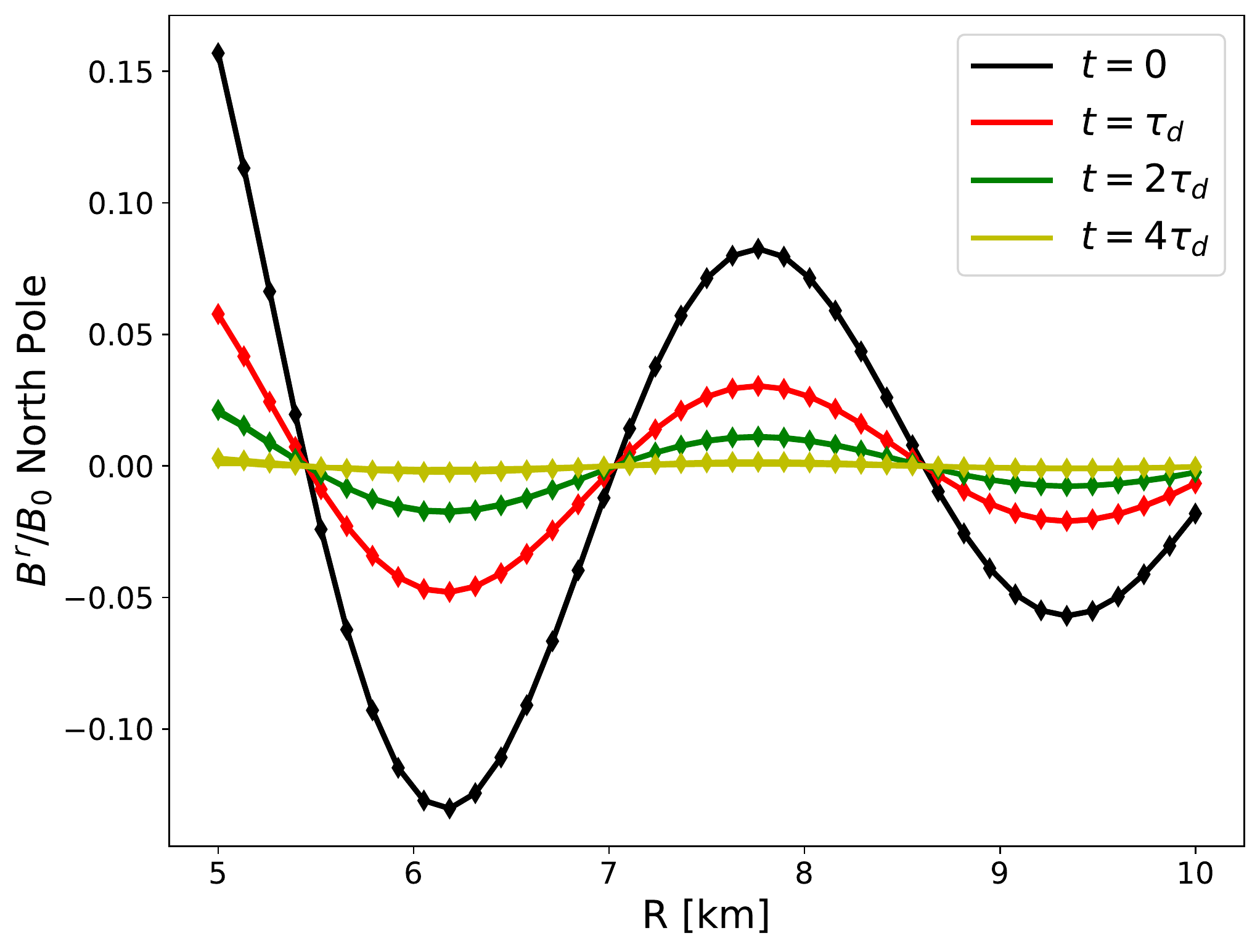}
\includegraphics[width=0.33\textwidth]{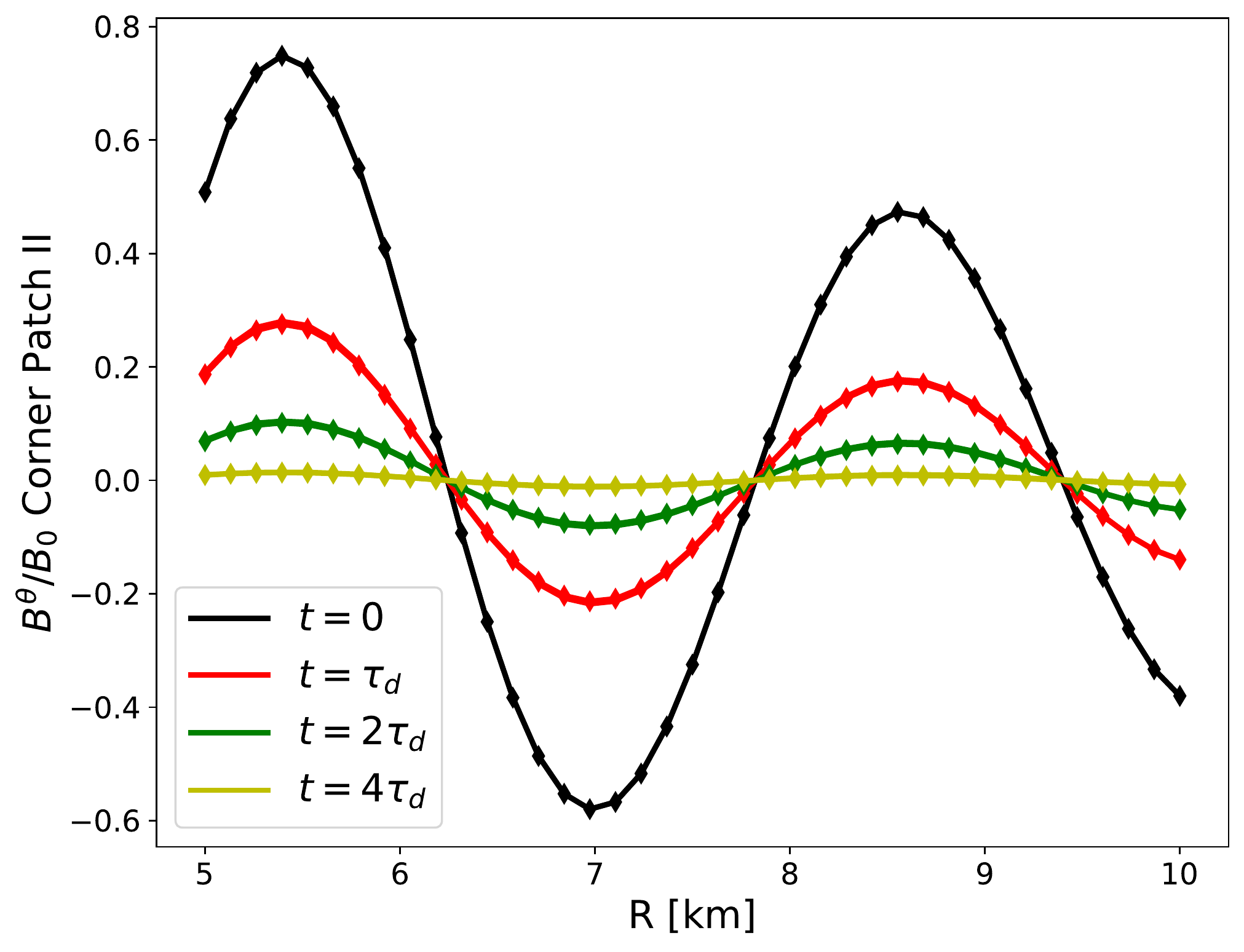} 
\includegraphics[width=0.33\textwidth]{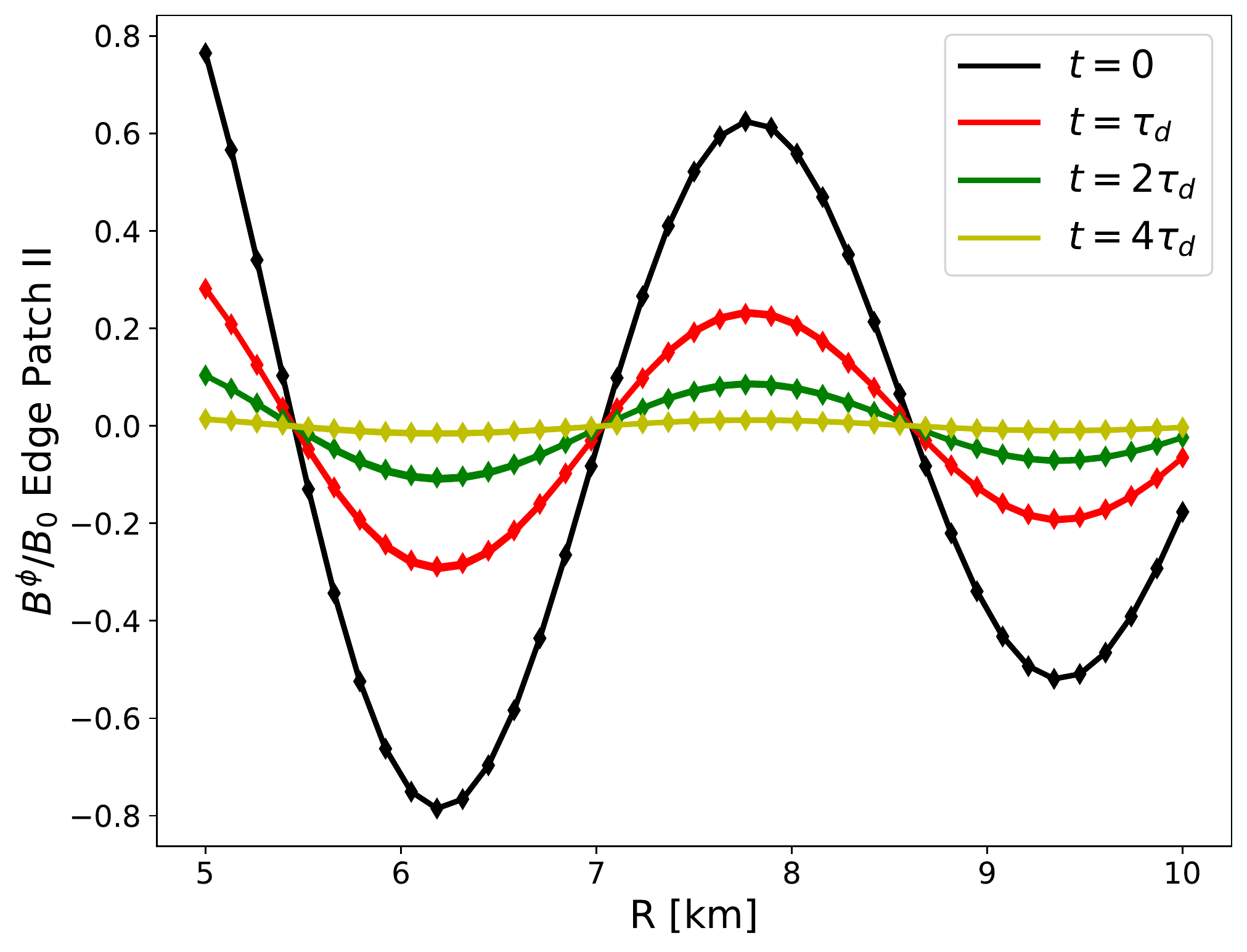}
\caption{Bessel test: radial profiles of the magnetic field components at different Ohmic timescales, with $\tau_d=0.25$ Myr. A comparison of the numerical (solid lines) and the analytical (diamonds) solutions for a model with wave-number in unit length, $\alpha = 2$ km$^{-1}$, in a spherical shell of radius $[R_c= 5:R=10]$ km.
% , with a resolution of $40\times43\times43$ per patch in the cubed-sphere coordinates. 
\textit{LHS panel:} $B^r/B_0$ radial profile at the north pole of the star. \textit{Central panel:} $B^\theta/B_0$ radial profile at the upper right corner of patch 2 (equatorial patch). \textit{right panel:} $B^\phi/B_0$ radial profile at the bottom left border of patch 2, with $B_0$ a normalization factor.}
\label{fig:Bessel test alpha=2}
\end{figure*}

Considering the spherical Bessel functions as initial conditions, and imposing the analytical solutions for $B^r$, $B^\xi$, and $B^\eta$ as boundary conditions, we follow the evolution of the modes during several diffusion timescales.

\begin{figure}
\includegraphics[width=.45\textwidth]{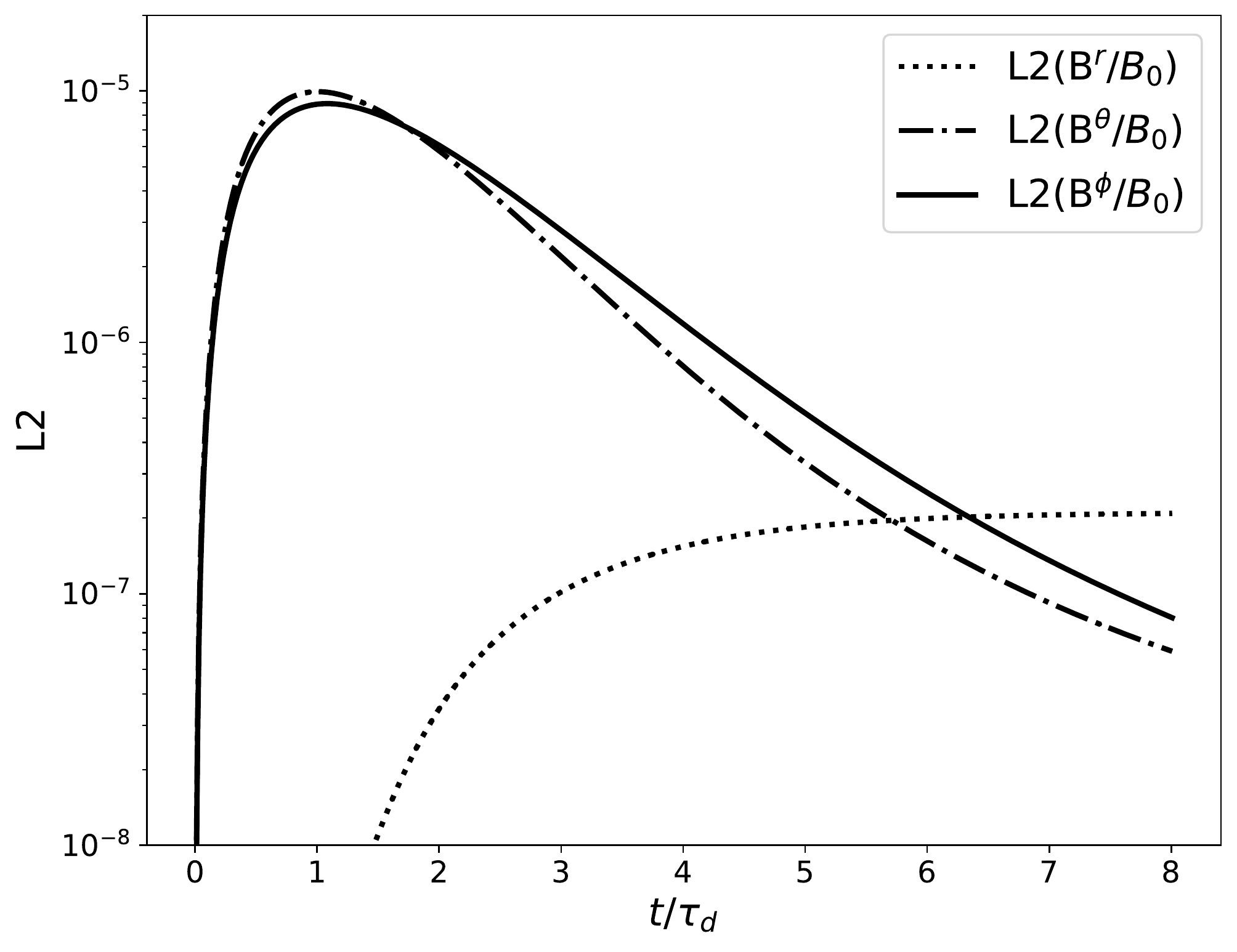}
\caption{The average absolute $L^2$ error as a function of $t/\tau_d$, 
for $B^r/B_0$ (dots), $B^\theta/B_0$ (dot-dashed lines) and $B^\phi/B_0$ (solid lines) components of the magnetic field. We have verified that the maximum absolute $L^2$ error is of the same order as the average one. The values are much smaller than the mean square of the initial magnetic field, which is of order $0.1 B_0^2$.  }
\label{fig: relative error of the Bessel test}
\end{figure} 

Fig. \ref{fig:Bessel test alpha=2} compares the numerical (solid lines) and analytical (diamonds) solutions of the magnetic field components for a magnetic field of order one, at different diffusion timescales, for a model with $\alpha = 2$ km$^{-1}$ , in a spherical shell defined by $r\in [5:10]$ km, with a resolution of $N_r= 40$ and $N_\xi=N_\eta= 43$ points per patch in the cubed-sphere coordinates. One can notice that the magnetic field has decreased below the visible scale in the figure around 4 $\tau_d$. Moreover the analytical and numerical results are indistinguishable in the graphic.

To quantify the deviation, we evaluate the average $L^2$ absolute error, in terms of deviation from the analytical solution, shown in Fig. \ref{fig: relative error of the Bessel test} for $B^r$ (dots), $B^\theta$ (dash-dotted lines) and $B^\phi$ (solid lines). The angular field components show a higher error than the radial one. That is due to the patchy grid employed. The $L^2$ error saturates after one diffusion timescale for the two angular field components, and after two diffusion timescales for the radial one. We have checked that by varying the resolution, the errors scale with $\Delta^2$, validating that the method is of second order.

\subsection{A comparison between the 2D and the 3D magnetic codes}
\label{subsec: axisymmetric - comparison 2D and 3D}

For the general case including the Hall term and with variable diffusivity and electron density, no analytical solution is available. However, since extensive results from 2D simulations are available,
a detailed comparison of the 3D magnetic code presented here and the 2D code \cite{vigano2012,vigano2021} developed by our group, helps to probe the validity of the results of the 3D code. 

For this comparative purpose, we employ analytical, fixed radial profiles for the magnetic diffusivity $\eta_b$ and the Hall prefactor $f_h$, in both codes. For the Hall prefactor $f_h$, we use the following fit adopted from \cite{vigano2021}
\begin{equation}
f_{h,\text{fit}} = 0.011 ~e^{k(\tilde{r}-\tilde{R_{c}})^b}
\left[ \frac{{\mathrm km}^2}{{\mathrm Myr}~ 10^{12} {\mathrm G}} \right]
    \label{eq: fh radial profile}
\end{equation}
where $k=10$, $b=1.8$, $\tilde{r}$ and $\tilde{R}_c$ are $r$ and $R_c$ given in km. This radial profile exhibits a super-exponential rise of about three orders of magnitude throughout the crust.

For the magnetic diffusivity $\eta_b$, we use the analytical radial profile 
\begin{equation}
\eta_b(r) = 6 \frac{(r-R_c)^{k_0}}{(R-R_c)^{k_0}} + 3 \frac{(R-r)^{k_1}}{(R-R_c)^{k_1}}  
\left[\mathrm{ \frac{km^2} {Myr}} \right],
    \label{eq: etab radial profile}
\end{equation}
with $k_0=3.5$ and $k_1=4$.

The initial magnetic field is an axisymmetric crustal-confined field with a poloidal dipole of $10^{14}$ G at the polar surface and a toroidal component consisting of a sum of a quadrupole and an octupole with a maximum value of $10^{15} $ G.

We use a grid resolution of $N_r = N_{\xi}= N_{\eta} = 30$ per patch (meaning 61 points from pole to pole and 120 along the equator). A similar resolution is used in the 2D code, e.g.,  $N_{\theta}=60$ and $N_r=30$.

\begin{figure*}
\centering
\includegraphics[width=0.33\textwidth]{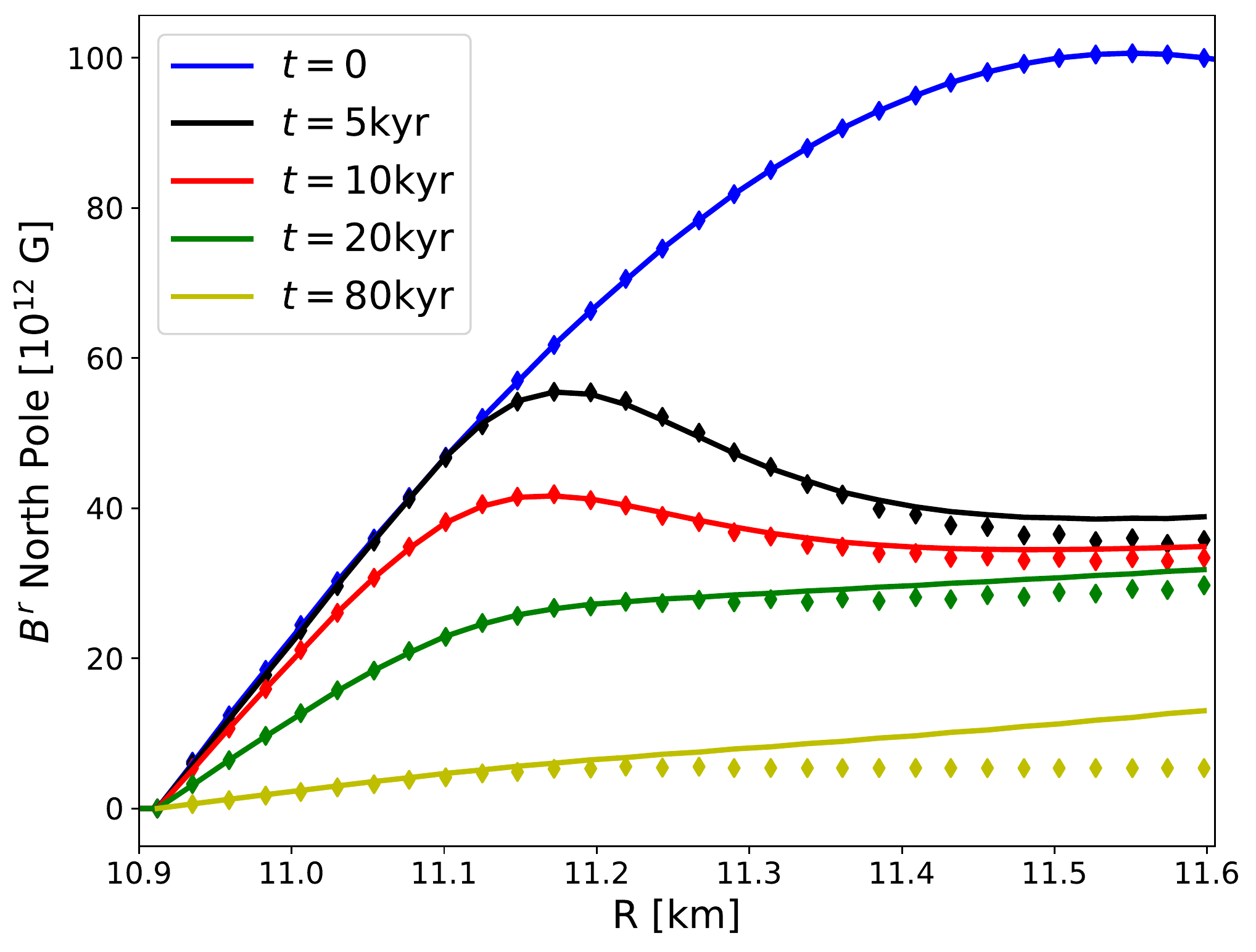}
\includegraphics[width=0.33\textwidth]{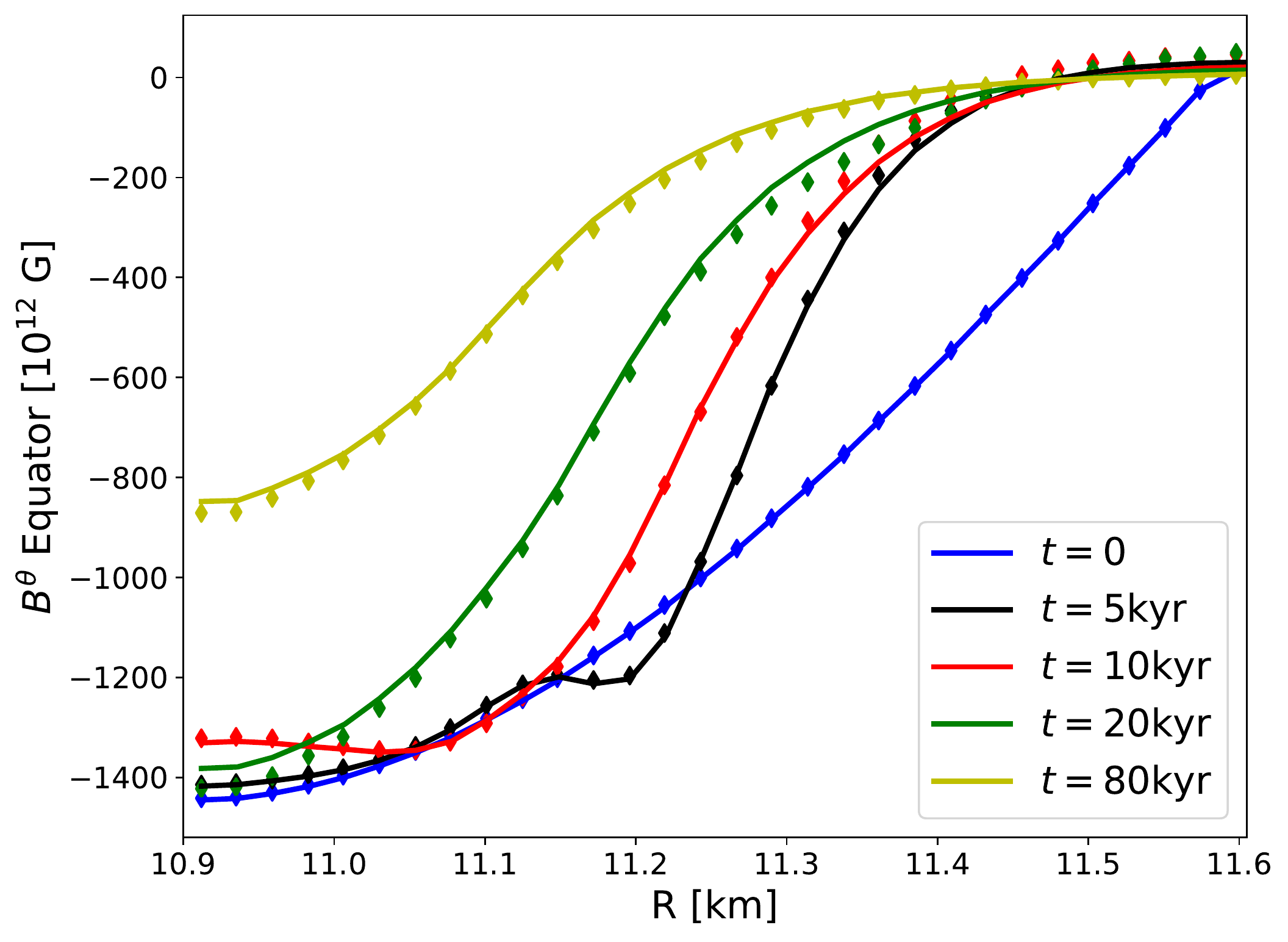} 
\includegraphics[width=0.33\textwidth]{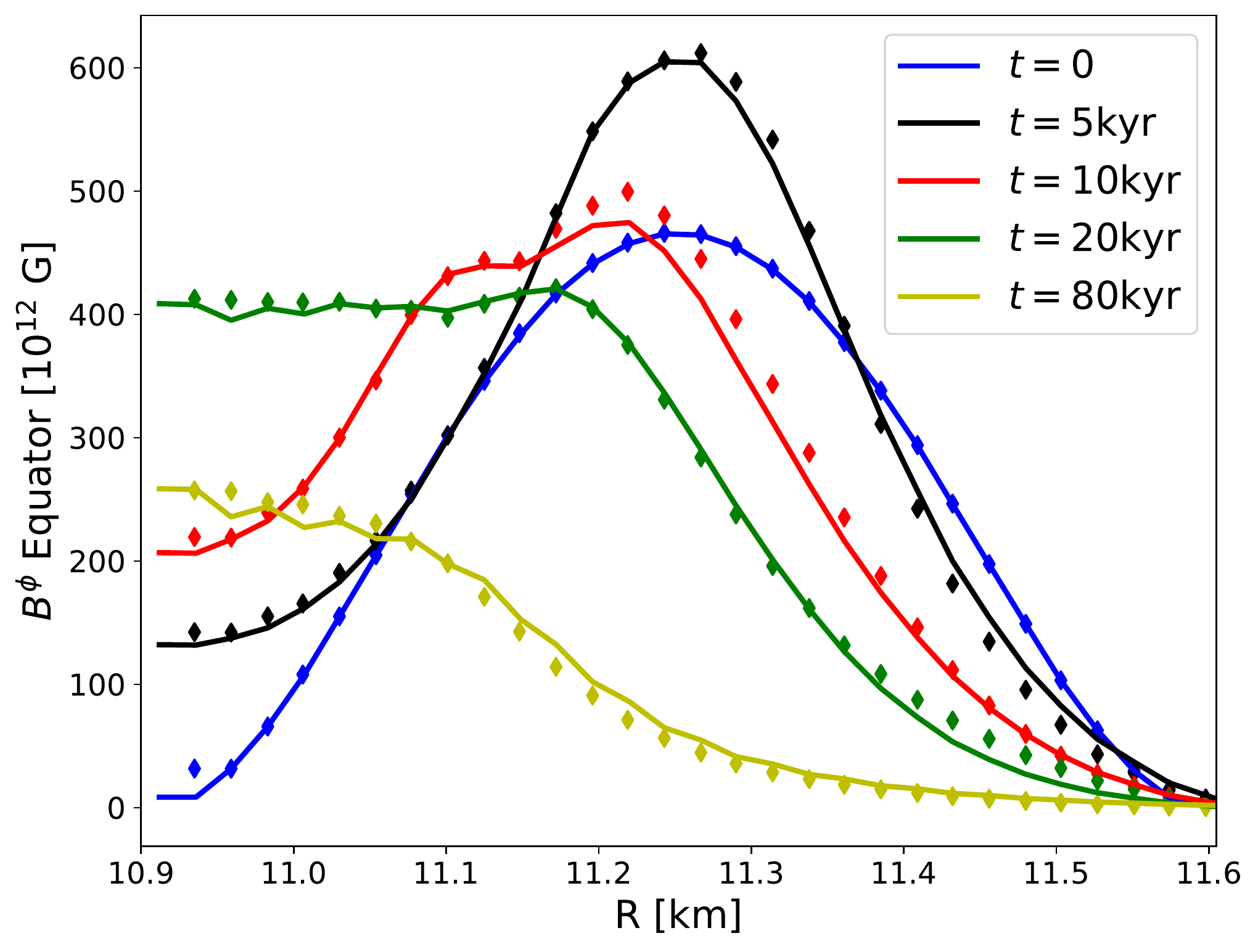}
\caption{ Radial profiles of the magnetic field components. On the left, we show the radial profile of the normal component of the magnetic field $B^r$ at the north pole of the star. In the center and on the right, we illustrate the radial profiles of the two angular components of the magnetic field $B^{\theta}$ and $B^{\phi}$ respectively, at the equator of the star. The solid lines correspond to the results of the 3D code, whereas the diamonds correspond to those of the 2D code. Different colors correspond to different evolution time.}
\label{fig: 2d-3d comparison Bfield profiles}
\end{figure*}

\begin{figure}
	\centering
	\includegraphics[width=0.45\textwidth]{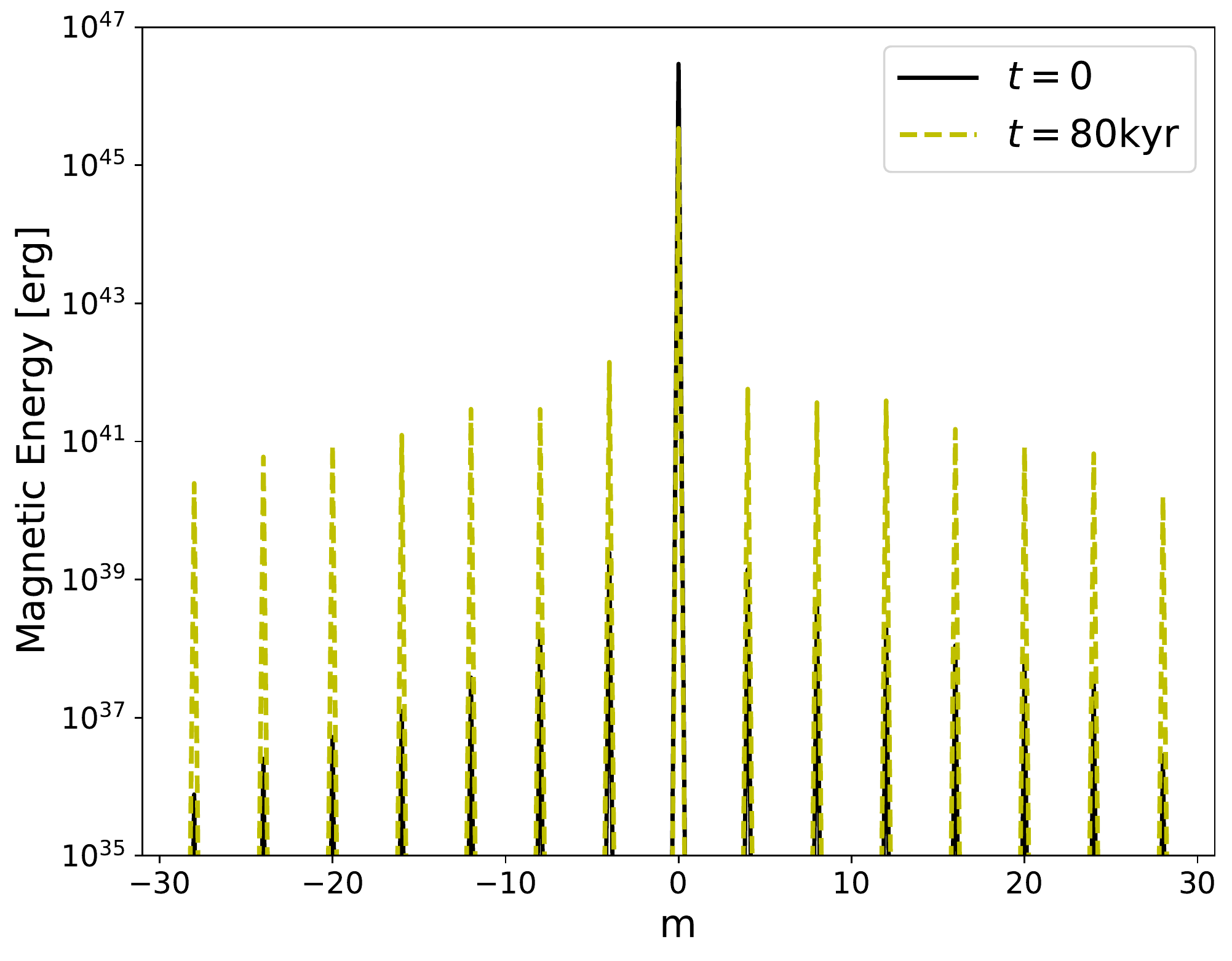}
	\caption{Evolution of the magnetic energy spectrum as a function of $m$ (summed over all $l$), for the axisymmetric case. We show $t=0$ (in black) and $t=80$ kyr (in yellow). Throughout the evolution, the power contained in all $m \neq 0$ modes never grows above $10^{-4}$ the power in the $m=0$ mode.}
	\label{fig: 3d axisymmetric m energy spectrum}
\end{figure}

\begin{figure*}
    \includegraphics[width=.33\textwidth]{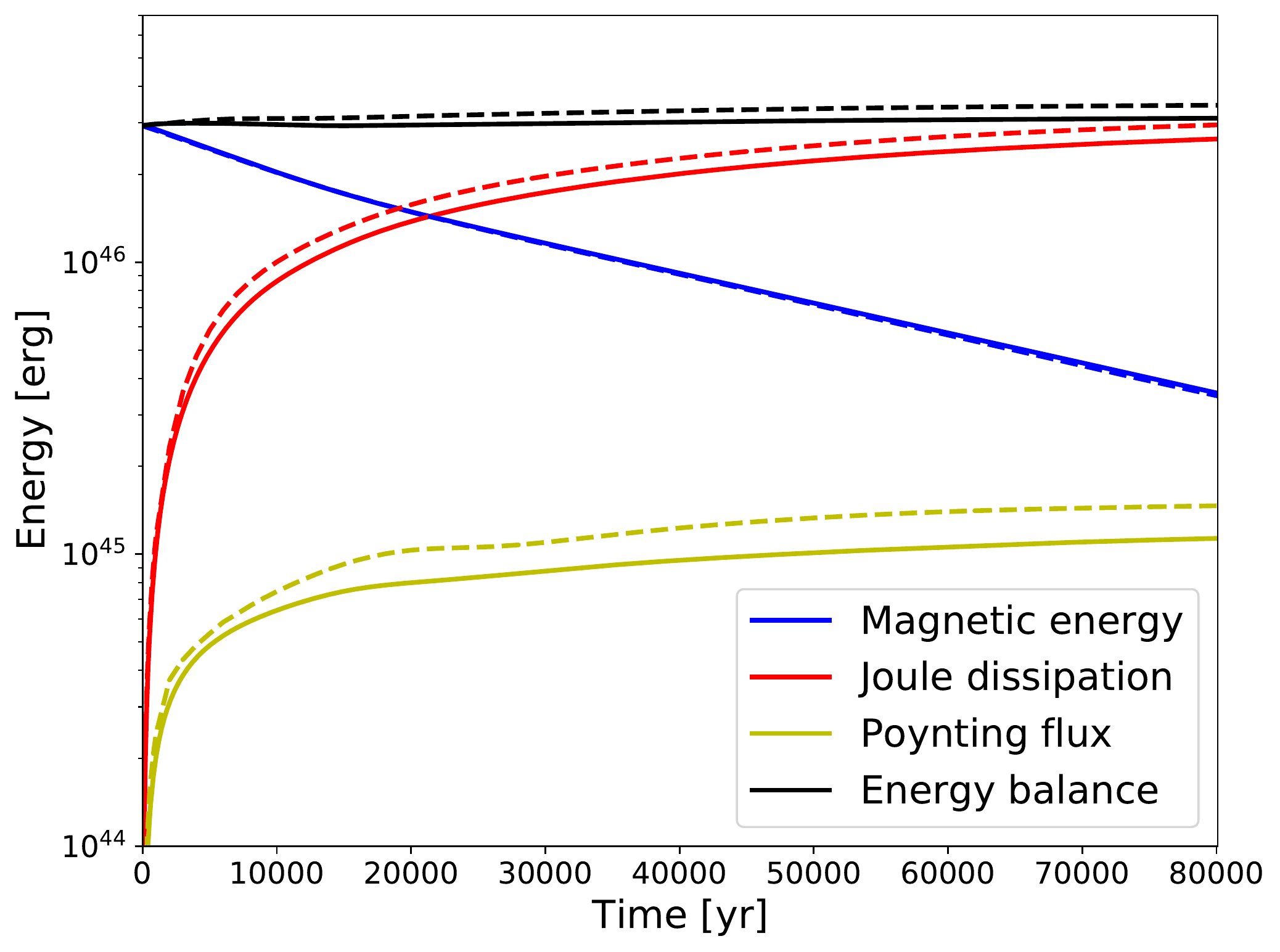}
        \includegraphics[width=0.33\textwidth]{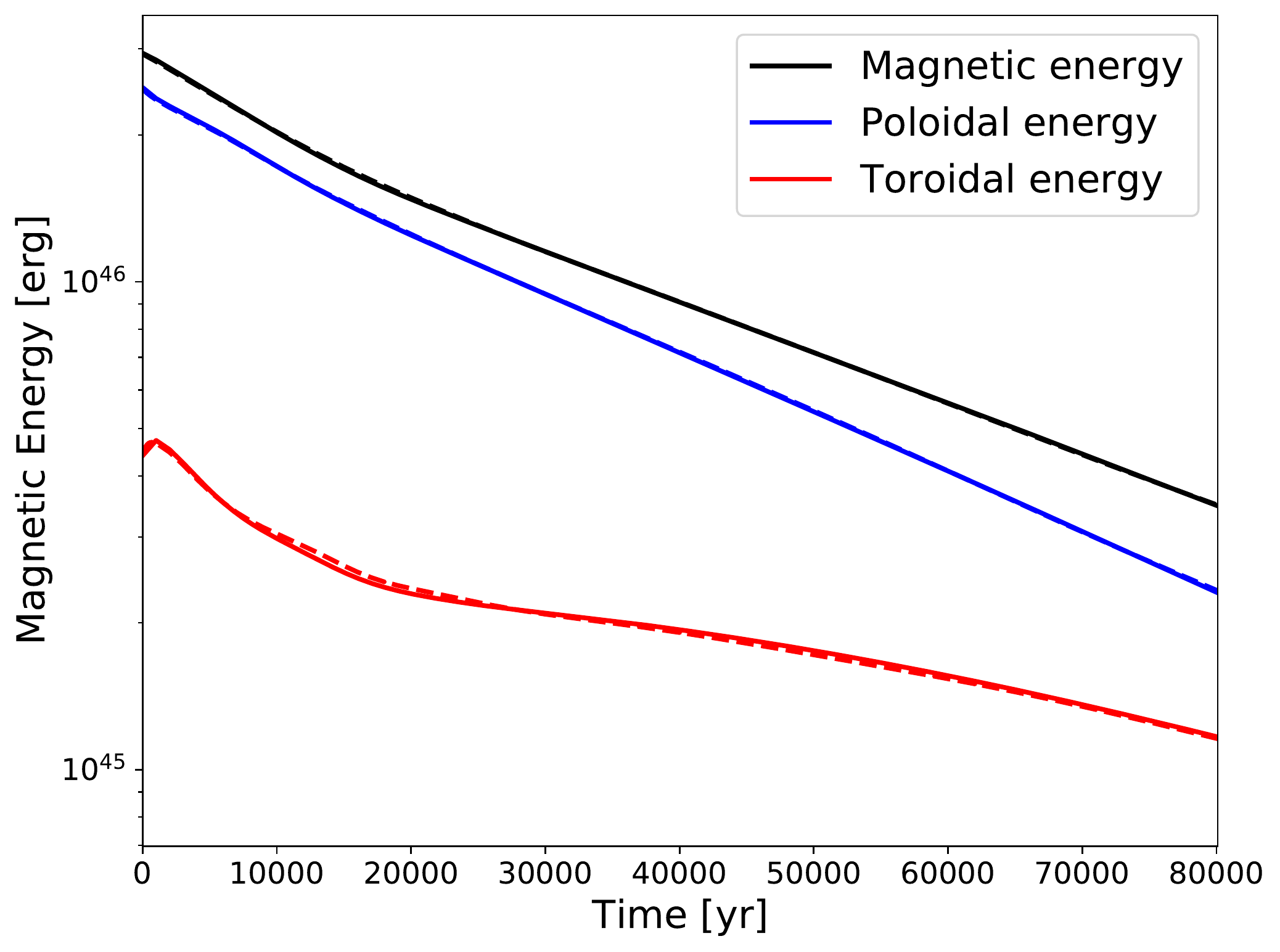}  \includegraphics[width=0.33\textwidth]{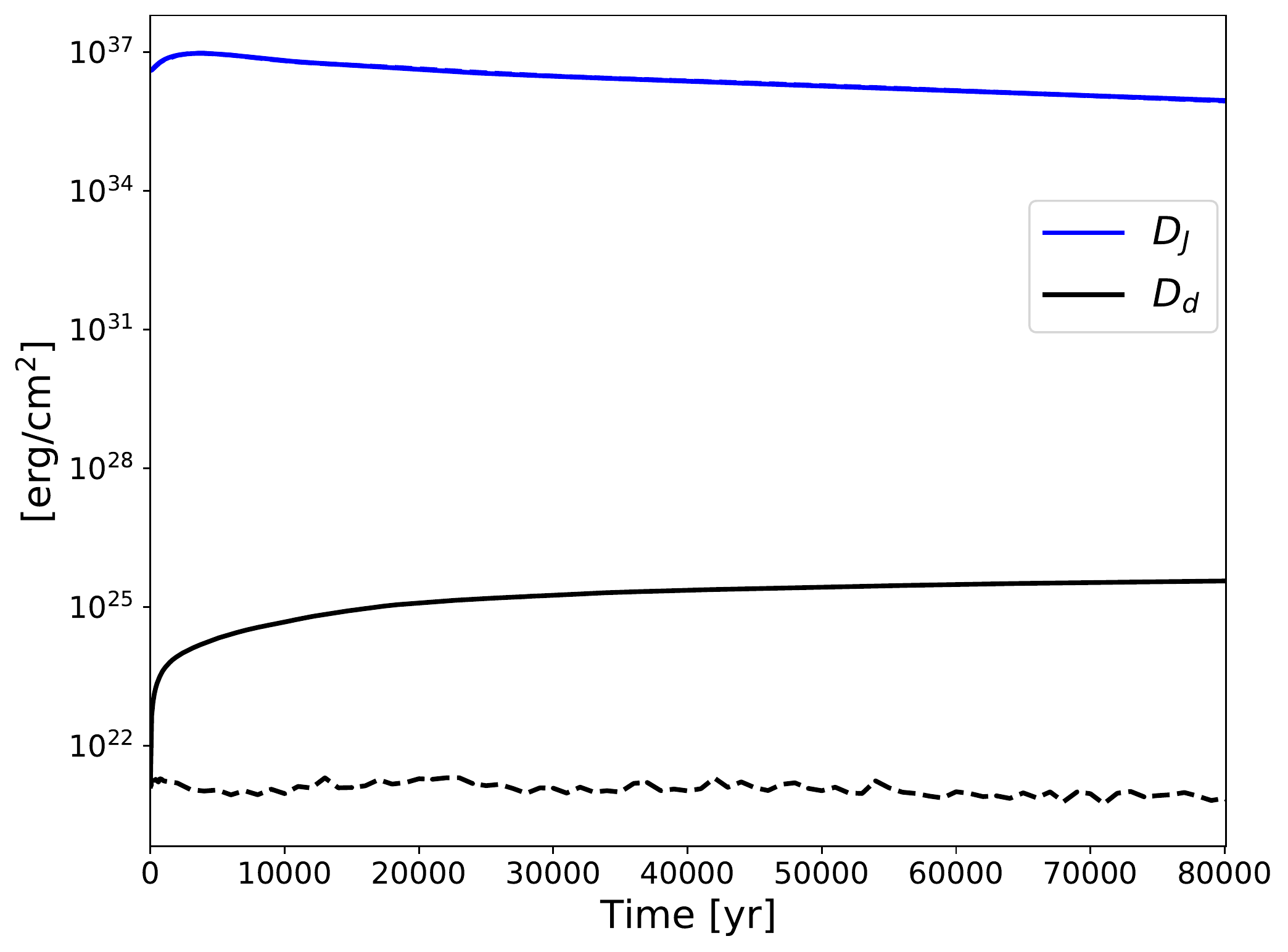}
\caption{ The results of the comparison between the 2D (dashed lines) and the 3D (solid lines) codes up to $t=80$ kyr. \textit{Left panel:} energy as a function of time. Energy balance in black, Joule dissipation in red, magnetic energy in blue and Poynting flux in yellow. \textit{Central panel:} total magnetic energy in black, poloidal magnetic energy in blue, toroidal magnetic energy in red. \textit{Right panel:} $D_d$ in black and $D_J$ in blue (eq. \ref{eq: divB volume} and eq. \ref{eq: J2-star} respectively).}
\label{fig: 2d - 3d comparison energy and divB}
\end{figure*}

The results of the comparison for an evolution up to $t=80$ kyr are displayed in Figs. \ref{fig: 2d-3d comparison Bfield profiles}, \ref{fig: 3d axisymmetric m energy spectrum}, and
\ref{fig: 2d - 3d comparison energy and divB}. The radial magnetic profiles for the three components of the magnetic field are displayed in Fig. \ref{fig: 2d-3d comparison Bfield profiles} at $t=0$, $5$, $10$, $20$ and $80$ kyr: $B^r$ at the north pole in the \textit{left panel}, $B^\theta$ at the equator in the \textit{central panel}, and $B^\phi$ at the equator in the \textit{right panel}. The 3D results are represented with solid lines, whereas the diamonds correspond to the 2D results. Throughout the evolution, the maximum magnetic Reynolds number is much greater than  unity, e.g., $R_m \sim 100$. Therefore, the Hall term dominates in the induction equation. The observed evolution is very similar. Local differences in the values of the components are typically less than a few percent, except for the radial component of the magnetic field at late times, which are likely due to the slightly different numerical implementation of the inner and outer boundary condition used in the two codes. 

An important point is assessing to which extent the 3D numerical code preserves axial symmetry. If we start with a pure $m=0$ mode, one should expect that this symmetry is kept to some small error, during the whole evolution. 
To give a quantitative measure of possible deviations, we study the energy spectrum (eq. \ref{eq: spectral magnetic energy}) by monitoring the evolution in time of each mode.

In Fig. \ref{fig: 3d axisymmetric m energy spectrum}, we plot in logarithmic scale the energy spectrum as a function of $m$ ($E_m \equiv \sum_{l}  E_{lm}$) at $t=0$ (in black) and after $t=80$ kyr (in yellow).
The spectral magnetic energy is concentrated at $m=0$ as expected. The rest of the modes are zero to the round-off error, except the modes with $m=\pm 4$, and higher harmonics, having anyway six or seven orders of magnitude less energy than the main one. They are caused by the discretization over the cubed-sphere grid, and in particular by the four patches that cover the tropical latitudes over the entire azimuthal direction. Their contribution to the energy spectrum is negligible. We remark that this unavoidable error introduced by the cubed-sphere grid is not increasing in time and it remains several orders of magnitude smaller than the magnetic energy contained in the $m=0$ mode after $80$ kyr of evolution. Moreover, it decreases for higher resolution.

In Fig. \ref{fig: 2d - 3d comparison energy and divB}, we show the different contributions to the total energy balance as a function of time (\textit{left panel}), the energy stored in the toroidal and poloidal components (\textit{central panel}), and a measure of the evolution of the divergence of the magnetic field (\textit{right panel}).
The solid lines correspond to 3D, whereas dashed lines to 2D. The total energy is conserved somewhat better in the 3D code, within $\sim 3\%$ in 3D and $\sim 15\%$ in 2D, after $80$ kyr of evolution. We attribute this minor differences in the energy balance to the use of spherical coordinates in 2D, which may cause more numerical errors close to the axis. 

As seen in the central panel, for this model, most of the magnetic energy is stored in the poloidal field while the toroidal field represents $\sim 15 \%$ of the total magnetic energy at $t=0$ and $\sim 34 \%$ at $t=80$ kyr. The increase in the relative fraction of the toroidal energy is caused by the non-linear term, which results in some redistribution of magnetic energy between poloidal and toroidal components.

In the \textit{right panel}, we compare the square of the divergence of the magnetic field integrated in the star volume (eq. \ref{eq: divB volume}) to the volume integrated $J^2$ (eq. \ref{eq: J2-star}). Both quantities have the same units, i.e., erg/cm$^2$ and the comparison is a good proxy for the level of conservation of the divergence constraint. The differences between the 2D and the 3D values of $D_d$ are most likely due to the different coordinates employed.
Nevertheless, $D_d$ is always several orders of magnitude lower than $D_J$ and is nearly constant in time. Generally speaking, we conclude that the results of the two codes agree within the expected grid/formalism-dependent numerical errors. 
% At the surface of the star, since we consider potential boundary conditions, the radial scalar function is known (check \ref{subsec: outer B.C} for details), and the magnetospheric magnetic energy for $r \geq R$ in c.g.s units and including relativistic correction is written as 

% \begin{equation}
%      E_\text{spectral}^{r\geq R} = \frac{1}{8\pi} \int e^{\nu} r^2 dr \sum_{lm} \bigg(\frac{R}{r}\bigg)^{2l+4} (b^m_l)^2 (l+1) (2l+1),
% \label{eq: spectral magnetic energy surface}
% \end{equation}
% with $b^m_l$ are the reconstructed weights of multipoles considering a potential boundary conditions (section. \ref{subsec: outer B.C}). 

%Note that eq. \ref{eq: spectral magnetic energy surface} is the expression of the vacuum magnetic energy in 3D, which could be used as a further diagnostic test. 

\section{Results}
\label{sec: results and simulations}

\subsection{Physical setup}

We now turn to the full problem: non-axisymmetric 3D simulations in a realistic NS crust, with a stratified electron number density and a temperature-dependent resistivity.
Although a detailed evolutionary model requires the simultaneous numerical solution of the heat diffusion equation coupled to the 3D magnetic evolution, in this paper we use the analytical approximation for the isothermal redshfited temperature ($Te^\nu$) evolution of \cite{yakovlev2011}, so that the physical temperature $T$ reads: 
\begin{equation}
    T(t) = 3.45 \times 10^8 \text{K} ~\left(1-\frac{2GM}{c^2R_\star}\right) \bigg[1+0.12 \bigg(\frac{R_\star}{10 \text{km}} \bigg)^2  \bigg] \bigg(\frac{t_c}{t} \bigg)^{1/6} e^{-\nu}
    \label{eq: isothermal cooling}
\end{equation}
where $t_c$ is some fiducial (normalization) time-scale.
For our model, $M=1.4M_\odot$ and $R_\star=11.7$ km and $t_c$ 
is set to the age of the Cas A supernova remnant ($330$ yr). 
It has been shown that this time dependence is accurate during the neutrino cooling stage \citep{yakovlev2011}. These simplification will suffice for our purpose in this work.
%The full coupling of thermal and magnetic evolution and the code for the thermal diffusion equation will be reported in an upcoming paper.

The electrical conductivity (needed to calculate $\eta_b$) is calculated locally at each timestep, considering the temperature, local density and composition. We use the same public code from Alexander Potekhin\footnote{\url{http://www.ioffe.ru/astro/conduct/}} \citep{potekhin2015}, which has been used in all previous 2D simulations by our group.

\subsection{Initial magnetic topology}

The magnetic field configuration of NSs at birth is largely unknown. Recent magneto-hydrodynamic simulations of the magneto-rotational instability in core-collapse supernovae \citep{obergaulinger2014,Aloy_2021, Reboul-Salze_2021} suggest a complex picture, in which the magnetic energy of the proto-NS spreads over a wide range of spatial scales. Such simulations find that most of the magnetic energy is contained in small or medium-scale size magnetic structures, both for the toroidal and the poloidal components. Note that this deviates substantially from the often-used, simple dipole+twisted torus configurations inferred by MHD-equilibrium studies \citep{ciolfi13}.

To assess the sensitivity of results to the uncertain initial conditions, we have considered three different magnetic field topologies, all confined to the crust (substantially similarly to \cite{aguilera2008}). The details of the radial dependence and how to construct a divergence-free magnetic field are given in Appendix \ref{appendix: initial conditions}. The numerical scheme maintains the local divergence up to machine error, by construction.

The different models studied in this section have an average initial magnetic field of $\sim 10^{14}$ G, corresponding to total magnetic energies of the order of $\sim 10^{45} $ erg.
They are summarized in Table \ref{tab: different models}. Most of the magnetic energy is contained in the toroidal component, except for the last model. They differ in the relative weights of multipoles of the initial configuration.
In the second and third models, the temperature is fixed to $10^9$ K and $2 \times 10^8$ K respectively, instead of evolving it. 
Note in general that we choose arbitrary combinations of a relatively small number of multipoles, in contrast with the expected smooth cascade over a wide range of them suggested by the above-mentioned proto-NS configurations. The total evolution time for the first three models of Table \ref{tab: different models}, e.g., L5, L5-T1e9 and L5-T2e8, is $70$ kyr. For L1 model, the total evolution time is $85$ kyr and for L10 model it is $100$ kyr. For some models, the total evolution time is limited by numerical instabilities appearing at late times when the temperature goes well below $10^8$ K (e.g., $T(100 \,{\rm kyr})\sim 10^7$ K from eq. \ref{eq: isothermal cooling}), and the magnetic Reynolds number grows. The appearance of instabilities also depends on the initial magnetic field strength and topology. This is similar to what we see in our 2D magneto-thermal code \citep{vigano2021}.
% Thus, they are far from a MHD or Hall-MHD equilibrium, which allows us to test whether (or in which cases) the Hall dynamics can redistribute the energy over all scales, regardless of the initial conditions.

For all the models displayed in Table \ref{tab: different models}, we consider a resolution of: $N_r=40$ and $N_\xi = N_\eta =43$ per patch which is equivalent to $172$ grid points around the equator and $87$ points along a meridian from pole to pole. Given the employed resolution, we look up to $l=30$. 
 
\begin{table*}
 \caption{Initial Models Considered. $l$ and $m$ are the initial non-zero multipoles considered in each model. $B_{avg}$ is the average initial magnetic field. $E_\text{mag}$ is the initial magnetic energy (all in the crust). $E_\text{tor}/E_\text{mag}$ is the fraction of the crustal toroidal energy. For all these models, we are confining the field lines to the crust of the star. We use the simplified cooling described in the text (eq. \ref{eq: isothermal cooling}), except in two models ("deactivated").}
 \label{tab: different models}
 \begin{tabular}{lcccccccccc}
  \hline
  Models & $l_\text{pol}$ & $m_\text{pol}$ &  $l_\text{tor}$ & $m_\text{tor}$ & $B_\text{avg} (t0)$  &  $B_\text{max} (t0)$  & $E_\text{mag} (t0)$&  $E_\text{tor}/E_\text{mag} $ & Simplified & $T_\text{fixed}$ \\
  &  &&    & &  [G] & [G] &  [erg]& $(t0)$ & Cooling &  [K] \\
  \hline
  L5 & $1,2,3,5$ &  $-1,0,1,2,3$&  $1,2,3,5$ &  $0,1,2,3$& $\sim 2 \times 10^{14}$ & $\sim 7 \times 10^{14}$& $ \sim 2\times 10^{45}$ & $\sim 63 \%$ &   activated & -\\
L5-T1e9 &  $1,2,3,5$ &  $-1,0,1,2,3$&  $1,2,3,5$ &  $0,1,2,3$& $\sim 2 \times 10^{14}$ & $\sim 7 \times 10^{14}$&$ \sim 2\times 10^{45}$ &  $ \sim 63 \%$  & deactivated & $10^{9}$\\
L5-T2e8 &  $1,2,3,5$ &  $-1,0,1,2,3$&  $1,2,3,5$ &  $0,1,2,3$& $\sim 2 \times 10^{14}$ & $\sim 7 \times 10^{14}$&$ \sim 2\times 10^{45}$ &  $ \sim 63 \%$  & deactivated & $2 \times 10^{8}$\\
  L1 & $1$ & $0$ & $1$ & $1$ & $\sim 3 \times 10^{14}$ & $\sim 6.5 \times 10^{14}$  &$\sim 4 \times 10^{45}$ & $\sim 95$\% & activated & -\\
  L10  & $1,6,7,10$   &  $ -5,-1,0,1,7,8 $   & $ 1,3,7,10 $   &  $-5,0,2,9$   & $\sim   10^{14}$   & $\sim 3 \times 10^{14}$ & $\sim 6 \times 10^{44}$& $\sim 10 \%$  & activated & -\\
  \hline
 \end{tabular}
\end{table*}

\subsection{L5 model}
\label{subsec: LSCC model}

The first model, named L5, has an average magnetic field of $~2 \times 10^{14}$ G on average and a maximum of $~7 \times 10^{14}$ G. The initial configuration consists of a large scale topology defined as a sum of multipoles up to $l=5$. Besides, this model is a Hall-dominant with a maximum magnetic Reynolds number $R_m \sim 200$, during the evolution. For a more quantitative analysis of the 3D magnetic evolution, we survey the magnetic energy spectrum to observe the redistribution of the magnetic energy over the different spatial scales.

\begin{figure*}
\centering
\includegraphics[width=0.45\textwidth]{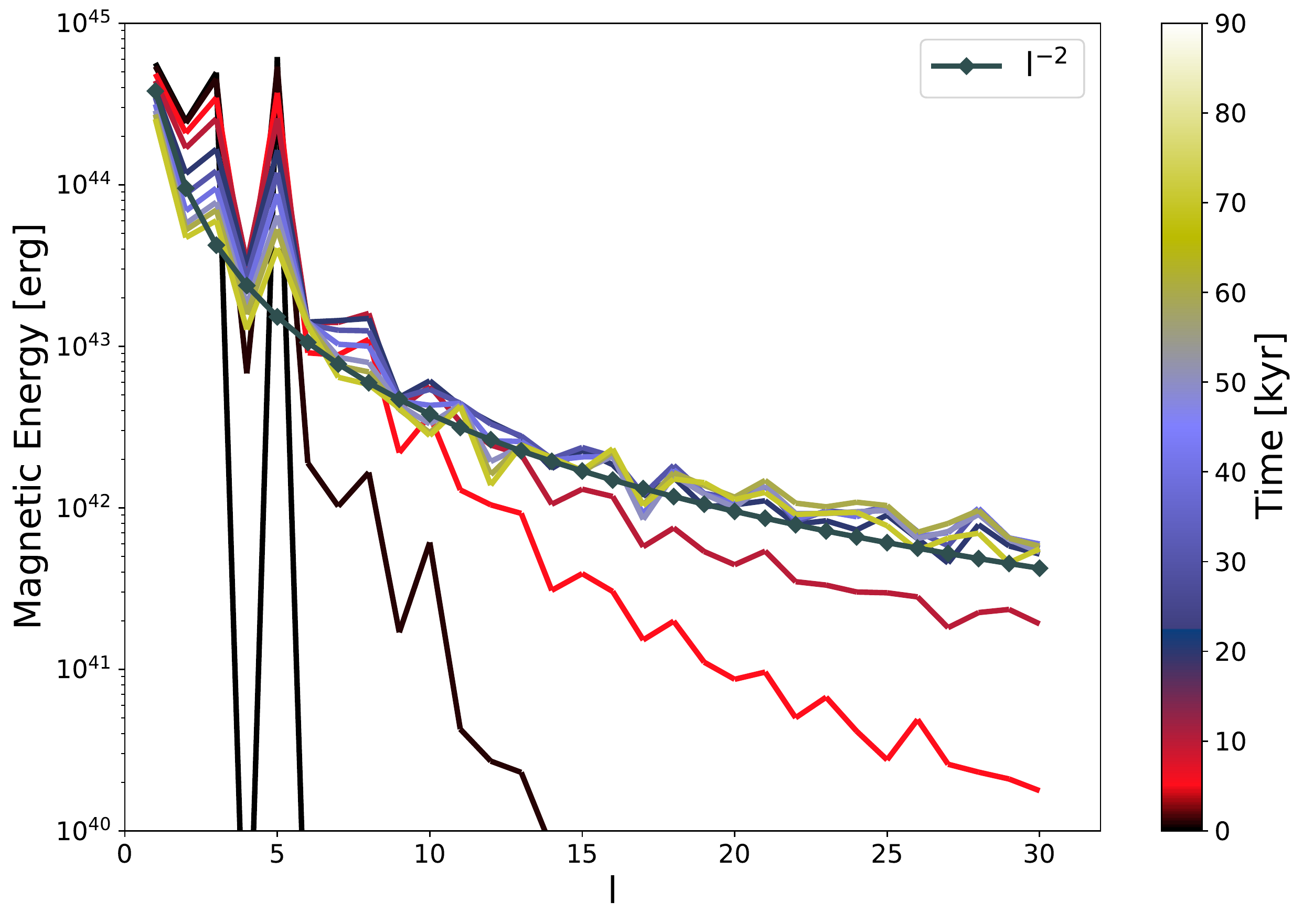}
\includegraphics[width=0.45\textwidth]{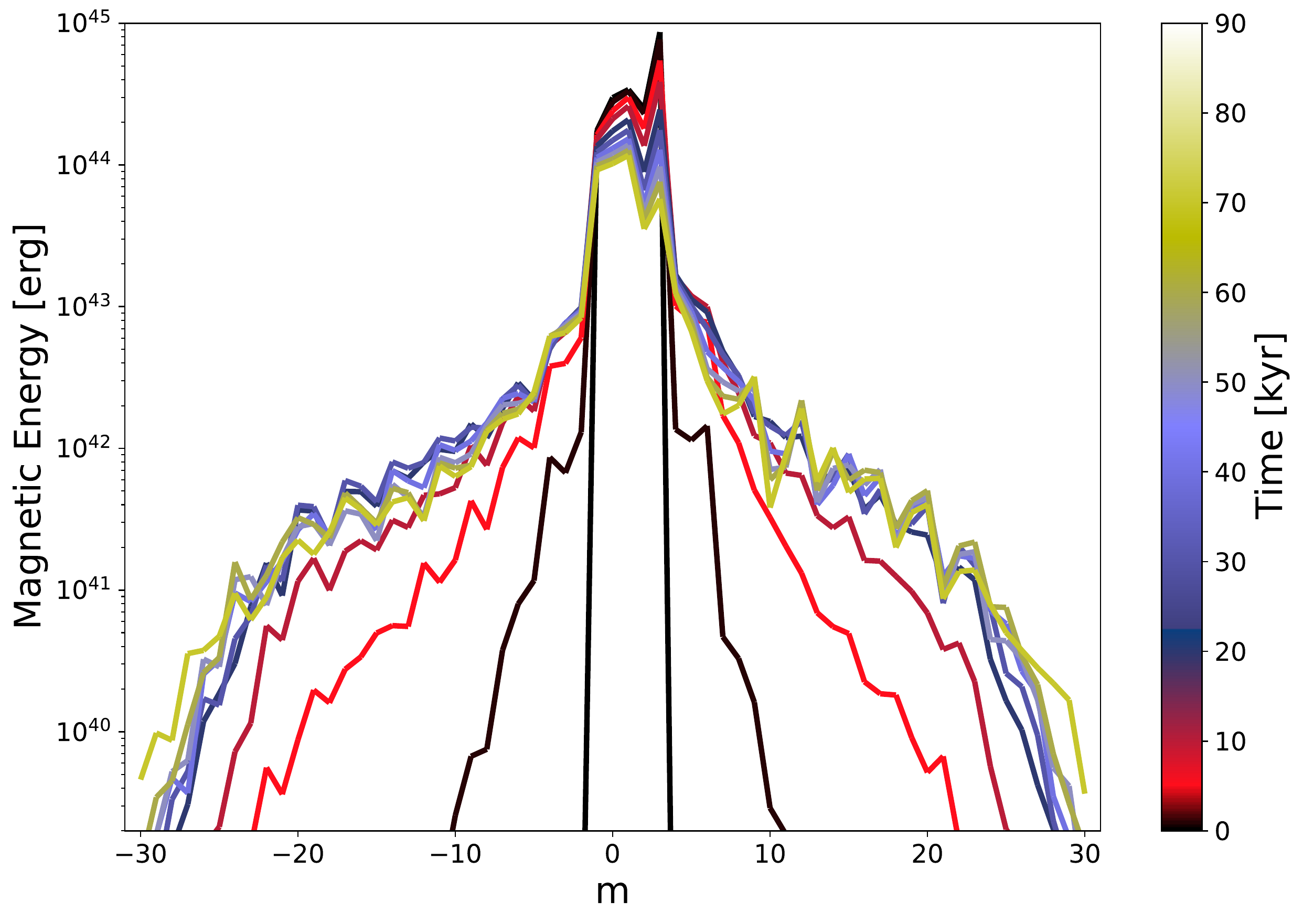}
\caption{\textit{L5 model}. \textit{Left panel:} $l$ energy spectrum. \textit{Right panel:} $m$ energy spectrum. The energy spectra are displayed at times $0, 1, 5, 10, 20, 30, 40, 50, 60$ and $70$ kyr (see color bars). The $l^{-2}$ slope corresponds to the Hall cascade equilibrium distribution of magnetic energy over a quite broad range of multipoles.}
\label{fig: energy spectrum LSCC}
\end{figure*}

In Fig. \ref{fig: energy spectrum LSCC}, we examine the $l$ energy spectrum (summing eq.~(\ref{eq: spectral magnetic energy}) over all $m$'s) in the left panel and the $m$ energy spectrum (summing it over all $l$'s) in the right panel, at different evolution times. At time zero, one can clearly distinguish the multipoles imposed initially. As soon as we start the evolution, part of the magnetic energy is transferred from the large-scale multipoles, into the smaller-scale ones. Moreover, we notice that higher order $m$ modes are excited in the system. At $1$ kyr, most of the magnetic energy is concentrated in the initially imposed multipoles. However, a fraction of the magnetic energy is already transferred to $l=4-10$. Following the evolution up to $5$ and $10$ kyr (red curves), the transfer of energy toward small-scales continues to fill in the entire spectrum.
At about $20$ kyr, the magnetic energy spectrum seems to have reached a quasi-stationary state, i.e., the Hall-saturation. Considering that the dissipation goes as $L^2/\eta_b$ ($L$ is the typical spatial scale of the field curvature), the small-scale structures dissipate faster than the large-scale ones. At the same time, the former are continuously fed by the latter, thanks to the Hall term in the induction equation. This is known as the Hall cascade, it consists in an equilibrium distribution of magnetic energy, over a quite broad range of multipoles, with an approximate $l^{-2}$ slope \citep{goldreich1992}.

% Since the dissipation goes as $L^2/\eta_b$ ($L$ is the typical spatial scale of the field curvature), the small-scale structures dissipate faster than the large-scale ones, but, at the same time, the former are continuously fed by the latter, thanks to the Hall term in the induction equation. It is the Hall cascade, that consists in an equilibrium distribution of magnetic energy, over a quite broad range of multipoles, with an approximate $l^{-2}$ slope \citep{goldreich1992}.
% At $50-70$ kyr, most of the magnetic energy has dissipated, i.e., $75-80\%$ respectively. 

Note that this cascade and saturation needs two main conditions: (i) a Hall-dominated dynamics, i.e. large enough magnetic field; (ii) an initial configuration that allows a full development of the Hall cascade. As a matter of fact, poloidal and toroidal fields are not symmetrically coupled: the odd multipoles of the former are more coupled to the even multipoles of the latter. 
In axial symmetry, this implies that one can maintain a perfect helicity-free configuration if the initial field is given by only $l=1,3,5...$ poloidal components and $l=2,4,6...$ toroidal components. In this special case, not all multipoles are excited, and only odd/even families will show up in the spectrum. However, in a general non-axisymmetric case with arbitrary combinations of initial multipoles, if the Hall term dominates, the relative weights of couplings between different modes are not so clear.

\subsection{The impact of temperature dependent microphysics}
\label{subsec: impact of microphysics}

To appreciate the role of temperature-dependent microphysics in the evolution of the magnetic field, we perform a comparison up to $70$ kyr, taking into consideration an identical magnetic field configuration, with (L5 model) and without (L5-T1e9 and L5-T2e8 models) temperature evolution. The microphysical coefficients for L5-T1e9 model are calculated at $T=10^9$ K, whereas the ones for L5-T2e8 model are calculated at $T= 2\times 10^8$ K.
Note that $T=10^9$ K corresponds to the temperature of a NS during the first years of its life, whereas $T= 2\times 10^8$ K, corresponds to the temperature at $\sim 10$ kyr in L5 model.

\begin{figure}
\includegraphics[width=.45\textwidth]{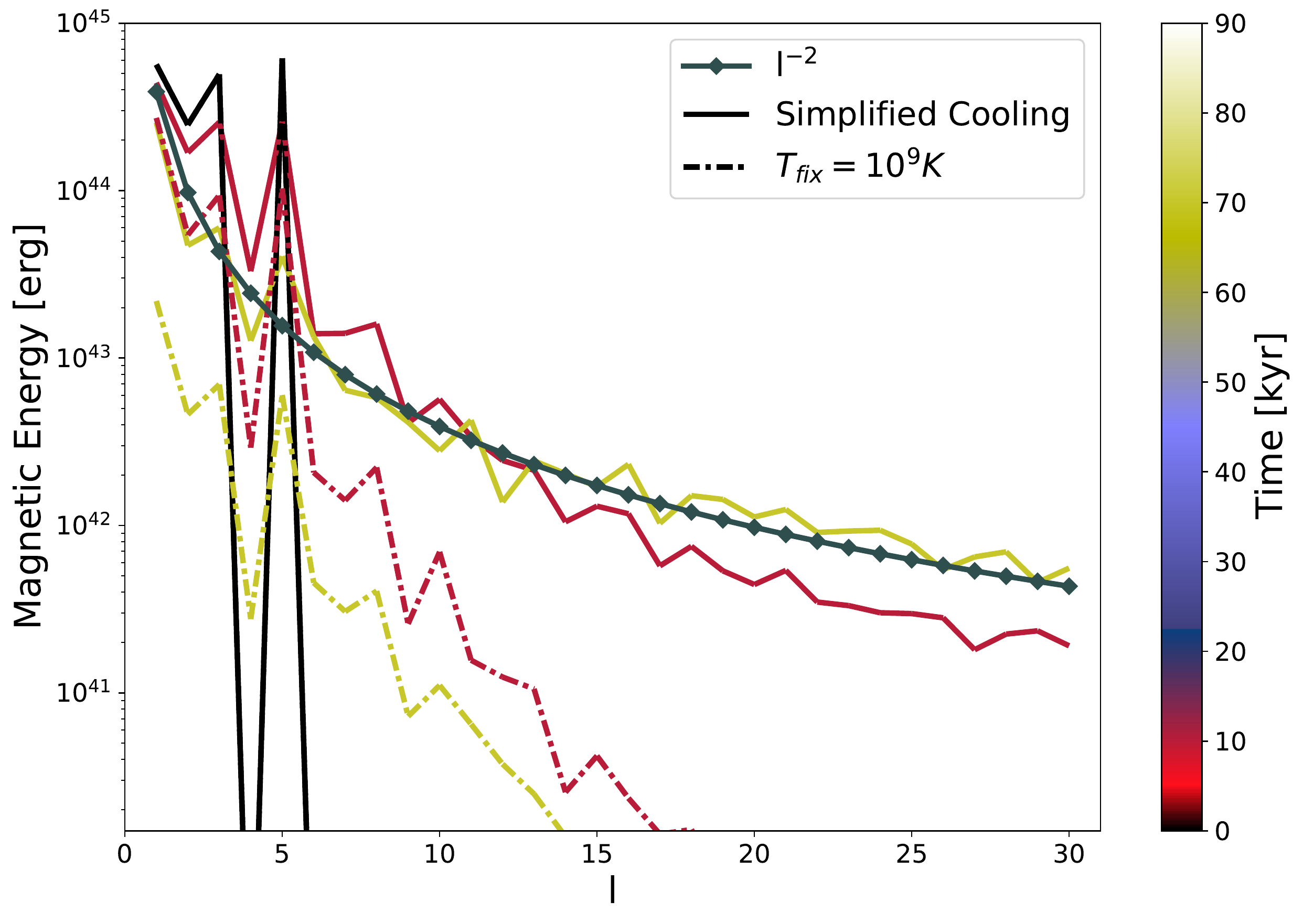}
\includegraphics[width=.45\textwidth]{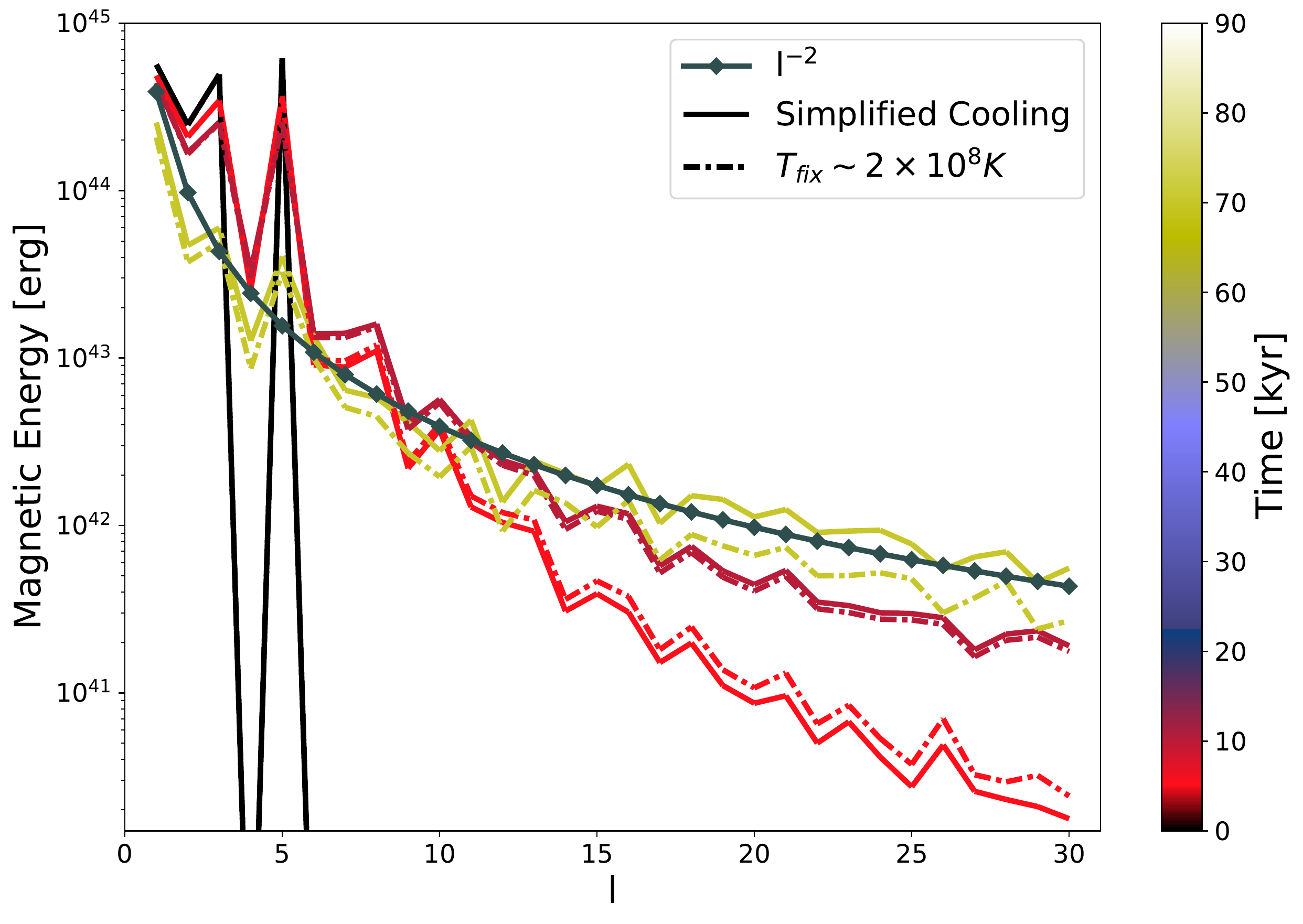}
\caption{A comparison up to $70$ kyr of the $l$ energy spectrum for L5 model, with temperature evolution (solid lines) and without temperature evolution (dash-dotted lines). The temperature is fixed to $T=10^9$ K in the \textit{upper panel}, and to $T=2\times 10^8$ K in the \textit{bottom panel}. In the \textit{upper panel}, the comparison is illustrated at $t=0$ (black), $t=10$ kyr (dark red) and $t=70$ kyr (yellow) and in the \textit{bottom panel} at $t=0$ (black), $t=5$ kyr (red), $t=10$ kyr (dark red) and $t=70$ kyr (yellow). The $l^{-2}$ slope corresponds to the Hall cascade equilibrium distribution of magnetic energy over a quite broad range of multipoles.}
\label{fig: Espectrum realistic vs analytical}
\end{figure}

The results of the comparison at different evolution times are illustrated in Fig. \ref{fig: Espectrum realistic vs analytical}. The upper panel corresponds to the comparison between L5 (solid lines) and L5-T1e9 (dash-dotted lines) models, whereas the bottom panel corresponds to the comparison between L5 (solid lines) and L5-T2e8 (dash-dotted lines) models. The three models overlap at initial time. A transfer of magnetic energy to small-scale structures occurs in all cases. Nevertheless, a distinguishable behaviour happens during the field evolution in the first case (upper panel of Fig. \ref{fig: Espectrum realistic vs analytical}). Model L5-T1e9 is mostly dissipating in time with negligible redistribution of the magnetic energy over the different spatial scales, i.e., the $l$ energy spectrum keeps the same shape at $t=10$ and $70$ kyr. About $70\%$ of the total magnetic energy has dissipated for L5-T1e9 model after $10$ kyr, but only $35\%$ of the total magnetic energy has dissipated for L5 model. Therefore, L5-T1e9 model is an Ohmic-dominant. Whereas, L5 model is a Hall-dominant. 

On the other hand, the evolution in time of L5 and L5-T2e8 models is pretty comparable (bottom panel of Fig. \ref{fig: Espectrum realistic vs analytical}). At about $5$ kyr, the transfer of energy is slightly more efficient for L5-T2e8 model. That is because the temperature value considered for L5-T2e8 model, e.g., $T=2\times10^8$ K, is lower than the temperature value at $t\sim 5$ kyr obtained using eq. \ref{eq: isothermal cooling}. Therefore, the magnetic Reynolds number is slightly higher for the L5-T2e8 model since the magnetic resistivity is lower for lower temperature. The $l$-energy spectrum of the two models appear pretty similar at about $10$kyr. Nevertheless, at about $70$ kyr, the results of the two simulations start to diverge again. L5-T2e8 has dissipated more than L5 model, since at this evolution stage, the magnetic Reynolds number is higher for L5 model. Nevertheless, both models, L5 and L5-T2e8 are Hall-dominant. 

These different behaviours of the time evolution of the energy spectrum highlight the impact of the temperature-dependent microphysics on our results. The difference in spectra is very important in the first case (upper panel of Fig. \ref{fig: Espectrum realistic vs analytical}), instead it is slight in the second case (bottom panel of Fig. \ref{fig: Espectrum realistic vs analytical}). Moreover, for a Hall-dominated field, the specific value of the magnetic diffusivity will only determine the resistive scale, i.e., the width of the inertial range where we see the Hall cascade.
Note, however, that in the second comparison (bottom panel of Fig. \ref{fig: Espectrum realistic vs analytical}), we set the diffusivity assuming $T=2\times 10^8$ K, which is not far from the average value of temperature during the first $50$ kyr. To obtain more realistic results, a 3D magneto-thermal code coupled with realistic microphysics is needed.

% Much larger differences could be obtained by, e.g., manually fixing very different values of $n_e$ and $\eta_b$ (or fixing much larger or smaller temperatures). 

% behaviour is slightly different behaviour of the evolution of the energy spectrum in time highlights: (i) for a Hall-dominant evolution, the magnetic diffusivity plays a minor role during the evolution, (ii) the magnetic resistivity is temperature-dependent. Therefore, a 3D magneto-thermal evolution coupled with realistic microphysics is needed to achieve realistic results. 

 \subsection{Different initial multipolar topology}
 \label{subsec: different initial multipoles}
 
 \subsubsection{Magnetic field lines}

To study the impact of adopting different topologies, we consider three different models with different initial multipoles. Besides the L5 model presented in section \ref{subsec: LSCC model}, L1 is a model with a pure dipolar field, i.e., $l=1$, in both the poloidal and toroidal components. L10 is characterized by having a wider combination of initial multipoles, up to $l=10$.
%The average magnetic field for the three models accounts to a few $10^{14}$ G with a magnetic energy of a few $\sim 10^{45} $ erg.
Throughout the evolution, the maximum magnetic Reynolds number reaches $\sim 200$ for L5 model, $\sim 150$ for L1 model and $\sim 50$ for L10 model.

 \begin{figure*}
\centering
\includegraphics[width=0.33\textwidth]{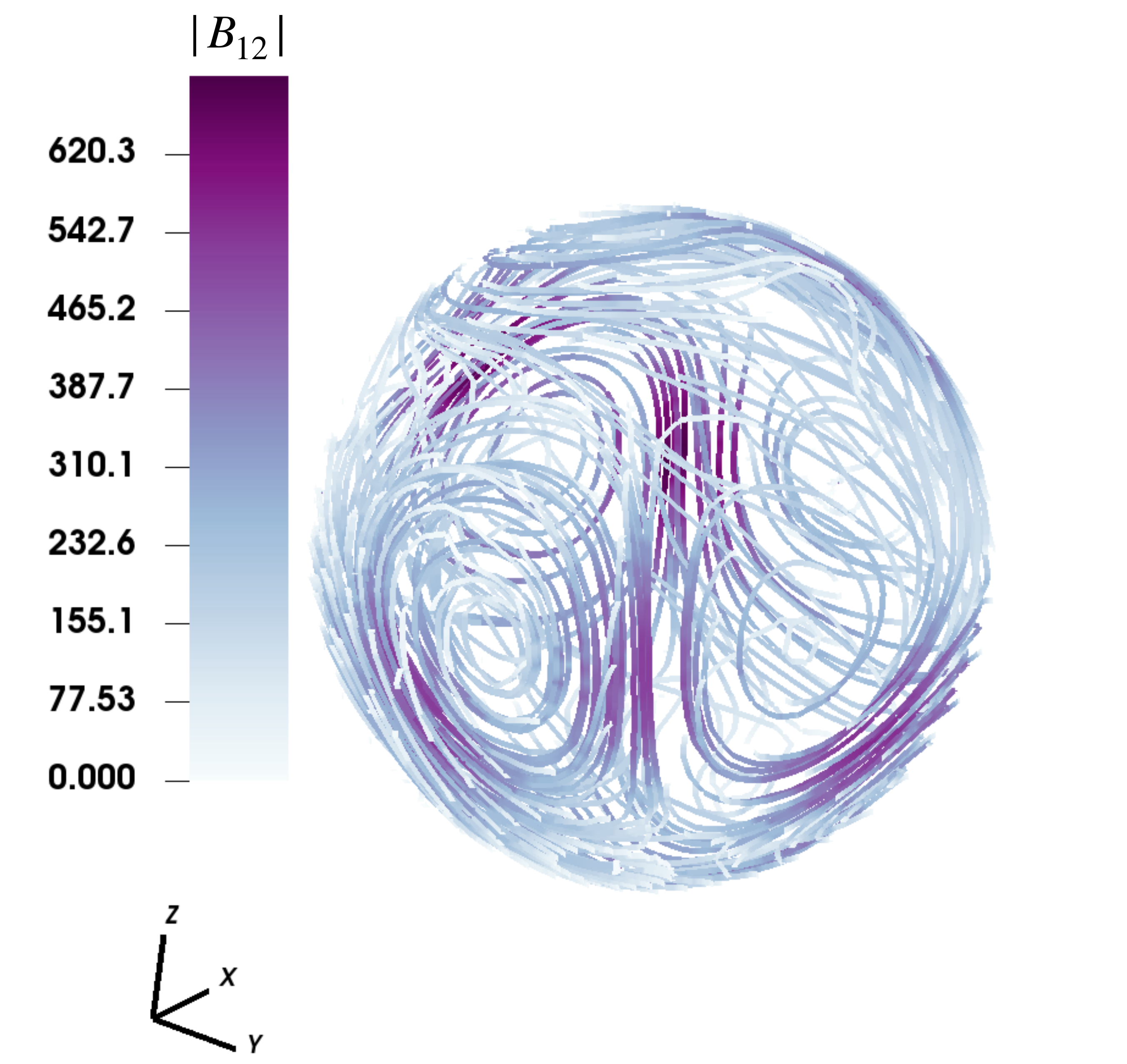}
\includegraphics[width=0.33\textwidth]{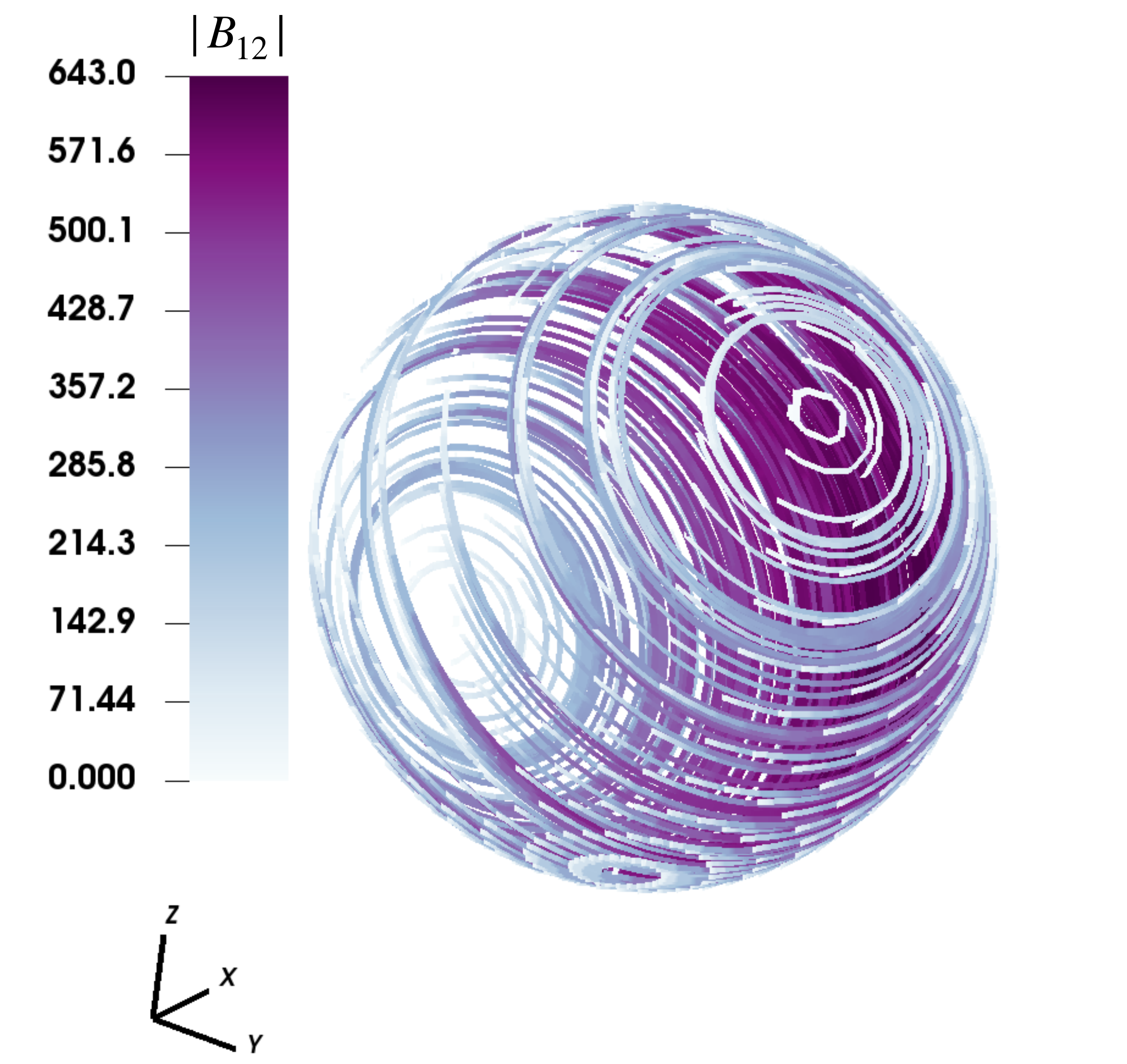}
\includegraphics[width=0.33\textwidth]{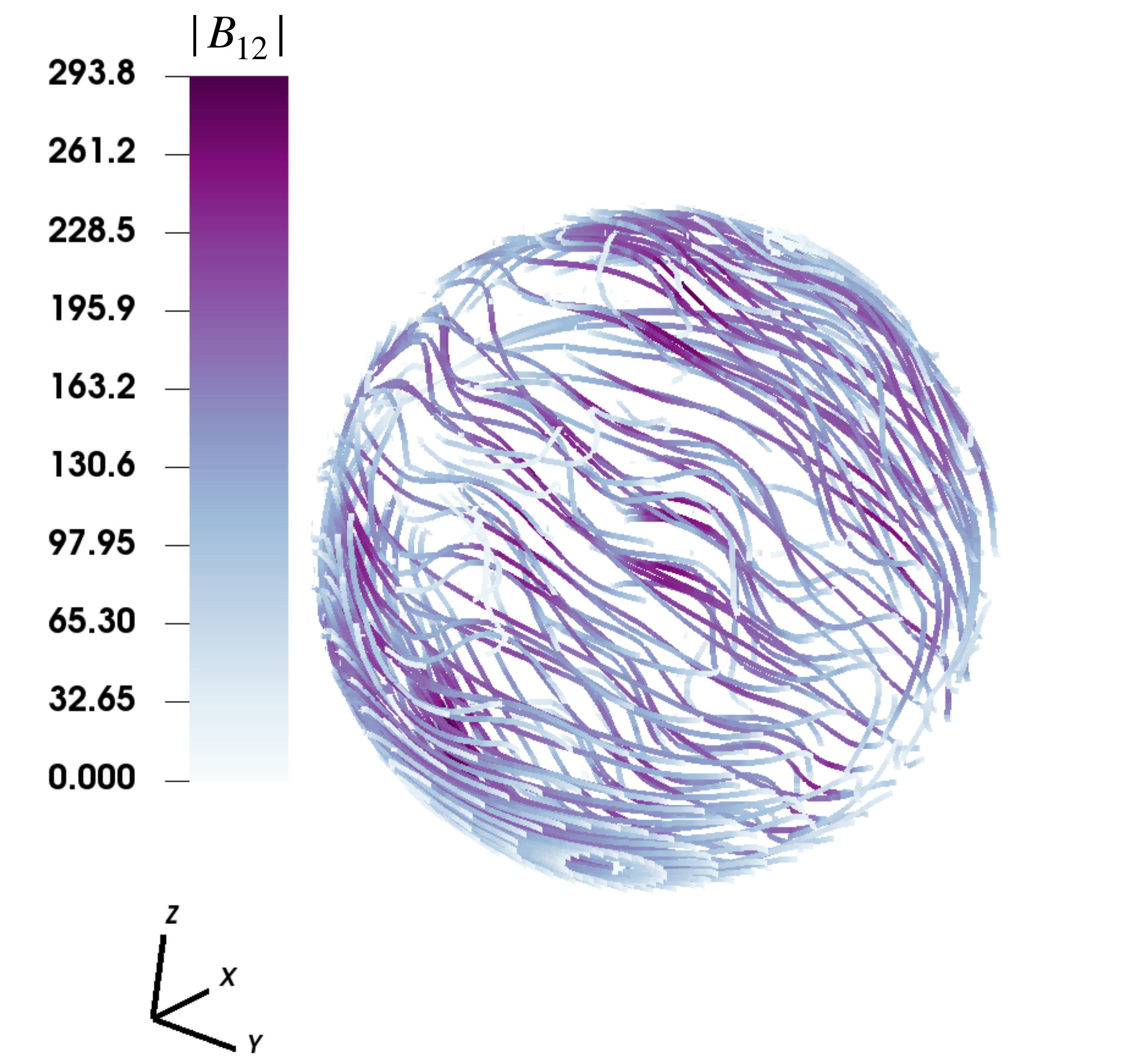} 
\includegraphics[width=0.33\textwidth]{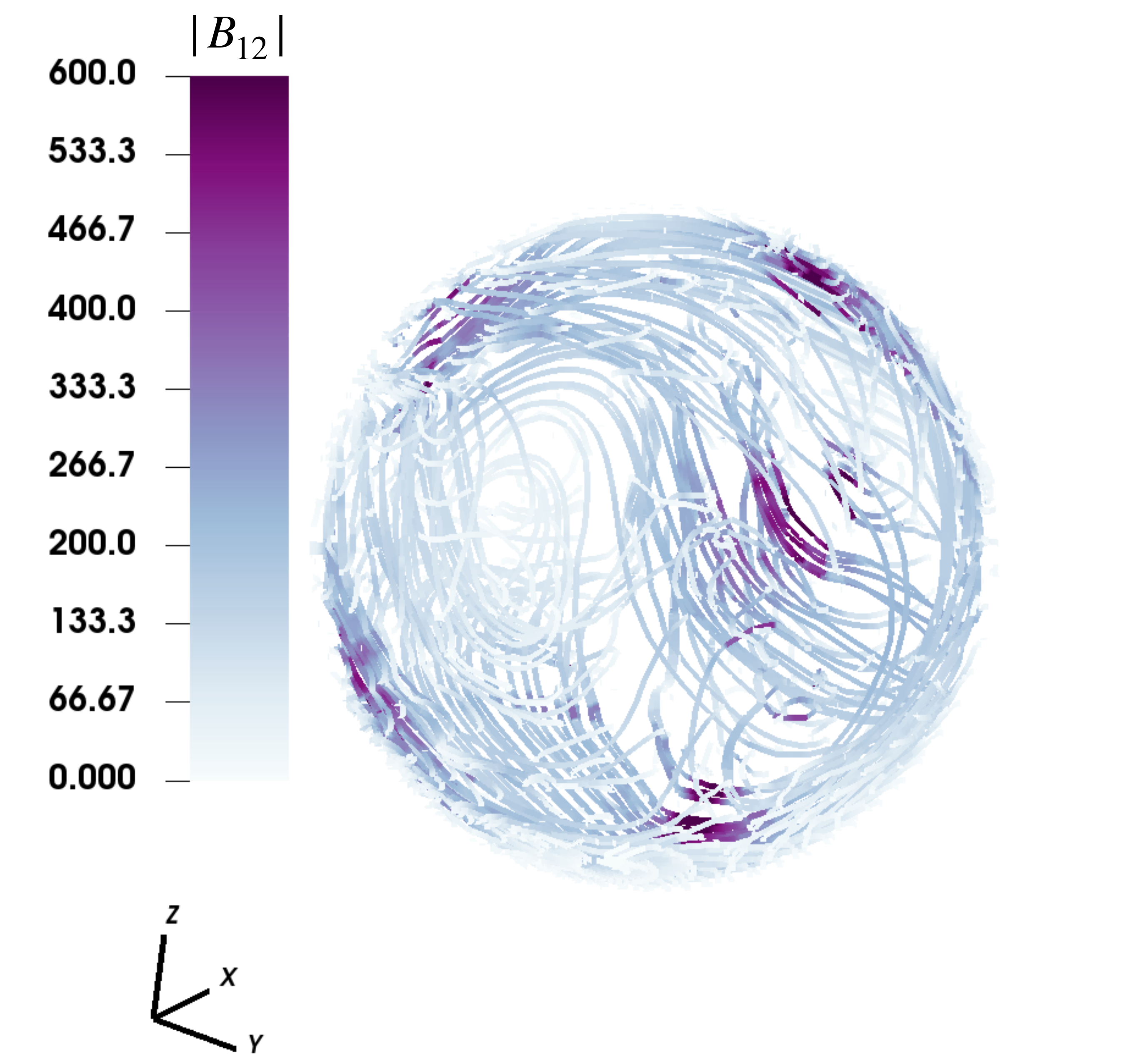}
\includegraphics[width=0.33\textwidth]{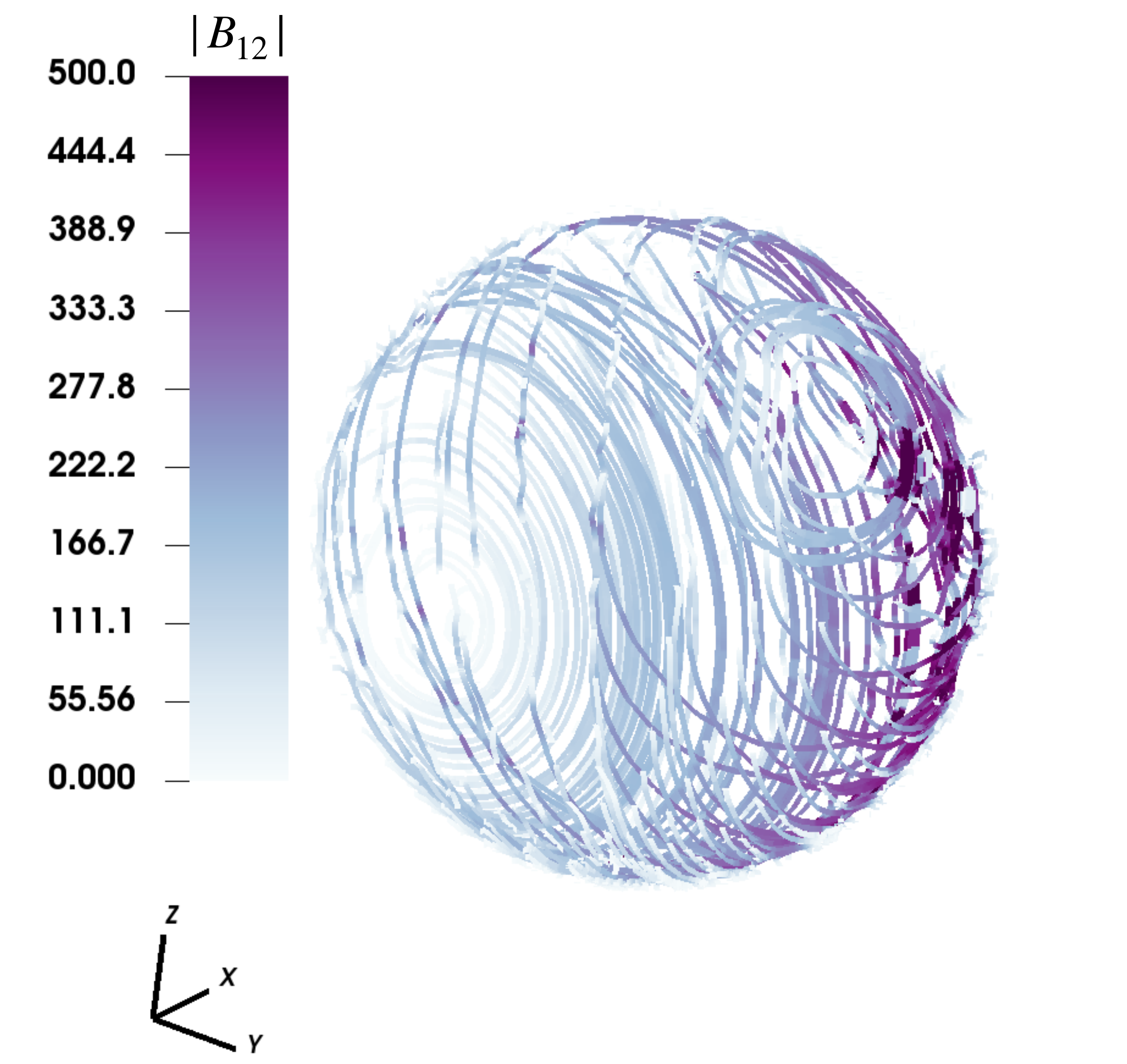}
\includegraphics[width=0.33\textwidth]{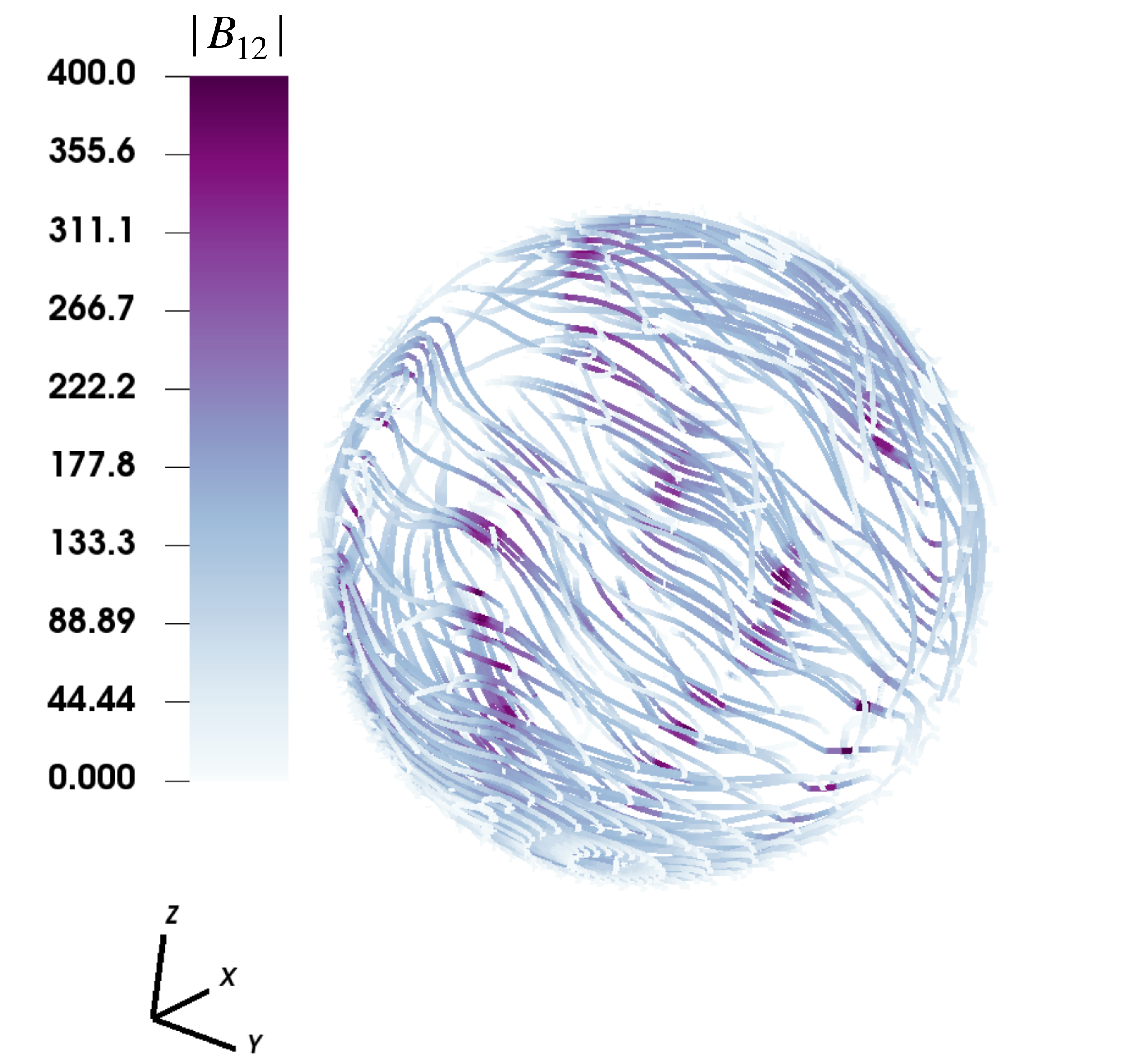}
\caption{Field lines in the crust of a NS, for L5 model on the left, L1 model in the center and L10 model on the right, at $t=0$ (upper panels) and after $t=50$kyr (bottom panels). The color scale indicates the local field intensity, in units of $10^{12}$ G.}
\label{fig: field lines visit Dip LSCC Multi}
\end{figure*}

In Fig. \ref{fig: field lines visit Dip LSCC Multi}, we display the magnetic field lines for L5 model (left panels), L1 model (central panels) and L10 model (right panels), at $t=0$ (upper row) and after $50$ kyr (bottom row). At $t=0$, one can clearly distinguish the three different magnetic field configurations adopted in these models. After a few Hall timescales, e.g., $\sim 50$ kyr, the field lines are more tangled but one can still recognize the initial magnetic field configurations: the star has not lost memory of the initial large scale topology.

\begin{figure*}
\centering
\includegraphics[width=0.45\textwidth]{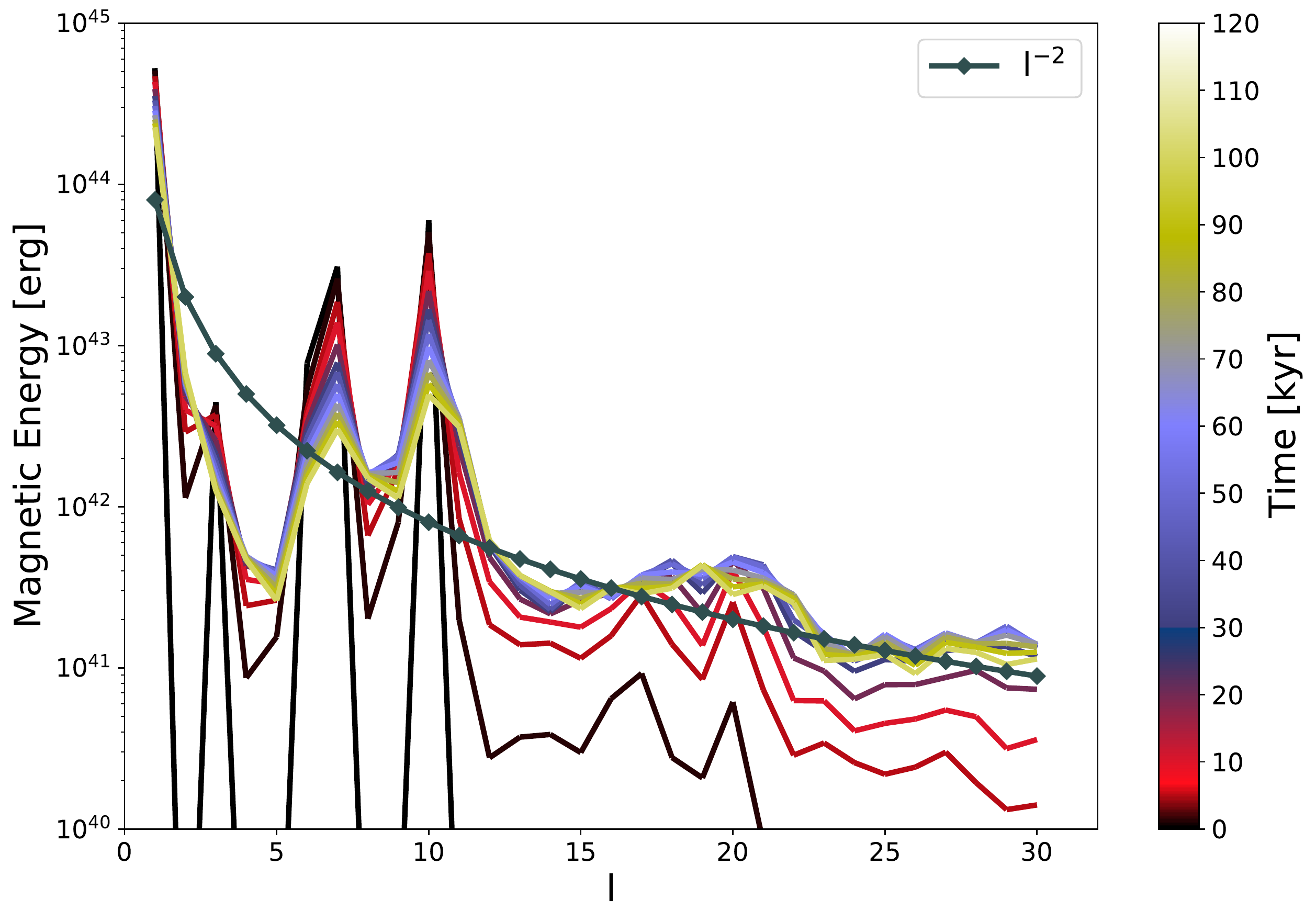}
\includegraphics[width=0.45\textwidth]{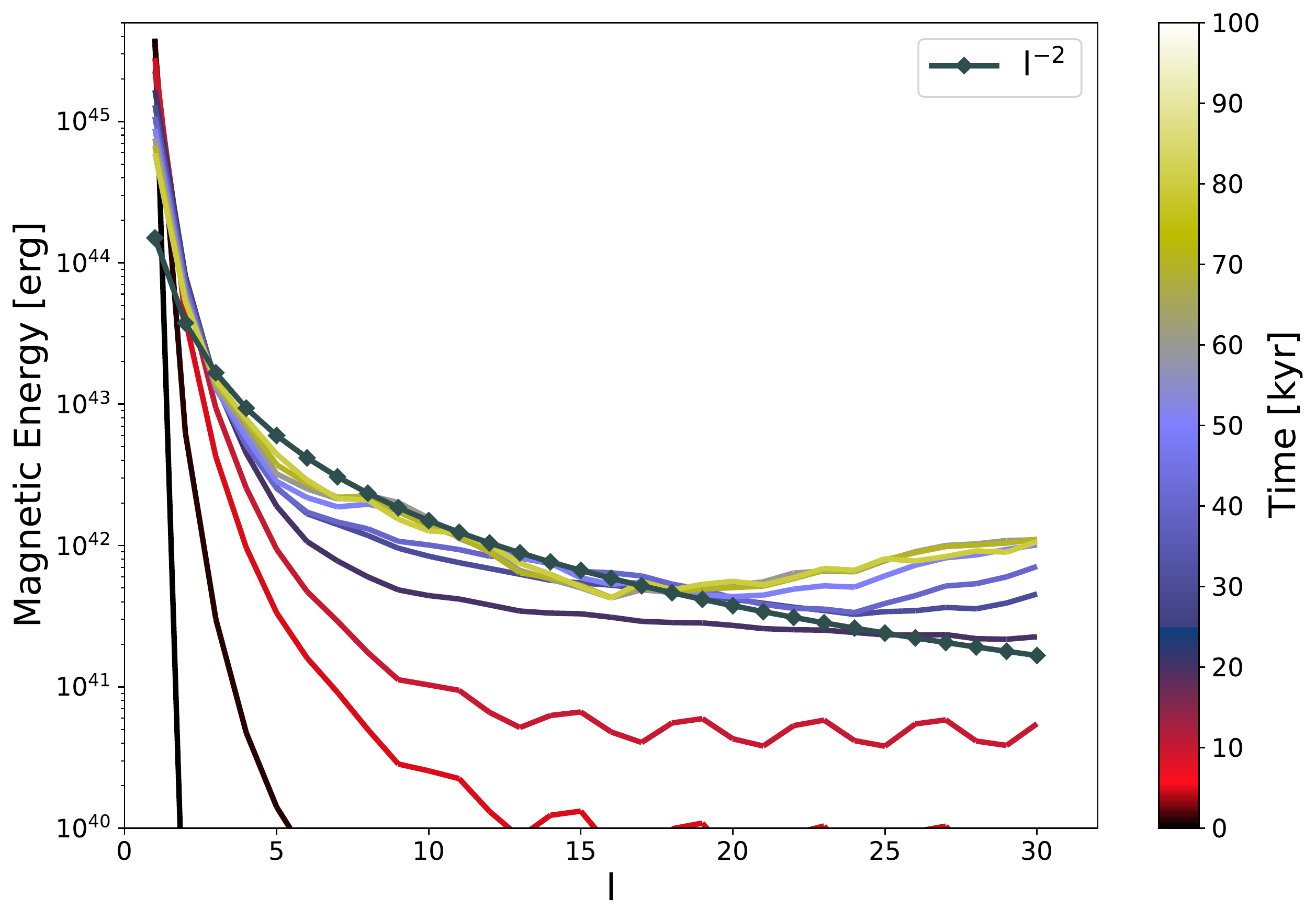}
\caption{$l$ energy spectrum up to $100 $ kyr for L10 model (left panel) and up to $85$ kyr for L1 model (right panel). See color bars to identify the ages. The $l^{-2}$ slope corresponds to the Hall cascade equilibrium distribution of magnetic energy over a quite broad range of multipoles.}
\label{fig: energy spectrum Dip Multi Scales models}
\end{figure*}

\subsubsection{Energy spectrum}

The evolution in time of the energy spectrum (as a function of $l$) is displayed in Fig. \ref{fig: energy spectrum Dip Multi Scales models}. On the left, we illustrate the L10 model, whereas on the right we show L1 model. Note that for the L10 model, one can infer the weights of the multipoles initially defined in the system by looking at the black energy spectrum in the left panel of Fig. \ref{fig: energy spectrum Dip Multi Scales models}. A redistribution of the magnetic energy over different spatial scales occurs in both cases. Nevertheless, L10 model tends to inject magnetic energy in $l=16-18$ and $l=20$, more than other modes. A similar, less evident bump in the energy spectrum appears at $l=27$ and $l=29$. These bumps are more evident at early stages of the evolution, e.g., up to $t=40$ kyr. 
The injected energy in these small-scale modes is insignificant with respect to the initially dominant modes in the system. Still this peculiar energy injection at small-scale structures could be a hint of Hall instability \citep{gourgouliatos2020} that will take place in such an initial field configuration for a higher magnetic Reynolds number. At later stages of the evolution, the magnetic energy is redistributed more homogeneously over the small-scale structures, and the lower part of the spectrum, e.g., from $l=12$ up to $l=30$ follows the $l^{-2}$ slope \citep{goldreich1992}. Nevertheless, the $l$-energy spectrum keeps a strong memory, at low $l$s, of the initial configuration for the whole evolution, e.g., $t=100$ kyr. This is because the largest scales have longer timescales, $\sim L^2/(f_h B)$, where $L$ is the length-scale of the field (related to $l$), therefore it is much harder to transfer energy out of/into them. Said in other words, the inertial range of the Hall cascade includes scales with sufficiently short timescales.

The transfer of magnetic energy over different spatial scales is smooth for L1 model compared to L10 model. For L1 model, the energy spectrum is well described by an $l^{-2}$ power-law up to $l=20$. For smaller scale multiples, e.g., $l > 20$, we notice an excess of energy. This injection of energy in the smallest structures grows in time and becomes evident at about $t\sim 70-80$ kyr, although it remains orders of magnitude lower than the dominant dipolar mode, $l=1$. We defer a deeper exploration of different initial magnetic field topologies and their astrophysical implications for future works.

\subsubsection{Poloidal and toroidal decomposition}

\begin{figure*}
\centering
\includegraphics[width=.43\textwidth]{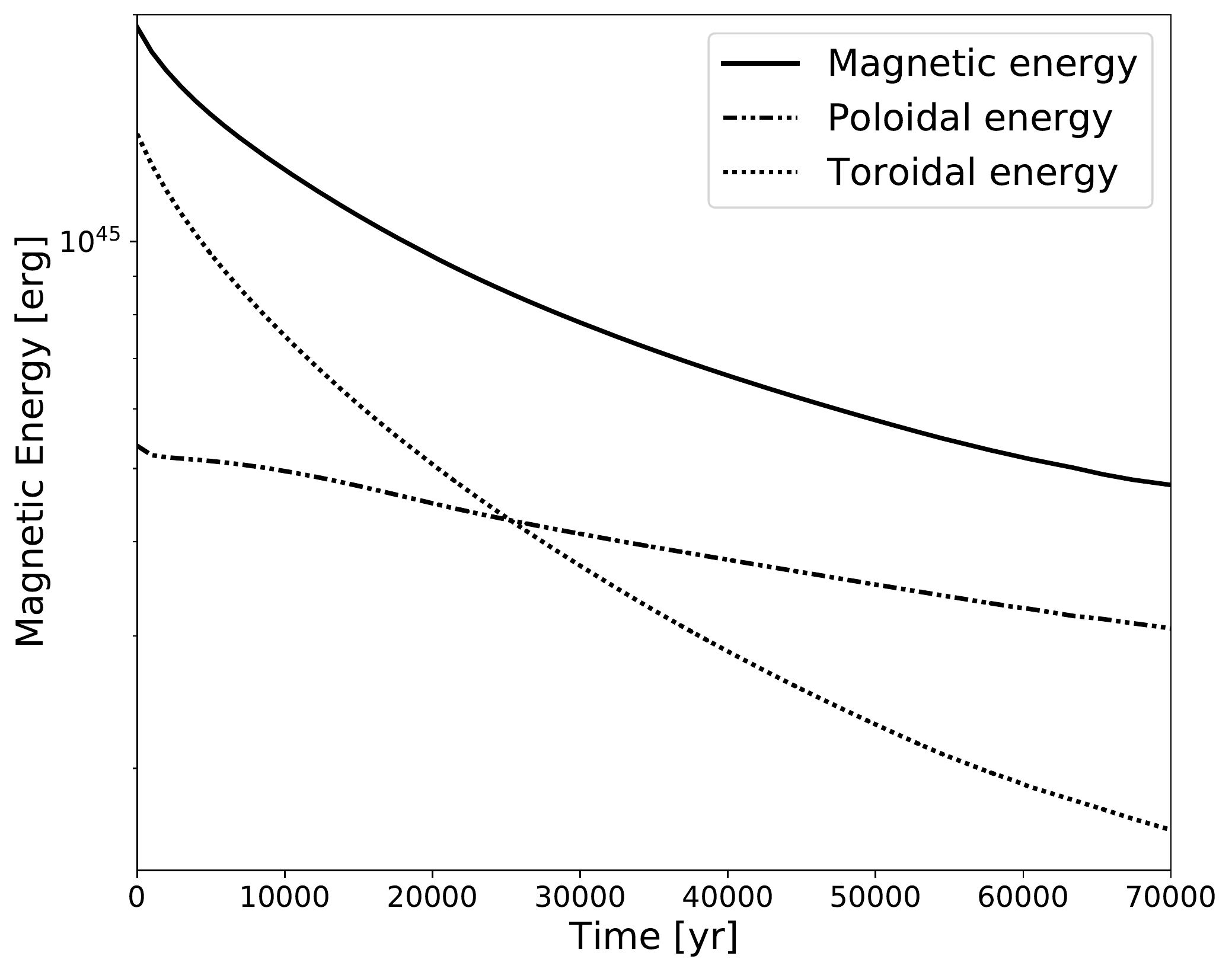}
\includegraphics[width=.5\textwidth]{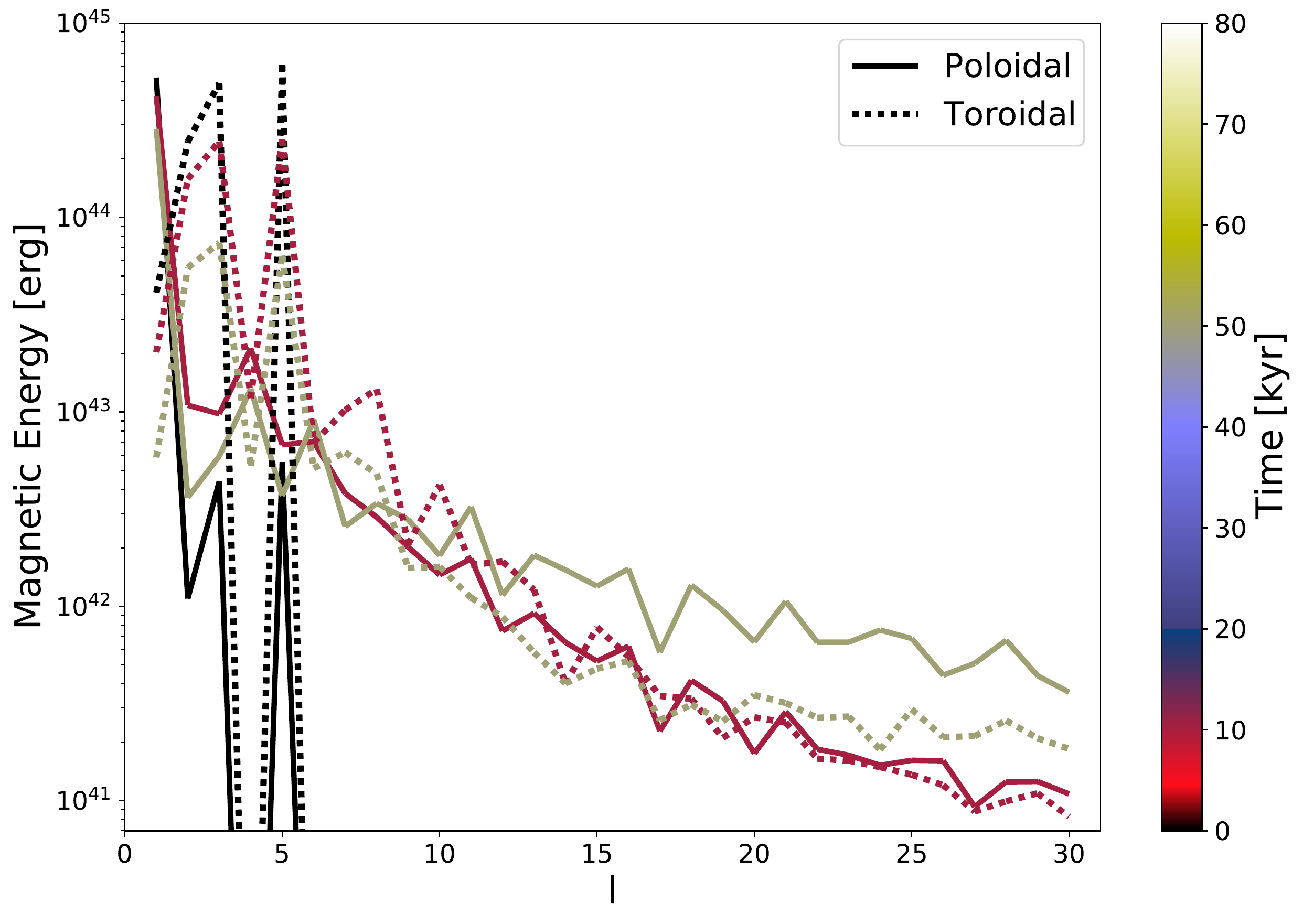}
\includegraphics[width=.43\textwidth]{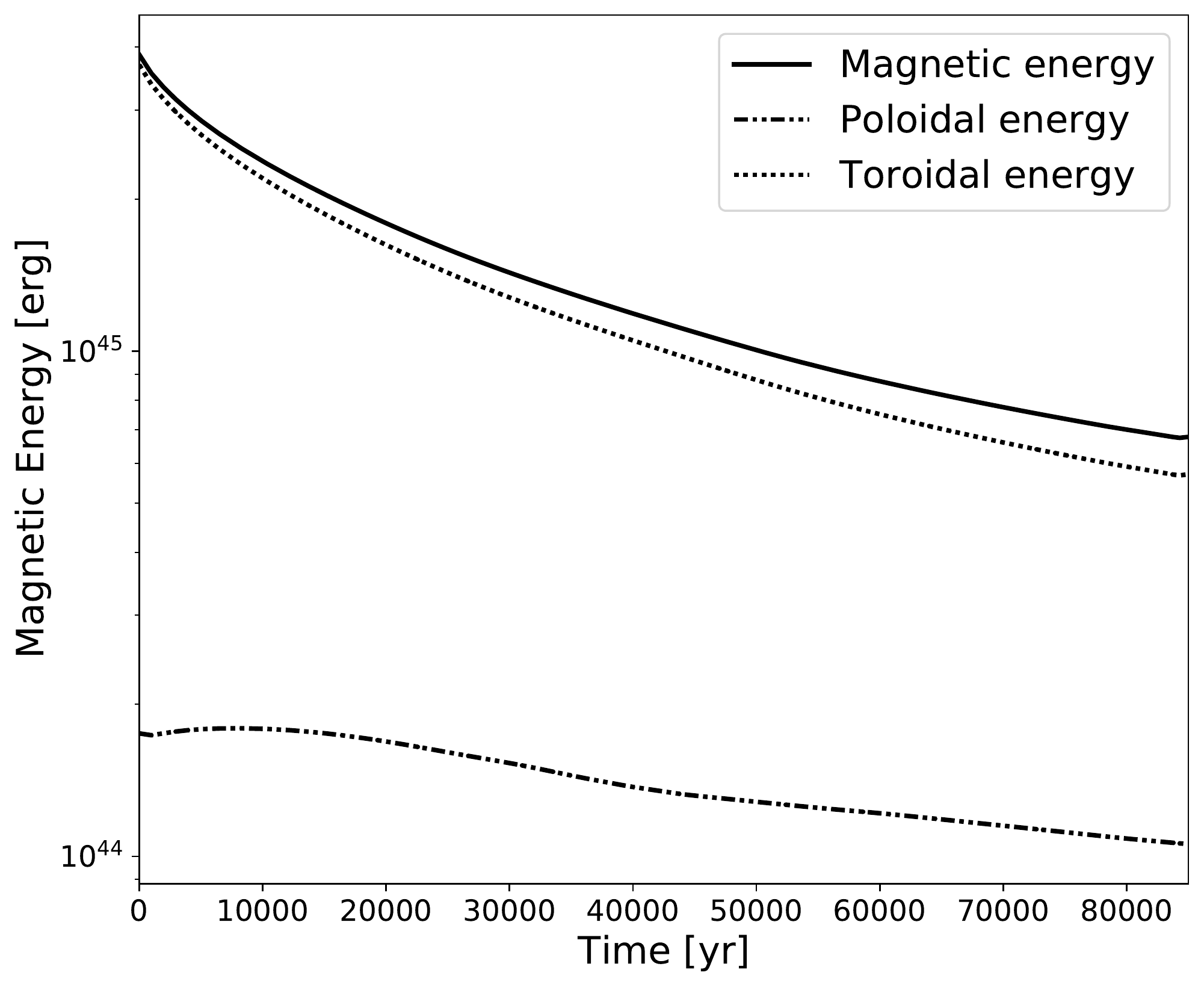}
\includegraphics[width=.5\textwidth]{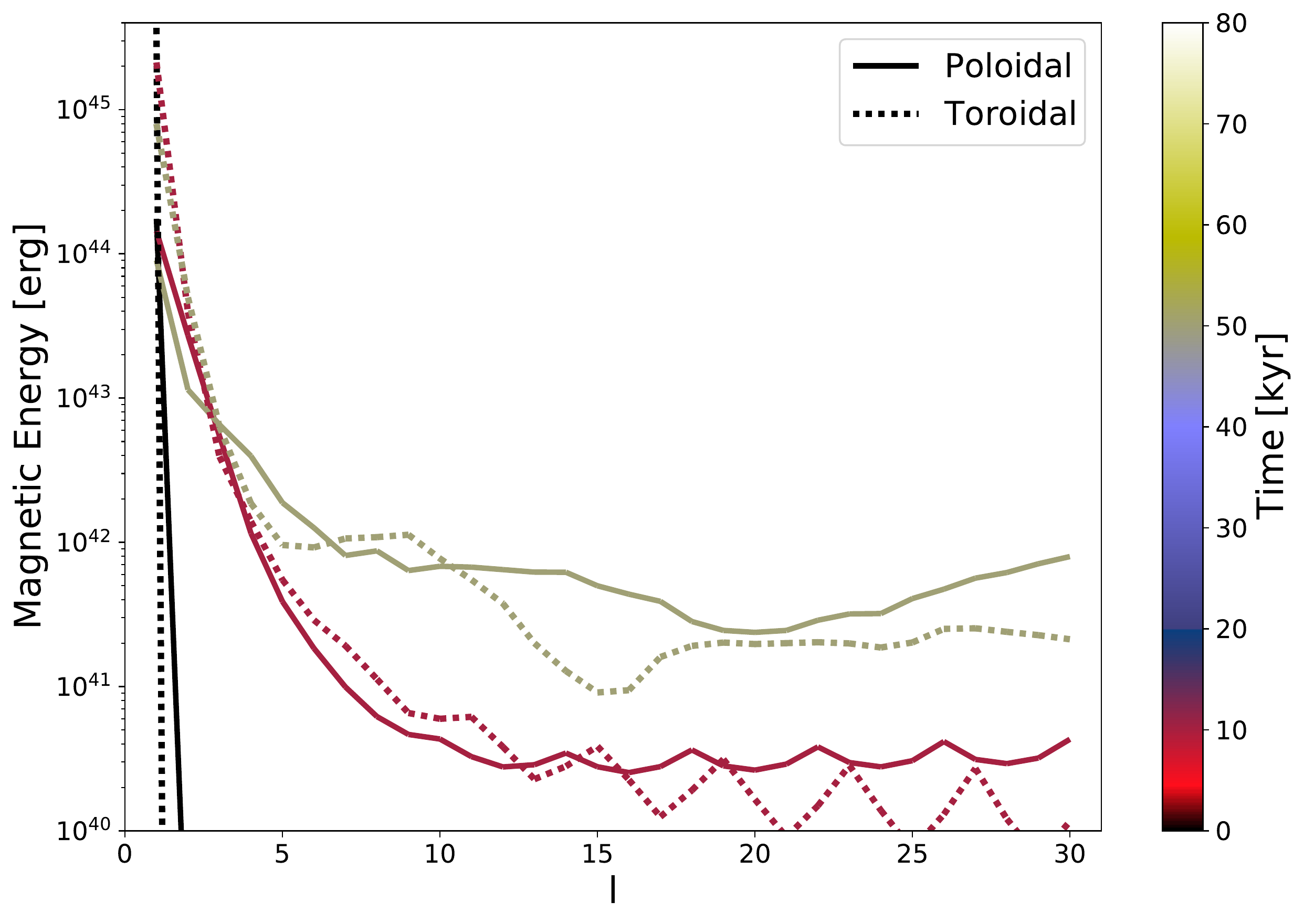}
\includegraphics[width=.43\textwidth]{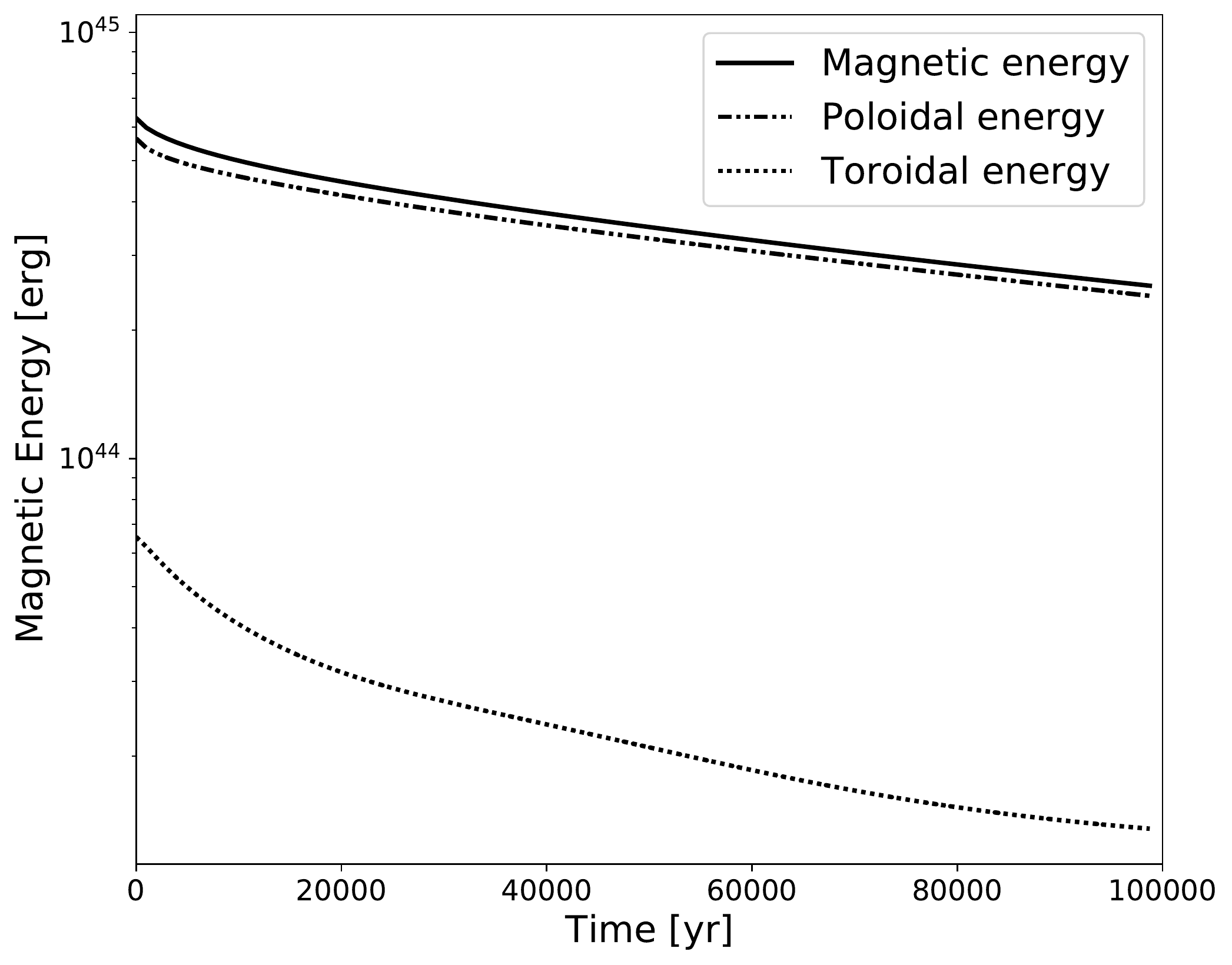}
 \includegraphics[width=.5\textwidth]{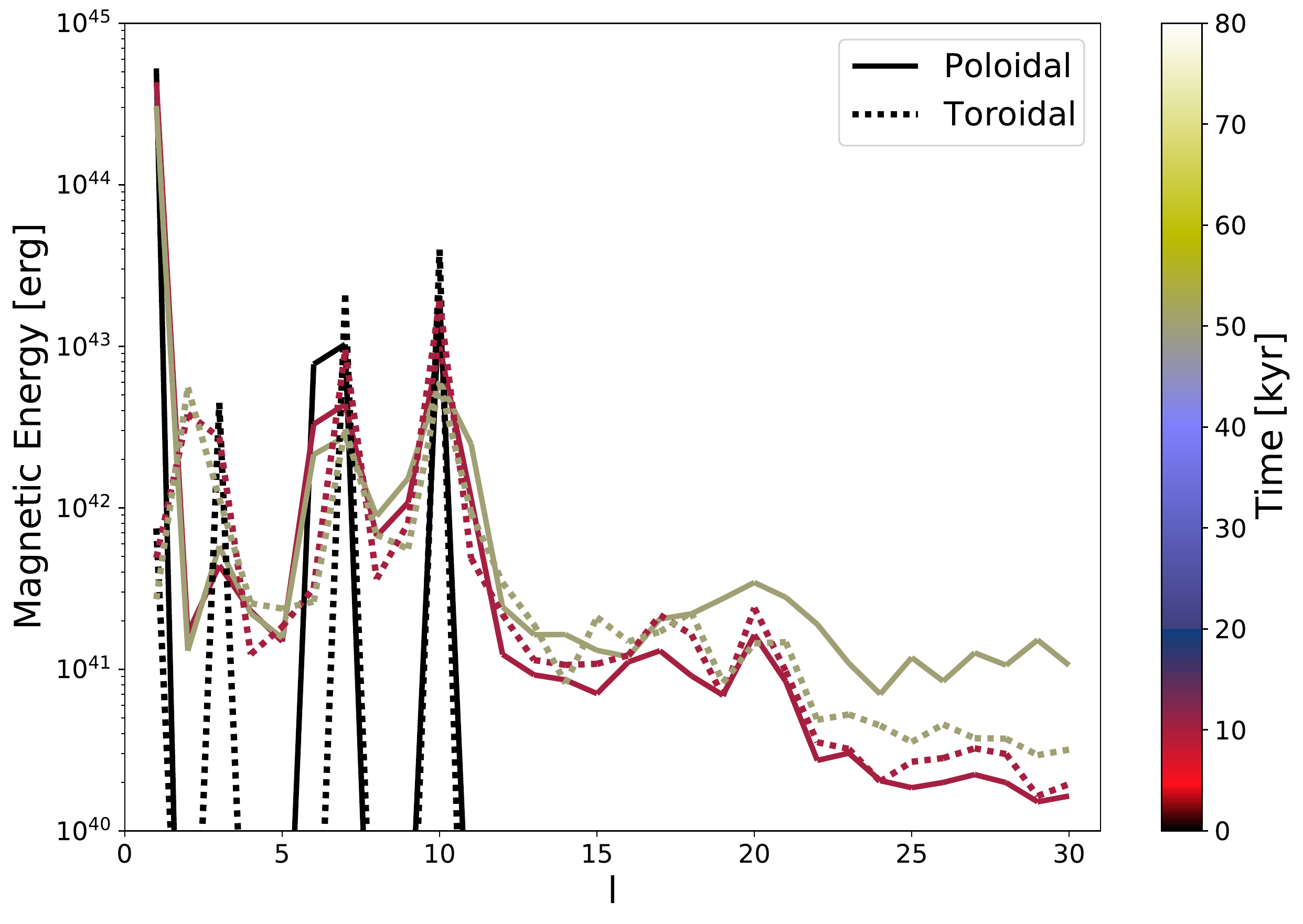}
\caption{Decomposition of the poloidal and toroidal magnetic energy. Poloidal magnetic energy is represented with dash-dotted lines, the toroidal energy with dots, and the total magnetic energy with solid lines. \textit{Upper panels:} \textit{L5} model. \textit{Central panels:} \textit{L1} model. \textit{Lower panels:} \textit{L10} model. \textit{Left Column:} Poloidal and toroidal decomposition of the magnetic energy as a function of time. 
\textit{Right column:} $l$ energy spectrum at $t=0$ (black), $t=10$ kyr (dark red), and after $50$ kyr (olive).}
\label{fig: Toroidal- Poloidal energy Spectrum LS Dip Multi}
\end{figure*}

Using the notation in eq. (\ref{eq: spectral magnetic energy}), we can decompose the magnetic energy into its poloidal and toroidal parts. In Fig. \ref{fig: Toroidal- Poloidal energy Spectrum LS Dip Multi}, we show the evolution of the poloidal and toroidal magnetic energy. 
% The poloidal energy is illustrated with dash-dotted lines, the toroidal energy with dots and the total magnetic energy with solid lines. In the top panels, we display the results of L5 model. The central panels correspond to L1 model and the bottom panels to L10 model. 
% On the left, we show the magnetic energy as a function of time, and on the right, we display the $l$ magnetic energy spectrum at $t=0,10$, and $50$ kyr.
At early evolutionary stages, the bulk of the magnetic energy of L5 model is stored in the toroidal field ($\sim 63 \%$), whereas the poloidal energy accounts for about $\sim 37\%$ of the total magnetic energy. Following the evolution, we note that the toroidal field tends to dissipate almost $5$ times faster than its poloidal counterpart, resulting in an inversion of the poloidal-toroidal ratio.
That can be explained because the toroidal energy is effectively redistributed in smaller-scale multipoles (which in turn dissipate faster) while most of the poloidal energy remains in the $l=1$ mode (upper right panel). Moreover, we observe that, after a few Hall timescales, the system reaches some sort of equipartition of the magnetic energy between the poloidal and the toroidal energy spectrum, as a result of the Hall-dominant evolution. For L1 model (central panels) most of the magnetic energy, e.g., $\sim 90\%$, is stored in the toroidal component. Instead, for L10 model (bottom panels), most of the magnetic energy is stored in the poloidal component. After $\sim 100$ kyr of evolution, the magnetic energy remains stored in the dominant mode, e.g., L1 maintains a toroidal-dominion, whereas L10 maintains a poloidal-dominion. Nevertheless, approximate equipartition of the magnetic energy between the poloidal and toroidal components is also reached at about $\sim 10$ kyr, but only at small scales. Large scales are not easily forgotten or created.

These results validate the fact that the system favours the redistribution of magnetic energy between poloidal and toroidal components for the purpose of stabilizing the evolution. Note, however, that attaining this saturated configuration (often called the Hall attractor and was first introduced by \cite{gourgouliatos2014}) takes some tens of kyr, which is the same timescale over which magnetars are usually active. During this stage, the spectra and topology do still depend on the initial configuration.

\begin{figure*}
\includegraphics[width=18cm, height=5.5cm]{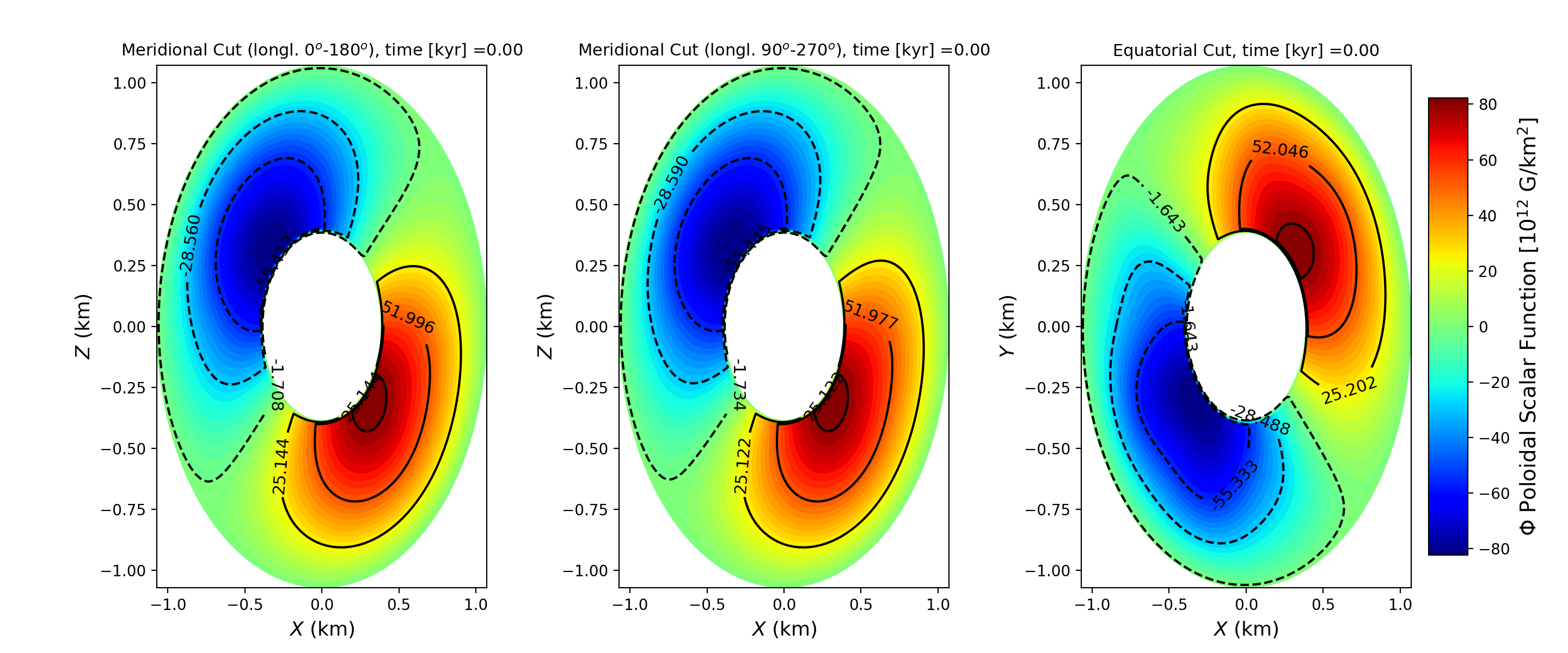}
\includegraphics[width=18cm, height=5.5cm]{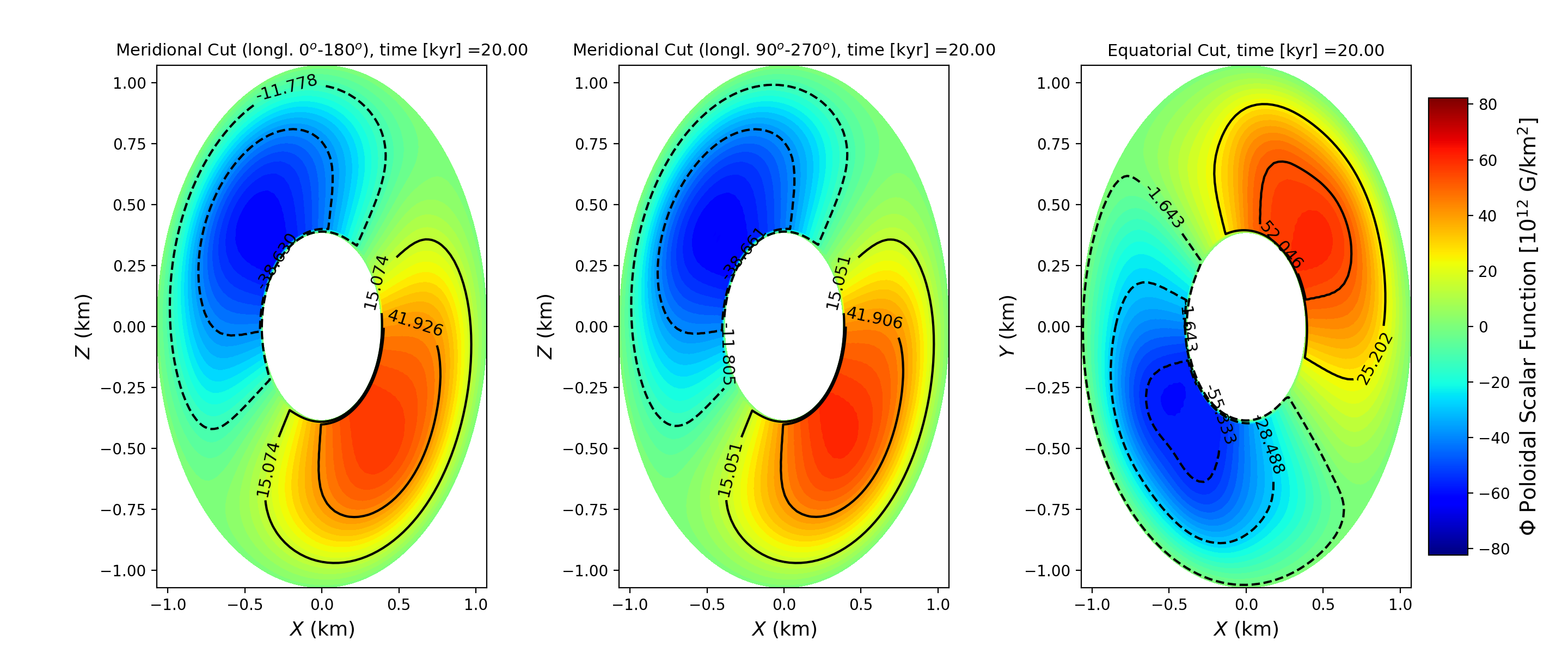}
\includegraphics[width=18cm, height=5.5cm]{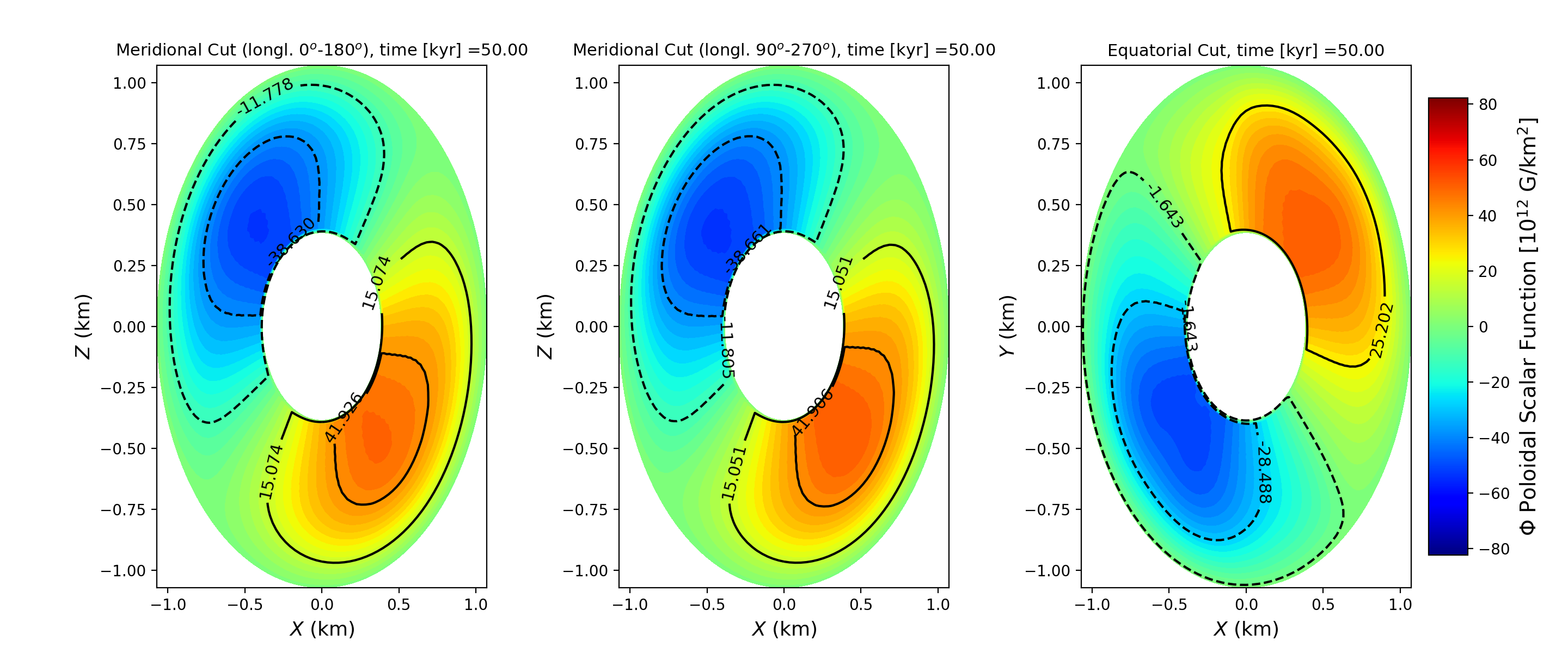}
\caption{\textit{Model L10}. Evolution of the $\Phi$ poloidal scalar function, at $0$, $20$ and $50$ kyr (from top to bottom). In the left panels, we show the meridional cuts at the longitudes $0-180^\circ$ (cutting through the center of patches I and III). In the central panels we illustrate the meridional cut at the longitudes $90-270^\circ$ (through the center of patches II and IV). In the right panels, we represent the equatorial 2D cuts. The crust is greatly enlarged for visualisation purposes.}
\label{fig: Phi Poloidal function Multi}
\end{figure*}

\begin{figure*}
\includegraphics[width=18cm, height=5.5cm]{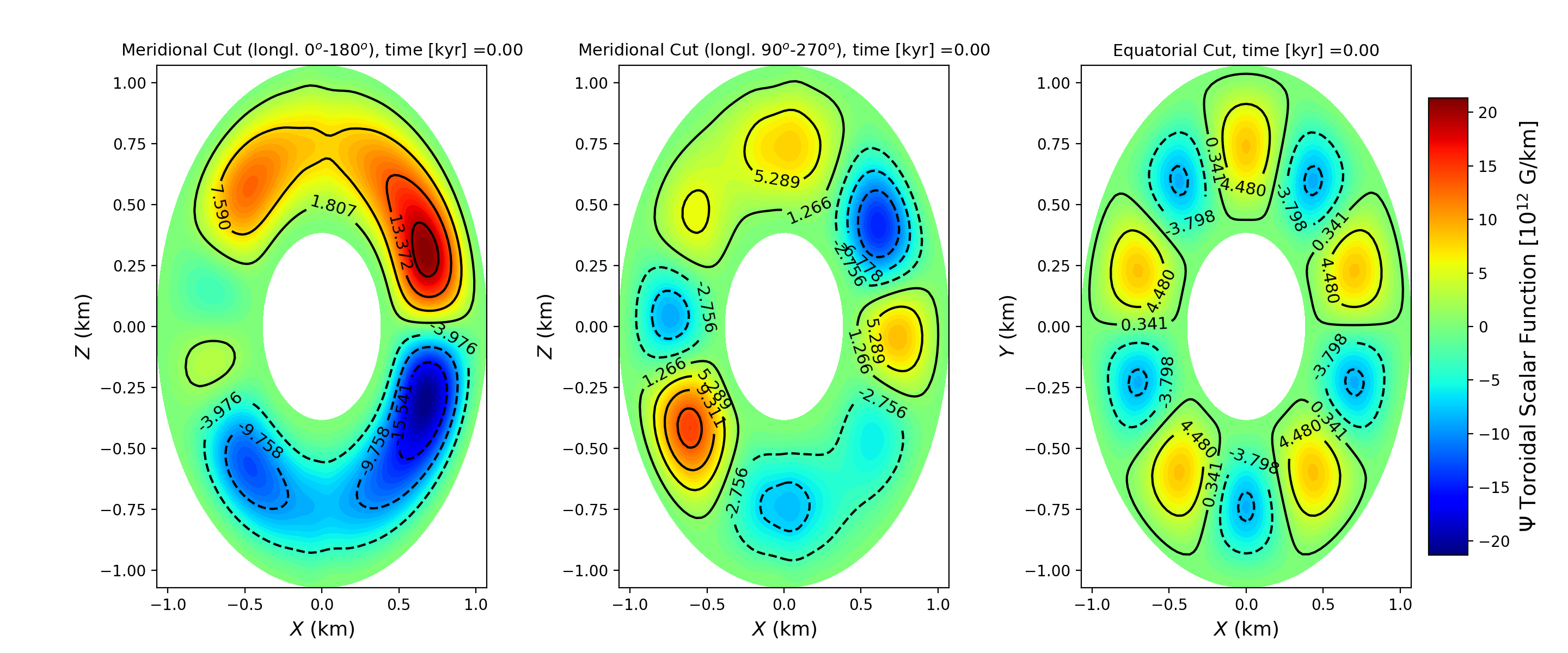}
\includegraphics[width=18cm, height=5.5cm]{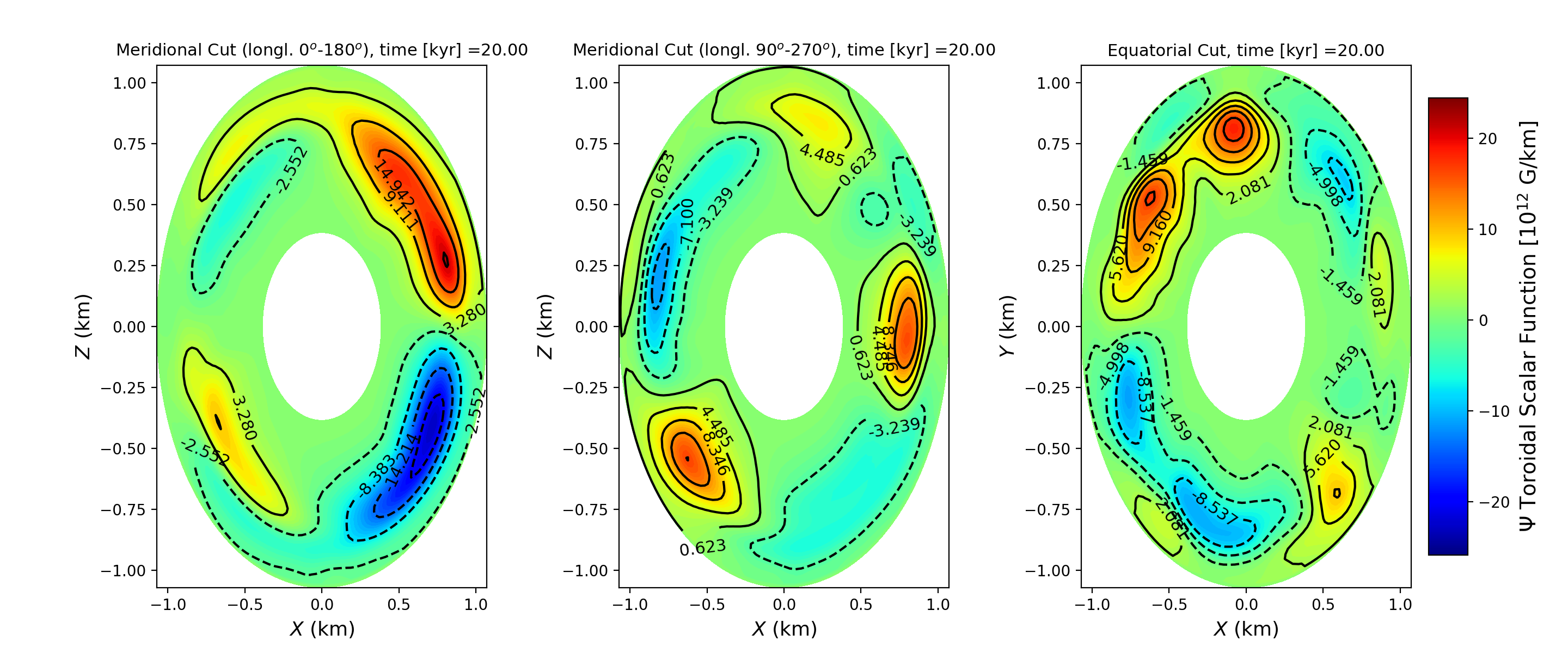}
\includegraphics[width=18cm, height=5.5cm]{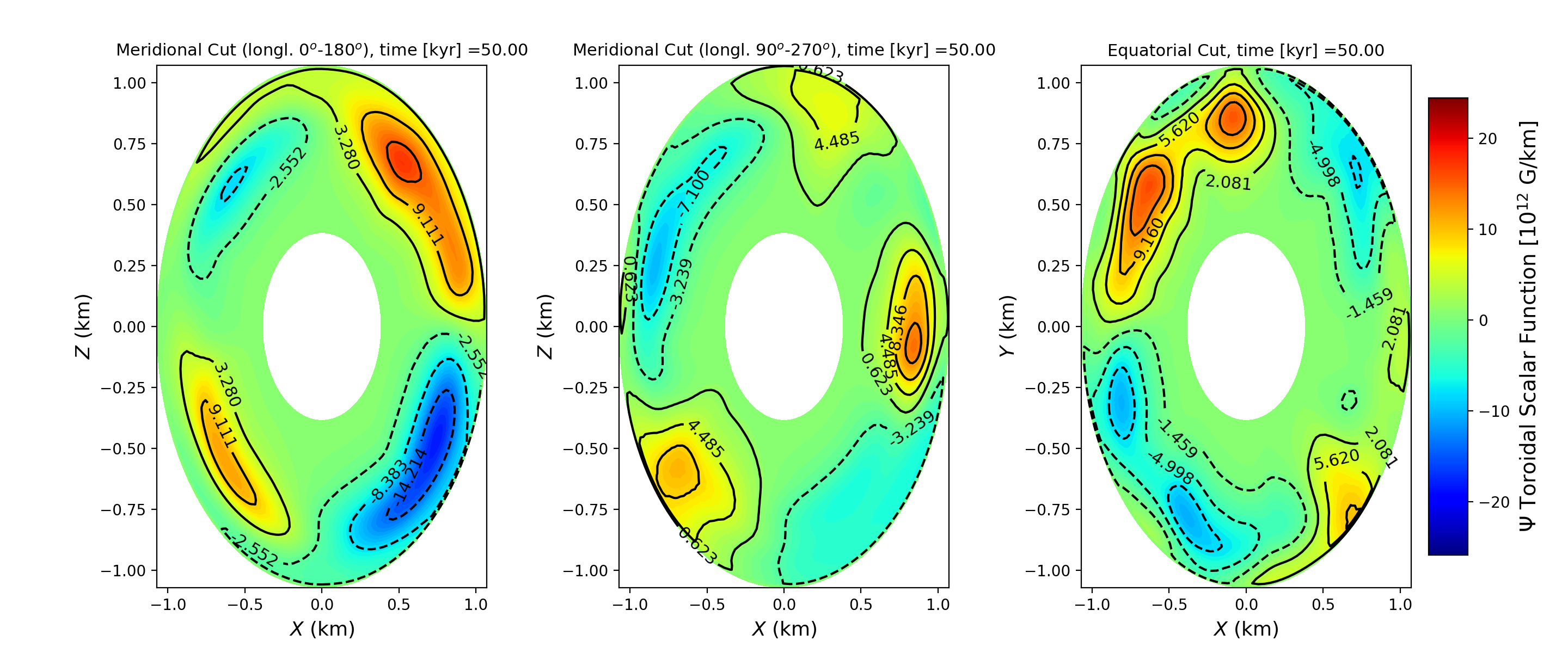}
\caption{\textit{Model L10}. The evolution of 2D cuts of the $\Psi$ toroidal scalar function. Same cuts and times as Fig. \ref{fig: Phi Poloidal function Multi}}
\label{fig: Psi Toroidal function Multi}
\end{figure*}

In Fig. \ref{fig: Phi Poloidal function Multi} and \ref{fig: Psi Toroidal function Multi}, we illustrate the meridional cuts at longitudes $0-180^\circ$ (left panels), $90-270^\circ$ (central panels), and equatorial cuts (right panels), of the poloidal $\Phi$ and toroidal $\Psi$ scalar functions for L10 model. The top panels correspond to the initial configuration, i.e., t=0, the central panels to $20$ kyr, whereas the bottom panels to $50$ kyr. Throughout the evolution, the poloidal function, which is initially dipole-dominated (lower right panel of Fig. \ref{fig: Toroidal- Poloidal energy Spectrum LS Dip Multi}), suffers only slight changes. On the contrary, the initially more complex toroidal scalar function is dominated by the $l=10$ mode. The latter presents some important rearrangements. Moreover, a drifting of the toroidal scalar function toward the surface of the star occurs, pointing up the need to couple this code with the evolution in the magnetosphere.

\section{Conclusion and outlook}
\label{sec: conclusion}

We have developed a new 3D code, \textit{MATINS}, for the magneto-thermal evolution of NS, of which we present here the magnetic field formalism and the first obtained results. The code is based on finite volume scheme applied to the cubed-sphere formalism, it is second-order accurate in space and fourth-order accurate in time. The cubed-sphere formalism is a peculiar gridding technique widely used in different fields of physics, and it allows to solve partial differential equations in spherical geometry avoiding the axis singularity problem: a common problem that emerges when adopting finite volume/difference scheme in spherical coordinates.

We have shown that \textit{MATINS} is stable and can follow to late times the evolution of the internal magnetic field in the crust of NSs. It conserves the total energy contained in the system and the divergence-free condition of the magnetic field. Moreover, it has been extensively tested, against analytical solution, e.g., the purely resistive test (section \ref{subsec: bessel test}) and numerical axisymmetric solutions replicable by our 2D code (section \ref{subsec: axisymmetric - comparison 2D and 3D}). 

\textit{MATINS} deals with realistic EoS and microphysics, important in particular for the local, temperature-dependent values of the conductivity. In this first magnetic oriented paper, we prescribe for simplicity an analytical formula for the evolution of the internal temperature, assumed to be homogeneous \citep{yakovlev2011}.

We have explored different initial field configurations (section \ref{appendix: initial conditions}) using this code. Our simulations (section \ref{sec: results and simulations}) confirm that for a strong enough magnetic field, e.g., $10^{14}-10^{15}$ G, the Hall cascade redistributes the energy across a wide range of scales, with a slope $\sim l^{-2}$. Moreover, an approximate equipartition of energy between the poloidal and toroidal components happens at small scales. Nevertheless, attaining this saturated configuration (often call Hall attractor) takes some tens of kyr, which is the same timescale over which magnetars are usually active. During this stage, the spectra and topology keep a strong memory of the initial large scales, which are much harder to be restructured or created. This indicates that the type of large-scale configuration attained during the neutron star formation is crucial to determine the magnetic field topology at any age of its evolution.

% during the life of a magnetar, the initially defined small-scale structure in the system lose the memory, on the contrary the large scale structure keep the memory, confirming the importance of the of the initial magnetic field configuration defined in the system.  

We also studied the difference (section \ref{subsec: impact of microphysics}) between having a fixed (i.e., no time-dependent) 
prescription for the temperature,
% prescription for the microphysics
or employing a more realistic scenario (i.e., simplified cooling), for an identical initial field topology. We remark that for a high enough temperature ($\sim 10^9$ K), the field evolution is Ohmic-dominant with negligible redistribution of the magnetic energy over the different spatial scales. Instead, for a lower temperature values ($\sim 10^8$ K), the field evolution is Hall-dominant, and the impacts of the magnetic resistivity on the topology and spectra are visible but minor. That is due to the fact that large scales are pretty insensitive on the exact value of the resistive scales. These differences feature the influence of the temperature-dependent microphysics on our results and point up the need of having a 3D magneto-thermal code coupled with realistic microphysics.

% due to the fact that the evolution here considered is Hall-dominated and the main, large scales are pretty insensitive on the exact value of the resistive scales. Such differences can become much more evident if more simplified, analytical treatment of the electron density profile and diffusivity are taken, or if the initial field is very turbulent. 

We are currently implementing the 3D anisotropic thermal evolution, taking into consideration its feedback on the magnetic evolution (and vice-versa). \textit{MATINS} code will be then compared and tested with observations, using for the first time the state-of-the-art microphysics (crucial to derive sound luminosities or temperature maps). Moreover, a detailed study exploring different initial field configurations and their physical interpretation is also planned. Nevertheless, it is important to highlight that a coupling of the internal crustal evolution to the magnetosphere and to the core of the neutron star is crucial for a complete study.

\section*{Acknowledgements}

We thank Borja Minano and Carlos Palenzuela for optimizing the code and Stefano Ascenzi for developing the 3D anisotropic thermal formalism for this code. We also thank the referee for her/his useful comments. CD and NR are supported by the ERC Consolidator Grant “MAGNESIA” No. 817661 (PI: Rea) and this work has been carried out within the framework of the doctoral program in Physics of the Universitat Aut\`onoma de Barcelona. This work was also partially supported by the program Unidad de Excelencia María de Maeztu CEX2020-001058-M. 
DV is supported by the European Research Council (ERC) under the European Union’s Horizon 2020 research and innovation programme (ERC Starting Grant "IMAGINE" No. 948582, PI: DV). JAP acknowledges support from the Generalitat Valenciana (PROMETEO/2019/071) and the AEI grant PID2021-127495NB-I00.

%%%%%%%%%%%%%%%%%%%%%%%%%%%%%%%%%%%%%%%%%%%%%%%%%%
\section*{Data Availability}
Data available on request.



\bibliographystyle{mnras}
\bibliography{mnras_template}

%\begin{thebibliography}{99}
%\end{thebibliography}

%%%%%%%%%%%%%%%%%%%%%%%%%%%%%%%%%%%%%%%%%%%%%%%%%%

%%%%%%%%%%%%%%%%% APPENDICES %%%%%%%%%%%%%%%%%%%%%

\appendix

\section{Cubed-sphere formalism}

\subsection{Coordinates transformations}\label{app:coordinates}

All evolution calculations are performed in the cubed sphere coordinates, with a few exceptions. The potential boundary conditions are imposed in spherical coordinates (section \ref{subsec: outer B.C}). Thus, a transformation from spherical to cubed-sphere coordinates is needed at each magnetic timestep. This transformation has been also used in some other cases, in particular when defining the initial magnetic field in spherical coordinates, e.g., Bessel test (section \ref{subsec: bessel test}). Instead, the transformation from cubed-sphere to spherical coordinates is applied to generate the output files.

\begin{figure}
	\centering
	\includegraphics[width=0.5\textwidth]{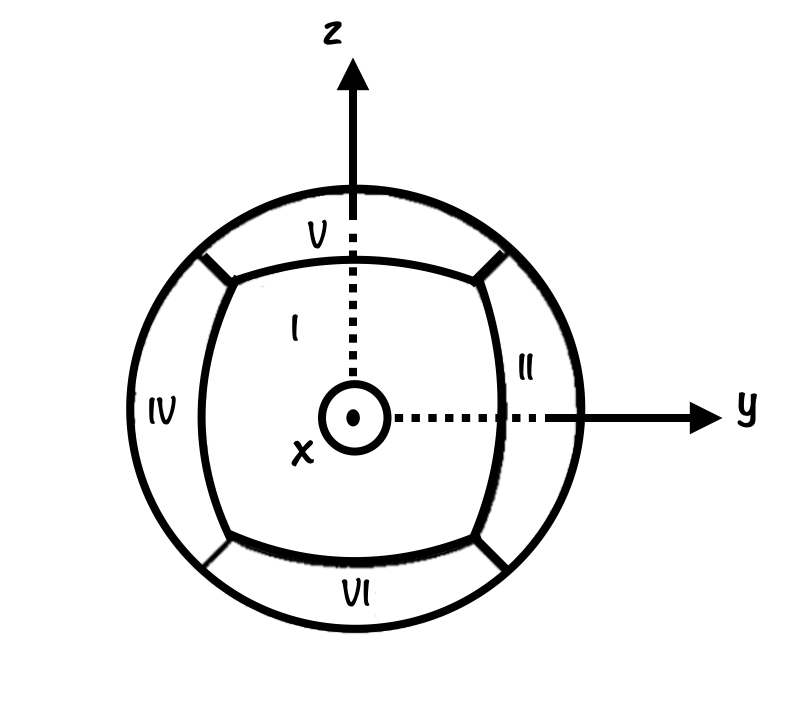}
	\caption{The cubed sphere represented with respect to Cartesian coordinates. The x-axis is directed outwards piercing the centers of patch I ($\hat{e}_x$) and III ($-\hat{e}_x$, behind, not visible), the y-axis goes to the right, piercing the centers of patch II ($\hat{e}_y$) and IV ($-\hat{e}_y$), the-z axis is directed upwards, piercing the centers of patch V ($\hat{e}_z$) and VI ($-\hat{e}_z$).}
	\label{fig: Cubed-sphere in cartesian axis}
\end{figure}

The spherical coordinates consist, as usual, of: $r\in [R_c,R_\star]$, the distance to the origin contained between the crust-core interface and the surface; $\theta \in [0,\pi]$, the co-latitude, also called polar or inclination angle, i.e. the angle with respect to the North pole ($x=y=0$, positive $z$ in Cartesian coordinates); $\phi \in [0,2\pi]$, the azimuth, i.e. the angle defined in the $x-y$ plane, starting from the $x$-axis. Each patch of the unit sphere is centered around a Cartesian axis, as shown in Fig.~\ref{fig: Cubed-sphere in cartesian axis}.
The transformations between the different coordinates are the same as in, e.g., \cite{ronchi96,lehner05}.

The coordinate directions of the patches are indicated in the exploded view of Fig.~\ref{fig:full_grid} and can be described qualitatively as follows:\footnote{The direction is indicated for brevity by $\rightarrow$, and is exact only at the center of each patch, with more non-trivial directions as they approaches the edges, where the angular deviation increases up to $\pi/4$ (at the three-patch common corners).}: 
\begin{itemize}
	\item Patch I: center in $x=1$ ($\theta=\frac{\pi}{2}, \Phi=0$); $\hat{e}_\xi \rightarrow \hat{e}_y$, $\hat{e}_\eta \rightarrow \hat{e}_z$.
	\item Patch II: center in $y=1$ ($\theta=\frac{\pi}{2}, \Phi=\frac{\pi}{2}$); $\hat{e}_\xi \rightarrow -\hat{e}_x$, $\hat{e}_\eta \rightarrow \hat{e}_z$.
	\item Patch III: center in $x=-1$ ($\theta=\frac{\pi}{2}, \Phi=\pi$); $\hat{e}_\xi \rightarrow -\hat{e}_y$, $\hat{e}_\eta \rightarrow \hat{e}_z$.
	\item Patch IV: center in $y=-1$ ($\theta=\frac{\pi}{2}, \Phi=\frac{3\pi}{2}$); $\hat{e}_\xi \rightarrow \hat{e}_x$, $\hat{e}_\eta \rightarrow \hat{e}_z$.
	\item Patch V: center in $z=1$ ($\theta=0$, $\Phi$ undefined); $\hat{e}_\xi \rightarrow \hat{e}_y$, $\hat{e}_\eta \rightarrow -\hat{e}_x$.
	\item Patch VI: center in $z=-1$ ($\theta=\pi$, $\Phi$ undefined); $\hat{e}_\xi \rightarrow \hat{e}_y$, $\hat{e}_\eta \rightarrow \hat{e}_x$.
\end{itemize}

With this notation, we have the following relations between cubed sphere, spherical and Cartesian coordinates, for each patch:\footnote{The transformation from cubed sphere to Cartesian coordinates is taken from sec. 4.1.3 of \cite{lehner05}. They use a grid equally spaced in $b\equiv X$ and $a\equiv Y$, not equally spaced in $\xi$ and $\eta$. Note also that their patches 0-5 correspond to I-VI here, in the same order. The $\theta(\xi,\eta)$ and $\phi(\xi,\eta)$ are derived in this work with simple trigonometric relations starting from the definitions $X(\theta,\phi)$ and $Y(\theta,\phi)$. For the polar patches, we have employed the identities $\sin(\arctan\beta) = \pm  \mid \beta\mid /\sqrt{1+\beta^2}$  (i.e., $"+"$ if $\beta>0$ and $"-"$ if $\beta<0$) and $\cos(\arctan\beta)=1/\sqrt{1+\beta^2}$.}

\begin{itemize}
  \item Patch I (Equator)
  \begin{eqnarray}
  X&=&y/x=\tan \phi \nonumber\\
  Y&=&z/x=1/\tan \theta \cos \phi\nonumber\\
  x&=& r/ \sqrt{\delta}, \hspace{1mm}y= rX/ \sqrt{\delta}, \hspace{1mm}z= rY/ \sqrt{\delta} \nonumber\\
  \theta &=& \arctan[(\cos\xi\tan\eta)^{-1}] = \arctan(C/Y) \nonumber\\
  \phi &=& \xi
  \end{eqnarray}

  \item Patch II (Equator)
  \begin{eqnarray}
  X&=& - x/y=-1/ \tan \phi \nonumber\\
  Y&=&z/y=1/\tan \theta \sin \phi \nonumber\\
  x&=& - r X/ \sqrt{\delta}, \hspace{1mm}y= r/ \sqrt{\delta}, \hspace{1mm}z= rY/ \sqrt{\delta} 
  \nonumber\\
  \theta &=&  \arctan(C/Y) \nonumber\\
  \phi &=& \xi + \frac{\pi}{2}
  \end{eqnarray}
  
  \item Patch III (Equator)
  \begin{eqnarray}
  X&=& y/x=\tan \phi \nonumber\\
  Y&=&-z/x=- 1/\tan \theta \cos \phi \nonumber\\
  x&=& - r/ \sqrt{\delta}, \hspace{1mm}y= - rX/ \sqrt{\delta}, \hspace{1mm}z= rY/ \sqrt{\delta}\nonumber\\
  \theta &=&  \arctan(C/Y) \nonumber\\
  \phi &=& \xi + \pi
  \end{eqnarray}

  \item Patch IV (Equator)
  \begin{eqnarray}
  X&=& - x/y=-1/ \tan \phi \nonumber\\
  Y&=& - z/y= - 1/\tan \theta \sin \phi\nonumber\\
  x&=& rX/ \sqrt{\delta}, \hspace{1mm}y= - r/ \sqrt{\delta}, \hspace{1mm}z= rY/ \sqrt{\delta}\nonumber\\
  \theta &=&  \arctan(C/Y) \nonumber\\
  \phi &=& \xi + \frac{3\pi}{2}
  \end{eqnarray}

  \item Patch V (North)
  \begin{eqnarray}
  X&=& y/z= \tan \theta \sin \phi \nonumber\\
  Y&=&-x/z= - \tan \theta \cos \phi\nonumber\\
  x&=& - rY/ \sqrt{\delta}, \hspace{1mm}y=  rX/ \sqrt{\delta}, \hspace{1mm}z= r/ \sqrt{\delta}\nonumber\\
  \theta &=&   \arctan\sqrt{ \delta - 1} \nonumber\\
  {\tt if } \hspace{1mm} (X>0,  Y<0) \hspace{2mm}
  \phi &=& - \arctan(X/Y) \hspace{2mm} ({\tt region}  \hspace{1mm}\alpha)
  \nonumber\\
  {\tt if } \hspace{1mm} (X>0,  Y>0) \hspace{2mm}
  \phi &=& \pi - \arctan(X/Y) \hspace{2mm} ({\tt region}  \hspace{1mm}\beta)
    \nonumber\\
  {\tt if } \hspace{1mm} (X<0,  Y>0) \hspace{2mm}
  \phi &=& \pi - \arctan(X/Y) \hspace{2mm} ({\tt region}  \hspace{1mm}\gamma)
    \nonumber\\
  {\tt if } \hspace{1mm} (X<0,  Y<0) \hspace{2mm}
  \phi &=& 2\pi - \arctan(X/Y) \hspace{2mm} ({\tt region}  \hspace{1mm}\delta)   \nonumber\\
  \end{eqnarray}
   
  \item Patch VI (South)
  \begin{eqnarray}
  X&=& - y/z= - \tan \theta \sin \phi \nonumber\\
  Y&=&-x/z= - \tan \theta \cos \phi \nonumber\\
 x&=& rY/ \sqrt{\delta}, \hspace{1mm}y=  rX/ \sqrt{\delta}, \hspace{1mm}z= -r/ \sqrt{\delta}\nonumber \\
   \theta &=&   \pi - \arctan(\sqrt{\delta}-1)  
   \nonumber\\
  {\tt if } \hspace{1mm} (X>0,  Y<0) \hspace{2mm}
  \phi &=& \pi + \arctan(X/Y) \hspace{2mm} ({\tt region}  \hspace{1mm}\alpha)
  \nonumber\\
  {\tt if } \hspace{1mm} (X>0,  Y>0) \hspace{2mm}
  \phi &=&\arctan(X/Y) \hspace{2mm} ({\tt region}  \hspace{1mm}\beta)
    \nonumber\\
  {\tt if } \hspace{1mm} (X<0,  Y>0) \hspace{2mm}
  \phi &=& 2\pi + \arctan(X/Y) \hspace{2mm} ({\tt region}  \hspace{1mm}\gamma)
    \nonumber\\
  {\tt if } \hspace{1mm} (X<0,  Y<0) \hspace{2mm}
  \phi &=& \pi + \arctan(X/Y) \hspace{2mm} ({\tt region}  \hspace{1mm}\delta)
  \end{eqnarray}
\end{itemize}
Note that along the equatorial-centered patches I-II-III-IV, the $\xi$ coordinate coincides with the $\phi$ coordinate in spherical, with a phase shift of $(0,\pi/2,\pi,3\pi/2)$ respectively, and the transformation into $\theta$ coordinate is the same in all the four patches (since they cover the same co-latitude).

For the polar patches, the transformation is less trivial. Remember also that the $\arctan$ function tend to $\pi/2$ (i.e., patch I-IV and $X>0$ in patch V and VI) and $3 \pi/2$ (i.e., $X<0$ in patch V and VI) if the argument tends to $\pm\infty$ (i.e., when the denominator $Y$ of the ratios $X/Y$ and $C/Y$ go to zero). 

\begin{figure}
	\centering
	\includegraphics[width=0.5\textwidth]{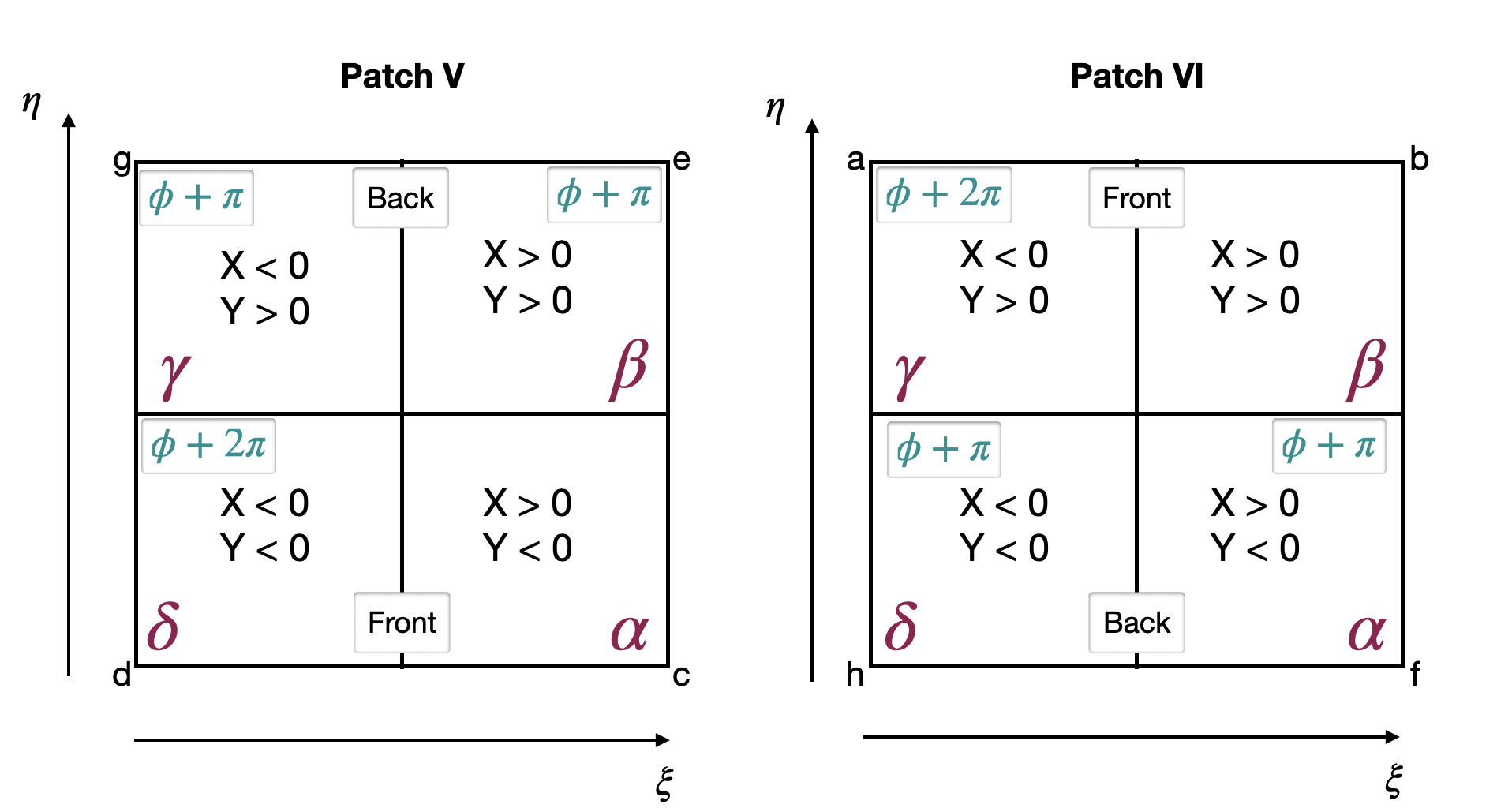}
	\caption{Subdivision of patch V and patch VI. Each of these patches is divided into four sub-regions, and each of these sub-regions has a different sign of the $X/Y$ ratio. This subdivision is crucial to properly define $\phi$ in the range $[0:2\pi]$.
% 	and consequently to obtain a positive $\theta$ in patch V and a negative $\theta$ in patch VI (i.e., $\theta_{VI} \equiv \pi+ \theta_{VI}$). 
	} 
	\label{fig: subregions patch V and patch VI}
\end{figure}

In order to define $\phi$ in the range $[0;2\pi]$ and $\theta$ in the range $[0;\pi]$, in patch V and patch VI, a subdivison of each of these patches is needed. This subdivision is crucial since the sign of $X/Y$ ratio changes in these subregions defined in patch V and patch VI of Fig. \ref{fig: subregions patch V and patch VI}. As a consequence, to guarantee that $\phi$ goes from $[0;2\pi]$, a $"+\pi"$ (i.e., subregion $\beta$ and $\gamma$ of patch V and subregion $\alpha$ and $\delta$ of patch VI) or a $"+2\pi"$ (i.e., subregion $\delta$ of patch V and subregion $\gamma$ of patch VI) must be added to the expression of $\phi$.

% \JP{in the sense that exactly along the axis the azimuthal and meridional direction are not defined. Therefore, the description of a non-radial vector (i.e., having non-zero tangential components), is not trivial and poses mathematical and numerical problems. }

\subsection{Jacobians}
\label{Appendix: Jacobians}

In order to transform vectors from spherical coordinates to cubed sphere coordinates, we need the Jacobian matrices. %Moreover, in the outer boundary conditions the reconstruction is done over the spherical components, then translated into cubed-sphere ones.
Hereafter we indicate only the $2\times 2$ Jacobian relating the transformation of the tangential components, since the radial coordinate is the same.

\begin{itemize}
	\item Patch I-IV (Equator)
	\begin{equation}
	\begin{pmatrix}
	A^\xi \\ A^\eta  
	\end{pmatrix} =
	\begin{pmatrix}
	0 & CD/\delta^{1/2}\\
	-1 & XY/\delta^{1/2}\\
	\end{pmatrix}
	\begin{pmatrix}
	A^\theta \\ A^\phi 
	\end{pmatrix}
	\label{eq: spherical to equator}
	\end{equation}
	
	\begin{equation}
	\begin{pmatrix}
	A^\theta \\ A^\phi 
	\end{pmatrix}
	=
	\begin{pmatrix}
	XY/CD & -1 \\
	\delta^{1/2}/CD & 0  \\
	\end{pmatrix}
	\begin{pmatrix}
	A^\xi \\ A^\eta  
	\end{pmatrix}
	\label{eq: equator to spherical}
	\end{equation}
	
	\item Patch V (North)
	\begin{equation}
	\begin{pmatrix}
	A^\xi \\ A^\eta  
	\end{pmatrix} =\frac{1}{\big(\delta-1 \big)^{1/2}} \begin{pmatrix}
	DX & -DY/\delta^{1/2}\\
	CY & CX/\delta^{1/2}  \\
	\end{pmatrix}
	\begin{pmatrix}
	A^\theta \\ A^\phi 
	\end{pmatrix} 
	\label{eq: spherical to patch 5}
	\end{equation}
	
	\begin{equation}
	\begin{pmatrix}
	A^\theta \\ A^\phi 
	\end{pmatrix}
	= \frac{1}{(\delta-1)^{1/2}}
	\begin{pmatrix}
	X/D & Y/C \\
	-Y \delta^{1/2} /D   & X \delta^{1/2} /C\\
	\end{pmatrix}
	\begin{pmatrix}
	A^\xi \\ A^\eta  
	\end{pmatrix}
	\label{eq: patch 5 to spherical}
	\end{equation}
	
	\item Patch VI (South)
	\begin{equation}
	\begin{pmatrix}
	A^\xi \\ A^\eta  
	\end{pmatrix} =\frac{1}{\big(\delta-1 \big)^{1/2}} \begin{pmatrix}
	- DX & DY/\delta^{1/2}\\
	- CY & -CX/\delta^{1/2}  \\
	\end{pmatrix}
	\begin{pmatrix}
	A^\theta \\ A^\phi 
	\end{pmatrix} 
	\label{eq: spherical to patch 6}
	\end{equation}
	
	\begin{equation}
	\begin{pmatrix}
	A^\theta \\ A^\phi 
	\end{pmatrix}
	=  \frac{1}{(\delta-1)^{1/2}}
	\begin{pmatrix}
	-X/D & - Y/C \\
	Y\delta^{1/2}/D   & - X\delta^{1/2}/C\\
	\end{pmatrix}  \begin{pmatrix}
	A^\xi \\ A^\eta  
	\end{pmatrix}
	\label{eq: patch 6 to spherical}
	\end{equation}
	
\end{itemize}
Remember that the quantities $X$, $Y$, $D$, $C$, $\delta$ are functions of $\xi$ and $\eta$, therefore the Jacobian depends on the location on the patch. Note also that in the equatorial patches vectors transform in the same way, due to the symmetry by construction of the four patches ($\xi$ and $\eta$ are directed in the same way in the four patches, so that their mutual interfaces are along the $\eta$ direction). This is not the case for the polar patches. 

On the axis, the angular components of the vectors in the spherical coordinates and the Jacobians above are ill-defined and thus they are not used. Therefore, when a spherical to cubed-sphere transformation is needed (boundary conditions, initial field given in spherical coordinates), the angular components in the cubed-sphere coordinates are averaged using the 8 closest neighbours in the tangential direction surrounding the axis point at a given radial layer.

At each patch edge, to go from the coordinate system of the adjacent patch to that of the original patch, we use a Jacobian matrix in order to calculate the vectors at the ghost cells and at the border. The Jacobian is built passing through spherical coordinates, e.g.: to go from the north patch to an equatorial patch, $\text{JAC}$ is a multiplication of the Jacobian needed to go from the north patch to spherical coordinates (eq. \ref{eq: patch 5 to spherical}) and the Jacobian needed to go from spherical coordinates to an equatorial patch in cubed-sphere coordinates (eq. \ref{eq: spherical to equator}); instead from an equatorial patch to Patch VI, $\text{JAC}$ is a multiplication of eq. \ref{eq: equator to spherical} and eq. \ref{eq: spherical to patch 6}.

% introduced in eq. \ref{eq: patch 5 to spherical}
% $\text{JAC}_\text{n-sph}$ (eq. \ref{eq: patch 5 to spherical}) and $\text{JAC}_\text{sph-eq}$ (eq. \ref{eq: spherical to equator});
% from an equatorial patch to Patch VI, $\text{JAC}$ is a multiplication of $\text{JAC}_\text{eq-sph}$ (eq. \ref{eq: equator to spherical}) and $\text{JAC}_\text{sph-s}$ (eq. \ref{eq: spherical to patch 6}) 

\subsection{Dot Product}
Considering the metric tensor defined in Eq. (\ref{eq: metric tensor}), the dot product is given by:
\begin{eqnarray}
      \boldsymbol{a}\cdot\boldsymbol{b}&=& \begin{pmatrix}
a^r &a^\xi  & a^\eta 
\end{pmatrix} \begin{pmatrix}
1 &0  & 0\\
 0 & 1 &  - \frac{XY}{CD} \\
0 &  - \frac{XY}{CD} & 1
\end{pmatrix} \begin{pmatrix}
b^r \\ b^\xi \\b^\eta \end{pmatrix} 
      \nonumber\\ 
   &=&    a^{r} b^{r}  + a^{\xi} b^{\xi}+ a^{\eta} b^{\eta} - \frac{XY}{CD} \big(a^{\xi} b^{\eta}+ a^{\eta} b^{\xi} \big).\nonumber\\ 
   \label{eq: dot product}
\end{eqnarray}
The mixing term $- \frac{XY}{CD} \big(a^{\xi} b^{\eta}+ a^{\eta} b^{\xi} \big)$, is due to the fact that $\hat{e}_\xi \cdot \hat{e}_\eta \neq 0$ since these two unit vectors are non-orthogonal. Therefore, the off-diagonal terms are different from zero.

\subsection{Cross Product}

The contravariant component of the cross product is given by
\begin{equation}
    (\boldsymbol{a} \times \boldsymbol{b})^{l} = a^{i} b^{j} g^{kl} \tilde{\varepsilon}_{ijk},
          \label{eq: contravariant cross product}
    \end{equation}
where $g^{kl}$ is the inverse of the metric, $\tilde{\varepsilon}_{ijk}= \sqrt{g} \hspace{0.5mm}\varepsilon_{ijk} $ is the covariant Levi-Civita tensor, $\varepsilon_{ijk}$ is the usual Levi-Civita symbol, and $\sqrt{g}$ is the square root of the determinant of the metric.
The contravariant component of the cross product is then written as follows:
\begin{eqnarray}
 \big(  \boldsymbol{a}\times \boldsymbol{b}\big)^l
      &=& 
         \frac{\delta^{1/2}}{ CD} \big( a^{\xi} b^{\eta} - a^{\eta} b^{\xi} \big) \hat{e}_r
              \nonumber\\
      && +
      \frac{1}{\delta^{1/2}} \bigg(CD \big( a^{\eta} b^{r} - a^{r} b^{\eta} \big)+ XY  \big( a^{r} b^{\xi} - a^{\xi} b^{r} \big) \bigg) \hat{e}_\xi
      \nonumber\\
      && + \frac{1}{\delta^{1/2}} \bigg( XY \big( a^{\eta} b^{r} - a^{r} b^{\eta} \big)+ CD  \big( a^{r} b^{\xi} - a^{\xi} b^{r} \big)  \bigg) \hat{e}_\eta. \nonumber\\
      \label{eq: contravariant cross product our metric}
\end{eqnarray}
The covariant components of the cross product are:
    \begin{equation}
   (\boldsymbol{a} \times \boldsymbol{b})_{l} =  \tilde{\varepsilon}_{uvl} a^{u} b^{v},
    \label{eq: covariant cross product}  
    \end{equation}
which, using the metric tensor of eq. (\ref{eq: metric tensor}), read
 \begin{eqnarray}
    \big(\boldsymbol{a}\times \boldsymbol{b}\big)_l &=& \frac{\delta^{1/2}}{CD}  \big(a^\xi b^\eta - a^\eta b^\xi \big) \boldsymbol{e}^r
        \nonumber\\
    && + \frac{\delta^{1/2}}{CD} \big(a^\eta b^r - a^r b^\eta \big)  \boldsymbol{e}^\xi
    \nonumber\\
    && + \frac{\delta^{1/2}}{CD} \big(a^r b^\xi - a^\xi b^r \big) \boldsymbol{e}^\eta.
    \label{eq: covariant cross product metric} 
 \end{eqnarray}
In our work, the covariant components of the cross product are used to calculate the covariant surface components (section \ref{subsec: Metric}) used in the curl operator in section \ref{subsec: finite Volume schemes}.

\section{Magnetic field formalism}

\subsection{Poloidal and toroidal decomposition}
\label{appendix: Poloidal and toroidal decomposition}
In MHD, different formalisms can describe the magnetic field. Here we describe the most common notations found in the literature. For any three-dimensional, solenoidal vector field $\boldsymbol{B}$ , like the magnetic field, we can always introduce the vector potential $\boldsymbol{A}$ so that
\begin{equation}
    \boldsymbol{B} = \boldsymbol{\nabla} \times \boldsymbol{A}
\end{equation}
$\boldsymbol{B}$ can be expressed by two scalar functions $\Phi(x)$ and $\Psi(x)$  that define its poloidal and toroidal components as follows:
  \begin{eqnarray}
      \boldsymbol{B}_\text{pol} &=&  
    %   \boldsymbol{\nabla} \times \big( \boldsymbol{\nabla} \times \Phi \boldsymbol{r}   \big)=  - \boldsymbol{r} \boldsymbol{\nabla}^2 \Phi + \boldsymbol{\nabla} \bigg(e^{-\lambda(r)}\frac{\partial (r \Phi)}{\partial r} \bigg)
    \boldsymbol{\nabla} \times \big( \boldsymbol{\nabla} \times \Phi \boldsymbol{k}   \big),
      \nonumber\\
        \boldsymbol{B}_\text{tor} &=& 
        % \boldsymbol{\nabla} \Psi  \times \boldsymbol{r}
        \boldsymbol{\nabla}\times \Psi \boldsymbol{k},
        \label{eq: poloidal toroidal field components}
  \end{eqnarray}
where $\boldsymbol{k}$ is an arbitrary vector. This decomposition is useful in problems where $\boldsymbol{k}$ can be taken to be normal to the physical boundaries, and the boundary conditions in the toroidal direction are periodic. Therefore, for a spherical domain, and using the cubed-sphere coordinates $(r, \xi, \eta)$, the suitable choice is $\boldsymbol{k} = \boldsymbol{r}$. 

Using the notation of \cite{geppert1991}, the basic idea is to expand the poloidal $\Phi$ and toroidal $\Psi$ scalar functions in a series of spherical harmonics at time zero in order to define the initial conditions. Expanding the two scalar functions $\Phi$ and $\Psi$ as a series of spherical harmonics we have:
  \begin{eqnarray}
    \Phi(t,r,\theta,\phi) &=& \frac{1}{r}\sum_{l,m} \Phi_{lm}(r,t) Y_{lm}(\theta,\phi)  \nonumber\\
    \Psi(t,r,\theta,\phi)  &=& \frac{1}{r}\sum_{l,m} \Psi_{lm}(r,t) Y_{lm}(\theta,\phi)
       \label{eq: phi and Psi scalar functions}
  \end{eqnarray}
where $l= 1, ...,l_{max}$ is the degree and $m=-l,...,l$ the order of the multipole. Note that in 3D, the toroidal field is a mix of the two tangential components of the magnetic field, whereas the poloidal field is a mix of the three components. This is less trivial than in 2D, where the toroidal part consists of the azimuthal component and the poloidal part consists of the two other components of the magnetic field.

Combining the poloidal and toroidal components of the magnetic field, one can express the three components of the magnetic field in spherical coordinates as 
\begin{eqnarray}
 B^{r} &=& \frac{1}{r^2} \sum_{lm} l(l+1)  \Phi_{lm}(r) Y_{lm}(\theta,\phi)   \nonumber\\
 B^{\theta} &=&   \frac{1}{r} \sum_{lm} \Phi'_{lm}(r) \frac{\partial Y_{lm}(\theta,\phi)}{\partial \theta} +  \frac{1 }{r sin\theta} \sum_{lm} \Psi_{lm}(r) \frac{\partial Y_{lm}(\theta,\phi)}{ \partial \phi}
 \nonumber\\
  B^{\phi} &=&    - \frac{1}{r}\sum_{lm} \Psi_{lm}(r) \frac{\partial Y_{lm}(\theta,\phi)}{ \partial \theta}  + \frac{1}{r sin\theta}  \sum_{lm}  \Phi'_{lm}(r) \frac{\partial Y_{lm}(\theta,\phi)}{\partial \phi} .  \nonumber\\
 \label{eq: magnetic field components spectral}
\end{eqnarray}
 with 
 \begin{equation}
\Phi'_{lm}= e^{-\lambda} \frac{\partial \Phi_{lm}}{\partial r} + \frac{1 - e^{-\lambda}}{r} \Phi_{lm}
\label{eq: radial derivative of the Phi scalar function}
 \end{equation}

To determine the spectral energy distribution (eq. \ref{eq: spectral magnetic energy}), one needs to reconstruct the three radial scalar functions, $\Phi_{lm}$, $\Phi'_{lm}$, and $\Psi_{lm}$ defined as:  
\begin{equation}
     \Phi_{lm}(r) =   \sum_{lm} \frac{1}{l(l+1) } \int dS_r B^r Y_{lm} (\theta,\phi),
\end{equation}

\begin{equation}
\Phi'_{lm}(r) =  \frac{1}{r} \sum_{lm} \frac{1}{l(l+1)} \int dS_r \bigg(B^{\theta}\frac{\partial Y_{lm}}{\partial \theta}  +  \frac{B^{\phi}}{sin\theta} \frac{\partial Y_{lm}}{\partial \phi}  \bigg), 
\end{equation}
and 
  \begin{equation}
      \Psi_{lm}(r) =  \frac{1}{r}\sum_{lm} \frac{1}{l(l+1)} \int dS_r \bigg( \frac{B^{\theta}}{sin\theta}\frac{\partial Y_{lm}}{\partial \phi}  -  B^{\phi} \frac{\partial Y_{lm}}{\partial \theta}  \bigg). 
\end{equation}

\subsection{Initial models}
\label{appendix: initial conditions}

The initial topology of the magnetic field can be constructed by choosing a set of spherical harmonics, which define our topology. For instance for a dipole, we use $Y_{l=1, m}$, for a quadrupole $Y_{l=2, m}$, whereas a multipolar topology can be simply constructed by summing several spherical harmonics $\sum_{lm} Y_{lm} (\theta,\phi)$. 

The set of spherical harmonics defines the angular part of the magnetic field topology. One has the freedom of choosing the desired set of spherical harmonics. In our study, we impose potential magnetic boundary conditions, and we use a set of radial scalar functions $\Phi_{lm}(r),\Psi_{lm}(r)$ that smoothly match the potential boundary conditions.

For simplicity, we impose the radial profile of the dipolar poloidal scalar function, $\Phi_{l=1,m}(r)$, as in eq. (8) of \cite{aguilera2008}:
\begin{equation}
    \Phi_{l=1,m}(r) = \Phi_0 \mu r (a + tan(\mu R_\star) b )
    \label{eq: funa function}
\end{equation}
where $\Phi_0$ is the normalization and 
\begin{equation}
    a= \frac{sin(\mu r)}{(\mu r)^2} - \frac{cos(\mu r)}{\mu r}, \hspace{5mm} b= -  \frac{cos(\mu r)}{(\mu r)^2} - \frac{sin(\mu r)}{(\mu r )},
\end{equation}
$\mu$ is a parameter related to the magnetic field curvature, that needs to be found for a given surface radius $R_\star$.

For higher-order multipoles of the poloidal scalar function $(l>1)$, and for all the toroidal scalar function contributions $\Psi_{lm}(r)$, we confine them inside the crust of a NS as follows
\begin{equation}
     \Phi_{l>1,m}(r), \Psi_{lm}(r) \propto - \big(R -r \big)^2 \big( r-R_c\big)^2 
\end{equation}
where the proportionality means that every multipole can have a different normalization (i.e., its initial weight).

%with R being the radius of a NS and in our case we fix $R=11$ km, and $R_c=10$ km and it corresponds to the crust-core interface. 
From $\Phi_{lm}(r)$ and $\Psi_{lm}(r)$, we build the magnetic field components, defined by eqs. (\ref{eq: poloidal toroidal field components}), using the curl operator in cubed-sphere coordinates, eqs. (\ref{eq: stokes theorem radial component})-(\ref{eq: stokes theorem eta component}). Such a construction ensure that the initial topology of the magnetic field is divergence-free up to machine error, and has no axis-singularity problem.

% Don't change these lines
\bsp	% typesetting comment
\label{lastpage}
\end{document}